\documentclass[aps,prd,showpacs,twocolumn,floatfix,nofootinbib]{revtex4-1}
\usepackage[dvipdfmx]{graphicx}
\usepackage{amsmath,amssymb,siunitx}
\usepackage{color}

\begin{document}
\title{Neutrino transport in black hole-neutron star binaries: Neutrino
emission and dynamical mass ejection}
\author{
Koutarou Kyutoku$^{1,2,3,4}$,
Kenta Kiuchi$^4$,
Yuichiro Sekiguchi$^{5,4}$,
Masaru Shibata$^{6,4}$,
and
Keisuke Taniguchi$^7$
}
\affiliation{
$^1$Theory Center, Institute of Particle and Nuclear Studies, KEK,
Tsukuba 305-0801, Japan\\
$^2$Department of Particle and Nuclear Physics, the Graduate University
for Advanced Studies (Sokendai), Tsukuba 305-0801, Japan\\
$^3$Interdisciplinary Theoretical Science (iTHES) Research Group, RIKEN,
Wako, Saitama 351-0198, Japan\\
$^4$Center for Gravitational Physics, Yukawa Institute for Theoretical
Physics, Kyoto University, Kyoto 606-8502, Japan\\
$^5$Department of Physics, Toho University, Funabashi, Chiba 274-8510,
Japan\\
$^6$Max Planck Institute for Gravitational Physics (Albert Einstein
Institute), Am M{\"u}hlenberg 1, Postdam-Golm 14476, Germany\\
$^7$Department of Physics, University of the Ryukyus, Nishihara, Okinawa
903-0213, Japan
}
\date{\today}

\begin{abstract}
 We study the merger of black hole-neutron star binaries by fully
 general-relativistic neutrino-radiation-hydrodynamics simulations
 throughout the coalescence, particularly focusing on the role of
 neutrino irradiation in dynamical mass ejection. Neutrino transport is
 incorporated by an approximate transfer scheme based on the truncated
 moment formalism. While we fix the mass ratio of the black hole to the
 neutron star to be 4 and the dimensionless spin parameter of the black
 hole to be 0.75, the equations of state for finite-temperature
 neutron-star matter are varied. The hot accretion disk formed after
 tidal disruption of the neutron star emits a copious amount of
 neutrinos with the peak total luminosity $\sim
 1$--\SI{3e53}{erg.s^{-1}} via thermal pair production and subsequent
 electron/positron captures on free nucleons. Nevertheless, the neutrino
 irradiation does not modify significantly the electron fraction of the
 dynamical ejecta from the neutrinoless $\beta$-equilibrium value at
 zero temperature of initial neutron stars. The mass of the wind
 component driven by neutrinos from the remnant disk is negligible
 compared to the very neutron-rich dynamical component, throughout our
 simulations performed until a few tens milliseconds after the onset of
 merger, for the models considered in this study. These facts suggest
 that the ejecta from black hole-neutron star binaries are very neutron
 rich and are expected to accommodate strong \textit{r}-process
 nucleosynthesis, unless magnetic or viscous processes contribute
 substantially to the mass ejection from the disk. We also find that the
 peak neutrino luminosity does not necessarily increase as the disk mass
 increases, because tidal disruption of a compact neutron star can
 result in a remnant disk with a small mass but high temperature.
\end{abstract}

\maketitle

\section{Introduction} \label{sec:intro}

Mass ejection from the merger of black hole-neutron star binaries has
been studied vigorously in recent years
\cite{rosswog2005,rantsiou_klr2008,kyutoku_ost2011,rosswog_pn2013,foucart_ddkmopsst2013,lovelace_dfkpss2013,kyutoku_is2013,deaton_dfookmss2013,foucart_etal2014,kawaguchi_knost2015,kyutoku_iost2015,kiuchi_skstw2015,just_bagj2015,foucart_dbdkhkps2017},
as well as that from binary-neutron-star mergers
\cite{hotokezaka_kkosst2013,bauswein_gj2013,sekiguchi_kks2015,palenzuela_lnlcoa2015,sekiguchi_kkst2016,radice_glror2016,foucart_orkps2016}. Neutron-rich
material ejected from black hole-neutron star binaries can accommodate
\textit{r}-process nucleosynthesis, which produces about a half of
neutron-rich nuclei heavier than the iron in the Universe
\cite{lattimer_schramm1974}. This nucleosynthesis is important not only
for cosmic chemical evolution but also for electromagnetic transient
radiation, which serves as electromagnetic counterparts to
gravitational-wave sources \cite{kyutoku_is2013}. In particular, the
decay of unstable \textit{r}-process elements heats up the ejecta, and
associated quasithermal radiation, so-called macronova/kilonova, will be
observed in infrared-optical bands on the time scale of a week
\cite{li_paczynski1998,kulkarni2005,metzger_mdqaktnpz2010,tanaka_hkwkss2014,kawaguchi_kst2016}. While
five gravitational-wave events from binary black holes were detected
successfully
\cite{ligovirgo2016,ligovirgo2016-4,ligovirgo2017,ligovirgo2017-2,ligovirgo2017-4},
the nature of many weak signal candidates including LVT151012
\cite{ligovirgo2016-5} remains elusive. If we could simultaneously
detect accompanying electromagnetic counterparts, gravitational waves
from the merger of black hole-neutron star binaries may be securely
identified as astrophysical.

The mass ejection and electromagnetic counterparts are also important to
reveal the nature of short-hard gamma-ray bursts (see
Refs.~\cite{lee_ramirezruiz2007,nakar2007,berger2014} for reviews). A
near-infrared transient detected in the afterglow of GRB 130603B is
consistent with models of the macronova/kilonova
\cite{tanvir_lfhhwt2013,berger_fc2013,hotokezaka_ktkssw2013}, supporting
the binary-merger hypothesis of short-hard gamma-ray bursts. While this
transient is detected only in a single epoch, future detailed
observations of the macronovae/kilonovae will give us valuable
information of their progenitors as well as mass ejection mechanisms,
and appropriate interpretation requires firm understanding of the ejecta
properties such as the mass and velocity. At the same time, if the
material is ejected along the polar axis of the remnant black hole and
surrounding accretion disk, the ejecta could also affect the propagation
of a possible gamma-ray burst jet via baryon loading and hydrodynamical
collimation
\cite{nagakura_hssi2014,murguiabertheir_mrdl2014,duffell_qm2015,just_ojbs2016,murguiabertheir_rmdrrtpl2017}. Thus,
the geometry of mass ejection is also a subject of investigation for
accurately understanding the short-hard gamma-ray burst.

Indeed, the first multimessenger detections of a binary-neutron-star
merger event, GW170817/GRB~170817A/AT~2017gfo, was announced during the
review process of this article
\cite{ligovirgo2017-3,ligovirgoem2017,ligovirgogamma2017}. The outlook
envisioned above is now broadly confirmed for binary neutron stars
except that GRB 170817A is not likely to be associated with an
ultrarelativistic jet \cite{mooley_etal2017,ruan_nhke2017}. Still, the
mechanism of mass ejection needs clarification to understand observed
electromagnetic emission. In anticipation of forthcoming detections of
black hole-neutron star binaries, detailed studies of their mass
ejection have also become increasingly important.

General relativity, radiation hydrodynamics, and neutrino transport are
all essential to study quantitatively the mass ejection from compact
binary mergers and subsequent nucleosynthesis. Recent
neutrino-radiation-hydrodynamics simulations in full general relativity
have shown that the electron fraction $Y_e$ of dynamical ejecta from
binary neutron stars can have a broad distribution with $Y_e \sim
0.05$--$0.5$ due to the strong shock heating and neutrino irradiation
\cite{wanajo_snkks2014,sekiguchi_kks2015,sekiguchi_kkst2016,radice_glror2016}. This
should be contrasted with Newtonian or approximately
general-relativistic simulations, by which the distribution is always
predicted to be concentrated in $Y_e \lesssim 0.1$, characteristic of
neutron stars in neutrinoless $\beta$-equilibrium at zero temperature
\cite{freiburghaus_rt1999,goriely_bj2011,korobkin_raw2012}. Broad
distribution of $Y_e$ is a key to reproduce the solar pattern of
\textit{r}-process abundances in a single event
\cite{wanajo_snkks2014,goriely_bjpj2015,roberts_ldfflnop2017} and may be
advantageous to explain the universality of the abundance pattern
observed in \textit{r}-process-enhanced metal-poor stars
\cite{sneden_cg2008,siqueiramello_etal2014}. Furthermore, neutrino
irradiation can also trigger neutrino-driven mass ejection, or
neutrino-driven wind, in the postmerger phase
\cite{dessart_obrl2009,perego_rckkal2014,martin_patkr2015,fujibayashi_sks2017}
(see also
Refs.~\cite{metzger_fernandez2014,just_bagj2015,sekiguchi_kks2015,foucart_hdoorklps2016,sekiguchi_kkst2016,radice_glror2016,foucart_orkps2016}). Even
if the dynamical ejecta do not have sufficiently broad distribution of
$Y_e$, such disk winds could play a crucial role in reproducing the
solar pattern \cite{just_bagj2015,wu_fmm2016}.

In this paper, we investigate the role of neutrino-matter interaction
during the merger process of black hole-neutron star binaries by
numerical-relativity simulations. While fully general-relativistic
simulations of the mass ejection from black hole-neutron star binaries
have been extensively performed adopting nuclear-theory-based equations
of state by various researchers
\cite{kyutoku_is2013,deaton_dfookmss2013,foucart_etal2014,kawaguchi_knost2015,kyutoku_iost2015,kiuchi_skstw2015,foucart_dbdkhkps2017},
the neutrino transport has been limited to the cooling
\cite{deaton_dfookmss2013,foucart_etal2014,foucart_dbdkhkps2017} except
for the study focusing only on the postmerger evolution
\cite{foucart_etal2015}. Our simulations incorporate both neutrino
cooling and heating in an approximate but self-consistent manner
throughout the inspiral-merger-postmerger phases in full general
relativity. Various nuclear-theory-based finite-temperature equations of
state are adopted to explore dependence of the merger outcome on
underlying microphysics. In this study, we focus particularly on the
neutrino emission and mass ejection from black hole-neutron star binary
mergers.

This paper is organized as follows. Section \ref{sec:method} describes
our numerical scheme and adopted models. Results of
neutrino-radiation-hydrodynamics simulations are presented in
Sec.~\ref{sec:result}, and they are compared with results of simulations
without neutrino absorption in Sec.~\ref{sec:nh} to single out the
effect of neutrino transport. Sections~\ref{sec:discussion} and
\ref{sec:summary} are devoted to discussions and a summary,
respectively. The gravitational constant and speed of light are denoted
by $G$ and $c$, respectively. The temperature is expressed in
\si{\mega\electronvolt}, implicitly multiplying the Boltzmann constant,
$k_\mathrm{B}$. The greek indices $\alpha$ and $\beta$ denote spacetime
components, and the latin index $i$ denotes space components.

\section{Numerical method} \label{sec:method}

\subsection{Equation of state and initial condition}

\begin{table}
 \caption{Key ingredients of the adopted equations of
 state. $M_\mathrm{max}$ is the maximum mass of a spherical neutron star
 at zero temperature. $M_{*,1.35}$, $R_{1.35}$, $\mathcal{C}_{1.35}$,
 and $\Lambda_{1.35}$ are the baryon rest mass $M_*$, circumferential
 radius $R$, and compactness $\mathcal{C} \equiv GM_\mathrm{NS} / (R
 c^2)$ with $M_\mathrm{NS}$ being the gravitational mass, and
 dimensionless tidal deformability $\Lambda$ of a $1.35 M_\odot$ neutron
 star, respectively.}
 \begin{tabular}{c|ccccc} \hline
  Model & $M_\mathrm{max} [ M_\odot ]$ & $M_{*,1.35} [ M_\odot ]$ &
  $R_{1.35} ( \si{km} ) $ & $\mathcal{C}_{1.35}$ & $\Lambda_{1.35}$ \\
  \hline \hline
  SFHo & 2.06 & 1.48 & 11.9 & 0.167 & 420 \\
  DD2 & 2.42 & 1.47 & 13.2 & 0.151 & 854 \\
  TM1 & 2.21 & 1.45 & 14.5 & 0.138 & 1428 \\
  \hline
 \end{tabular}
 \label{table:eos}
\end{table}

We adopt three nuclear-theory-based finite-temperature equations of
state for the neutron-star matter to span a plausible range of
nuclear-matter properties following our previous work on binary neutron
stars \cite{sekiguchi_kks2015}. Specifically, the so-called SFHo
\cite{steiner_hf2013}, DD2 \cite{banik_hb2014}, and TM1
\cite{hempel_fsl2012} equations of state are employed in this study. Our
choice is also the same as that for black hole-neutron star binary
mergers performed in a conformal flatness approximation of
Ref.~\cite{just_bagj2015}. The equations of state provide thermodynamic
quantities such as the pressure and entropy as functions of the
rest-mass density $\rho$, electron fraction $Y_e$, and temperature
$T$. Important properties of neutron stars computed with the adopted
equations of state are summarized in Table~\ref{table:eos} (see also
Ref.~\cite{sekiguchi_kks2015} for microphysics parameters). The radii of
$1.35M_\odot$ neutron stars at zero temperature are
\SI{11.9}{\kilo\meter}, \SI{13.2}{\kilo\meter}, and
\SI{14.5}{\kilo\meter} for SFHo, DD2, and TM1, respectively. While TM1
is marginally inconsistent with existing constraints
\cite{hebeler_lps2010,tews_lok2017},\footnote{TM1 is not favored from
the observation of GW170817 \cite{ligovirgo2017-3} either, which was
recently.} we include this model for exploring the dependence of merger
outcomes on equations of state in a wide range. All the equations of
state can support observed $2M_\odot$ neutron stars
\cite{demorest_prrh2010,antoniadis_etal2013}.

Initial data of black hole-neutron star binaries are given by
quasiequilibrium states computed in the puncture framework. Numerical
computations are performed using a multidomain spectral method library,
{\small LORENE},\footnote{http:\slash\slash{}www.lorene.obspm.fr\slash}
and the details are described in
Refs.~\cite{kyutoku_st2009,kyutoku_ost2011}. Neutrinoless
$\beta$-equilibrium states at \SI{0.1}{\mega\electronvolt} are assumed
for the initial neutron stars. The electron fraction typically takes
values of $0.05$--$0.1$ for the inner crust and the outer core, from
which dynamical ejecta arise \cite{kyutoku_iost2015}.

In this work, we fix all the parameters except for the neutron-star
equations of state for simplicity. The masses of black holes and neutron
stars are chosen to be $M_\mathrm{BH} = 5.4 M_\odot$ and $M_\mathrm{NS}
= 1.35 M_\odot$, respectively. Thus, the mass ratio $Q \equiv
M_\mathrm{BH} / M_\mathrm{NS}$ is 4. This relatively small value is
chosen so that the neutron star can be disrupted before the plunge in
the presence of a moderately large spin of the black hole
\cite{kyutoku_ost2011,kyutoku_iost2015}, while the mass is kept
astrophysically realistic \cite{kreidberg_bfk2012}. The dimensionless
spin parameter of the black hole is chosen to be $\chi \equiv c
J_\mathrm{BH} / ( G M_\mathrm{BH}^2 ) = 0.75$ with $J_\mathrm{BH}$ being
the angular momentum of the black hole, and the orientation is set to be
parallel to the orbital angular momentum of the binary. The initial
orbital angular velocity $\Omega$ of the binary is chosen to be $G m_0
\Omega / c^3 = 0.056$, with which binaries spend $\sim 3$--4 orbits
before tidal disruption of neutron stars.

\subsection{Radiation-hydrodynamics simulation}

Our numerical simulations are performed by a fully general-relativistic
neutrino-radiation-hydrodynamics code developed in
Refs.~\cite{sekiguchi_kks2015,sekiguchi_kkst2016}. The Einstein
evolution equations are solved in the Baumgarte-Shapiro-Shibata-Nakamura
formalism \cite{shibata_nakamura1995,baumgarte_shapiro1998} employing a
moving puncture gauge condition
\cite{campanelli_lmz2006,baker_cckm2006,marronetti_tbgs2008}. Radiation-hydrodynamics
equations for neutrinos are solved explicitly in time by an approximate
neutrino-transfer scheme based on the truncated moment formalism
\cite{thorne1981,shibata_kss2011} with cooling source terms computed by
a general-relativistic leakage scheme
\cite{sekiguchi2010,sekiguchi_shibata2011} and heating sources terms due
to neutrino capture processes (see also
Ref.~\cite{fujibayashi_sks2017}). Specifically, we decompose the
energy-momentum tensor and associated local conservation equations into
two parts. One part denoted by $T^{\alpha\beta}$ represents the sum of
the fluid and trapped neutrinos, for which the basic equation is given
by
\begin{equation}
 \nabla_\beta T^{\alpha \beta} = - Q_\mathrm{cool}^\alpha +
  Q_\mathrm{heat}^\alpha .
\end{equation}
The other part denoted by $T^{\alpha\beta}_{\nu,\mathrm{S}}$ represents
streaming neutrinos, and the basic equation is
\begin{equation}
 \nabla_\beta T_{\nu,\mathrm{S}}^{\alpha \beta} = Q_\mathrm{cool}^\alpha
  - Q_\mathrm{heat}^\alpha .
\end{equation}
Here, $Q^\alpha_\mathrm{cool}$ and $Q^\alpha_\mathrm{heat}$ are the
rates of neutrino emission (cooling) and absorption (heating),
respectively. The energy-momentum tensor of the fluid and trapped
neutrinos is written by the ideal-fluid form with the four velocity
$u^\alpha$ \cite{sekiguchi2010}, and that of streaming neutrinos is
given assuming an M1 closure relation with a variable Eddington factor
for handling gray regimes \cite{shibata_kss2011}. We also solve the
evolution of chemical composition incorporating both the neutrino
emission and absorption, applying a $\beta$-equilibrium limiter on the
electron fraction to avoid an unstable oscillation associated with stiff
source terms \cite{sekiguchi2010}. The floor density of artificial
atmosphere is set to be $\approx \SI{2e3}{\gram\per\cubic\cm}$ with $Y_e
= 0.47$ and $T = \SI{0.1}{\mega\electronvolt}$ in this work.

The emission and absorption rates are determined by an
optical-depth-weighted sum of the values for the diffusion and
free-streaming limits \cite{sekiguchi2010}. Our emission processes
include electron/positron captures on free nucleons, those on heavy
nuclei, pair annihilation, nucleon bremsstrahlung, and plasmon
decay. Precise forms of the emission rates are provided in
Ref.~\cite{sekiguchi_kks2012}. For the absorption, we consider neutrino
captures on free nucleons. The absorption rate is computed using the
cross section given in Ref.~\cite{sekiguchi_kks2012} assuming neutrinos
to obey the Fermi-Dirac distribution at the fluid temperature with
chemical potential that reproduces the energy density obtained from time
evolution. In fact, this assumption on the temperature is not always
well motivated, and we discuss this issue in Sec.~\ref{sec:nh}. We
collectively denote muon/tau neutrinos/antineutrinos as heavy-lepton
neutrinos $\nu_x$, all of which experience only the same set of
neutral-current interactions.

\begin{table}
 \caption{Grid setup of each model. $L$ is the box size of the largest
 computational domain. $\Delta x$ is the grid spacing at the finest
 computational domain. $R_\mathrm{diam} / \Delta x$ is the number of
 grids assigned to the semimajor diameter of the neutron star along the
 binary separation.}
 \begin{tabular}{c|ccc} \hline
  Model & $L$ (\si{km}) & $\Delta x$ (\si{m}) & $R_\mathrm{diam} /
  \Delta x$ \\
  \hline \hline
  SFHo & \num{17600} & 250 & 70 \\
  DD2 & \num{18040} & 270 & 75 \\
  DD2-low & \num{18432} & 400 & 52 \\
  TM1 & \num{19507} & 300 & 76 \\
  \hline
 \end{tabular}
 \label{table:model}
\end{table}

Our code implements a fixed-mesh-refinement algorithm to simultaneously
cover a large spatial region and resolve the compact
objects. Specifically, our computational domains consist of seven
cuboids centered at an approximate center of mass of the system, and
each of the cuboids has $(2N+1,2N+1,N+1)$ Cartesian grid points in
$(x,y,z)$ directions with the equatorial symmetry imposed at $z=0$. By
denoting the region covered by the largest and coarsest domain by
$[-L:L] \times [-L:L] \times [0:L]$, the grid spacing of the largest
domain is $(\Delta x)_0 = L/N$. The size and grid spacing of the
adjacent domain are halved simultaneously, and thus our smallest and
finest domain covers $[-L/2^6:L/2^6] \times [-L/2^6:L/2^6] \times
[0:L/2^6]$ with its grid spacing $\Delta x \equiv (\Delta x)_6 = L/(2^6
N)$. Precise values of $L$ and $\Delta x$ are presented in Table
\ref{table:model} for each model. For the DD2 model, we also perform a
low-resolution simulation denoted by DD2-low to check the resolution
dependence of our results (see Sec.~\ref{sec:result_res}).

\section{Result} \label{sec:result}

In this section, we describe the results of neutrino
radiation-hydrodynamics simulations of black hole-neutron star binary
mergers. Throughout this paper, we define the time of merger,
$t_\mathrm{merge}$, by the time when the baryon rest mass of $0.01
M_\odot$ is swallowed by the black hole (strictly speaking, the apparent
horizon) following our previous work
\cite{kyutoku_st2010,kyutoku_st2010e,kyutoku_ost2011,kawaguchi_knost2015,kyutoku_iost2015}. The
time of merger defined in this manner is earlier by up to
\SI{1.5}{\milli\second} than the time when half of the material is
swallowed (see, e.g., Refs.~\cite{foucart_dkt2011,foucart_etal2014}).

\subsection{Merger dynamics and remnant disk} \label{sec:result_disk[}

The inspiral and merger phases are essentially unaffected by the
neutrino transport (see Ref.~\cite{shibata_taniguchi2011} for reviews of
purely hydrodynamic simulations). Our initial data are chosen to
complete $\sim 3$--4 inspiral orbits taking $\sim
\SI{10}{\milli\second}$ before the onset of merger. For the binary
parameters adopted in this study, the neutron star is always disrupted
by the tidal force of the black hole before the plunge. While most of
the disrupted material immediately feeds the black hole, an outer part
forms a tidal tail spiraling around the central remnant. The outermost
part of the tail is ejected dynamically from the system, and we discuss
the mass ejection separately in Sec.~\ref{sec:result_ej}. The
irrelevance of neutrinos during this phase is expected and has been
found \cite{deaton_dfookmss2013,foucart_etal2014}, because no heating
mechanism raises the neutron-star temperature from the initial value
(except for the interaction with the artificial atmosphere, which is
found to be dynamically unimportant in our simulations).

\begin{figure*}
 \includegraphics[width=.95\linewidth]{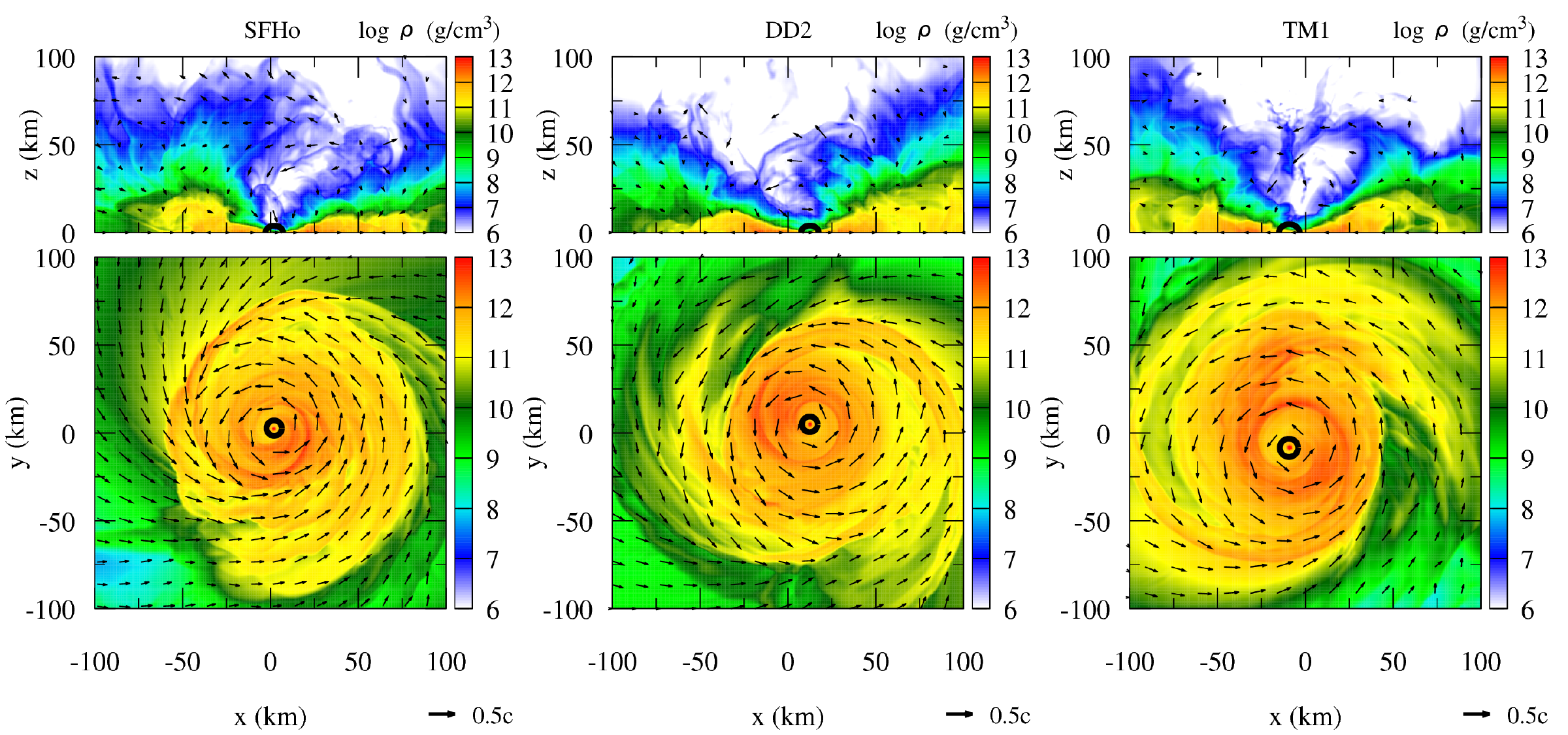} \caption{Rest-mass
 density profiles in the central region at \SI{10}{\milli\second} after
 the onset of merger, which approximately corresponds to the peak of
 neutrino luminosity. The black circles show the apparent horizons. The
 velocity vector $v^i \equiv u^i / u^t$ is overplotted.}
 \label{fig:snapdiskrho}
\end{figure*}

\begin{figure*}
 \includegraphics[width=.95\linewidth]{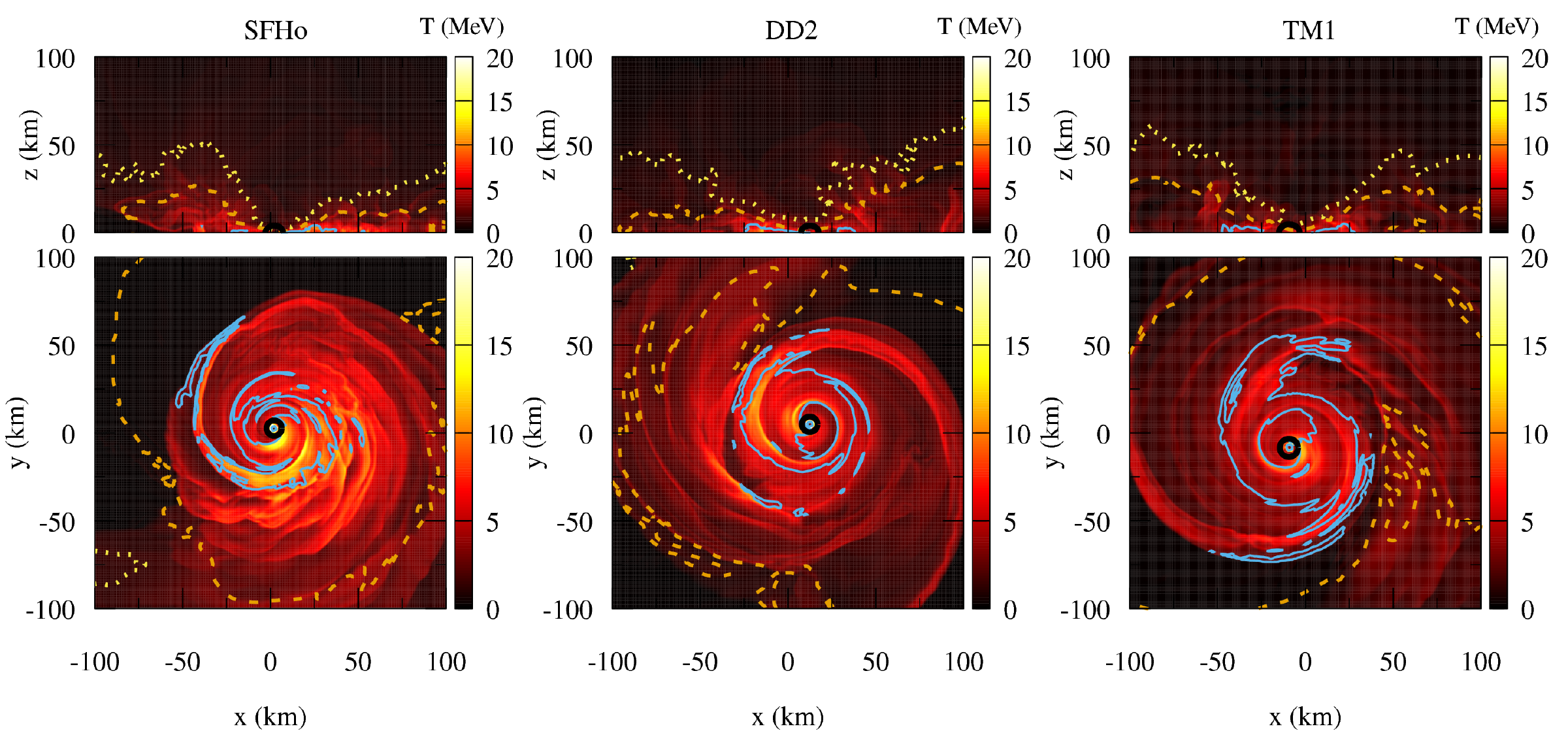} \caption{The same as
 Fig.~\ref{fig:snapdiskrho} but for the temperature. We also show
 isodensity contours for $\rho = \num{e8}$ (yellow dotted), \num{e10}
 (orange dashed), and \SI{e12}{\gram\per\cubic\centi\meter} (light blue
 solid).} \label{fig:snapdiskt}
\end{figure*}

\begin{figure*}
 \includegraphics[width=.95\linewidth]{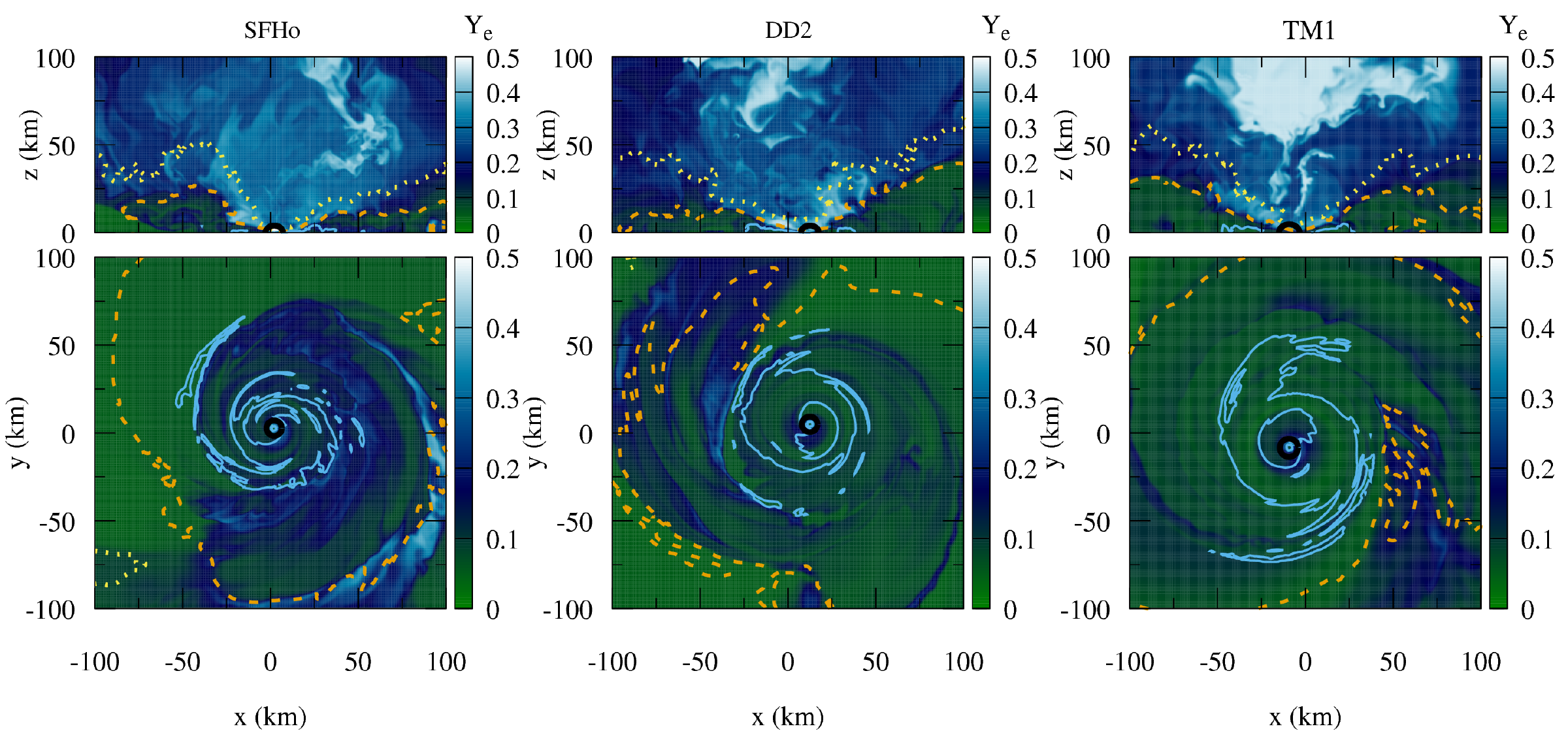} \caption{The same as
 Fig.~\ref{fig:snapdiskrho} but for the electron fraction. We also show
 isodensity contours for $\rho = \num{e8}$ (yellow dotted), \num{e10}
 (orange dashed), and \SI{e12}{\gram\per\cubic\centi\meter} (light blue
 solid). The mass of material with a high electron fraction in the polar
 region is very small.} \label{fig:snapdiskye}
\end{figure*}

Once the tidal tail collides itself and the shock heating sets in, a
remnant disk is formed with typical temperature of $\gtrsim$
\SI{10}{\mega\electronvolt} and neutrinos begin to emerge. Figures
\ref{fig:snapdiskrho}, \ref{fig:snapdiskt}, and \ref{fig:snapdiskye}
show the rest-mass density, temperature, and electron fraction,
respectively, in the central region at \SI{10}{\milli\second} after the
onset of merger. The rest-mass density inside the remnant disk exceeds
\SI{e12}{\gram\per\cubic\centi\meter} for the models considered in this
study. The electron fraction in the dense part is $\lesssim 0.2$ at this
time even though the temperature is as high as
\SI{10}{\mega\electronvolt}. Figure \ref{fig:snapdiskye} shows that the
electron fraction in the polar region is moderately high, but the mass
of such material is very small as the isodensity contours imply.

Remnant disks formed after black hole-neutron star binary mergers show
qualitative differences from equilibrium tori commonly adopted as
initial conditions of black hole-disk simulations
\cite{fernandez_metzger2013,fernandez_kmq2015,just_bagj2015,siegel_metzger2017}. The
angular momentum profile is close to Keplerian particularly in the
innermost region \cite{foucart_ddkmopsst2013,foucart_etal2014}, and the
disk is geometrically thinner than an equilibrium torus with constant
specific angular momentum. We also find that the remnant disk formed
from a compact neutron star such as SFHo exhibits an extended region
with $Y_e \approx 0.2$, which is higher than $Y_e = 0.1$ commonly
adopted as initial conditions of black hole-disk simulations (see also
Ref.~\cite{fernandez_fkldr2017}). These features should have various
implications for the postmerger dynamics as we discuss in
Sec.~\ref{sec:discussion}.

\subsection{Neutrino emission} \label{sec:result_nu}

\begin{figure}
 \begin{tabular}{c}
  \includegraphics[width=.95\linewidth]{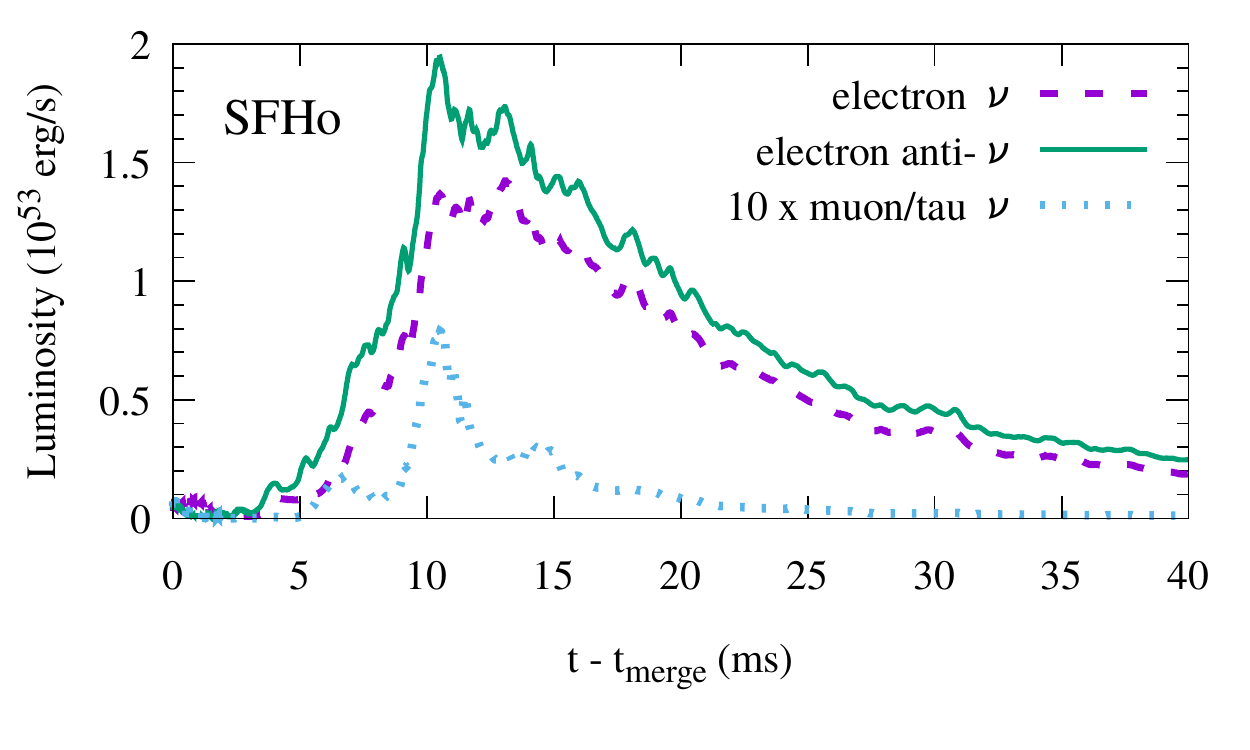} \\
  \includegraphics[width=.95\linewidth]{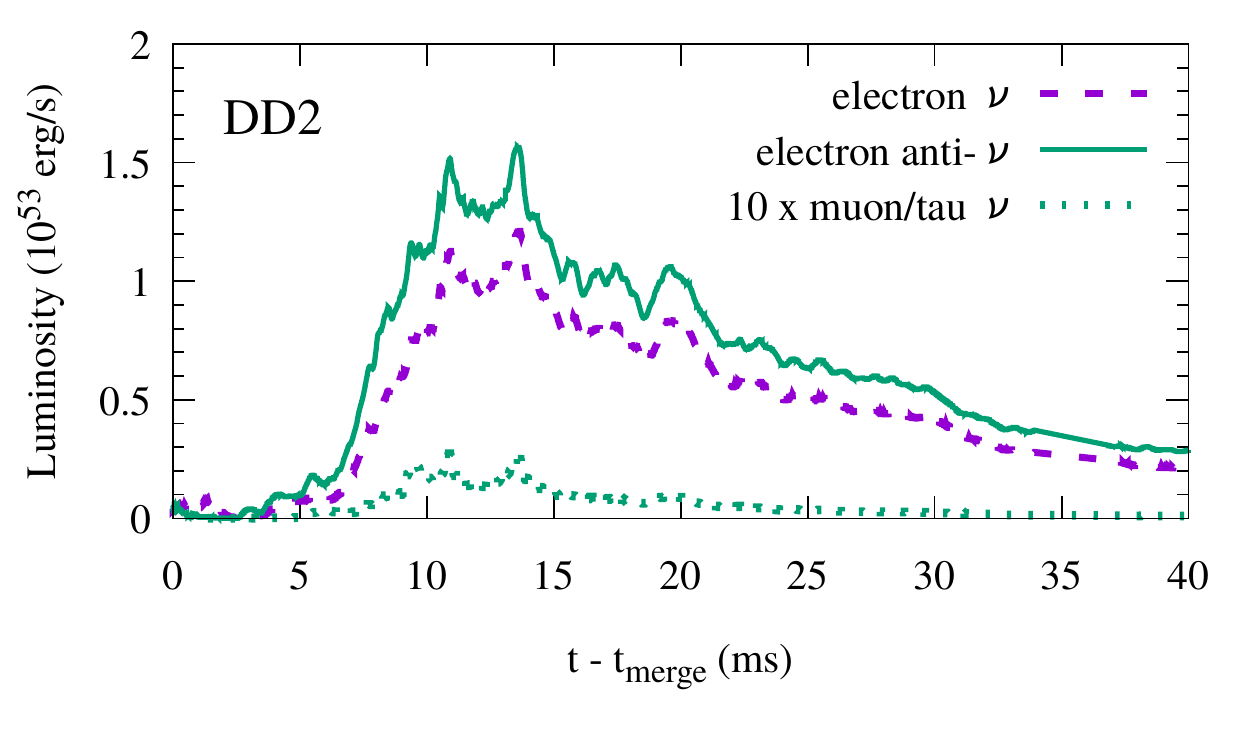} \\
  \includegraphics[width=.95\linewidth]{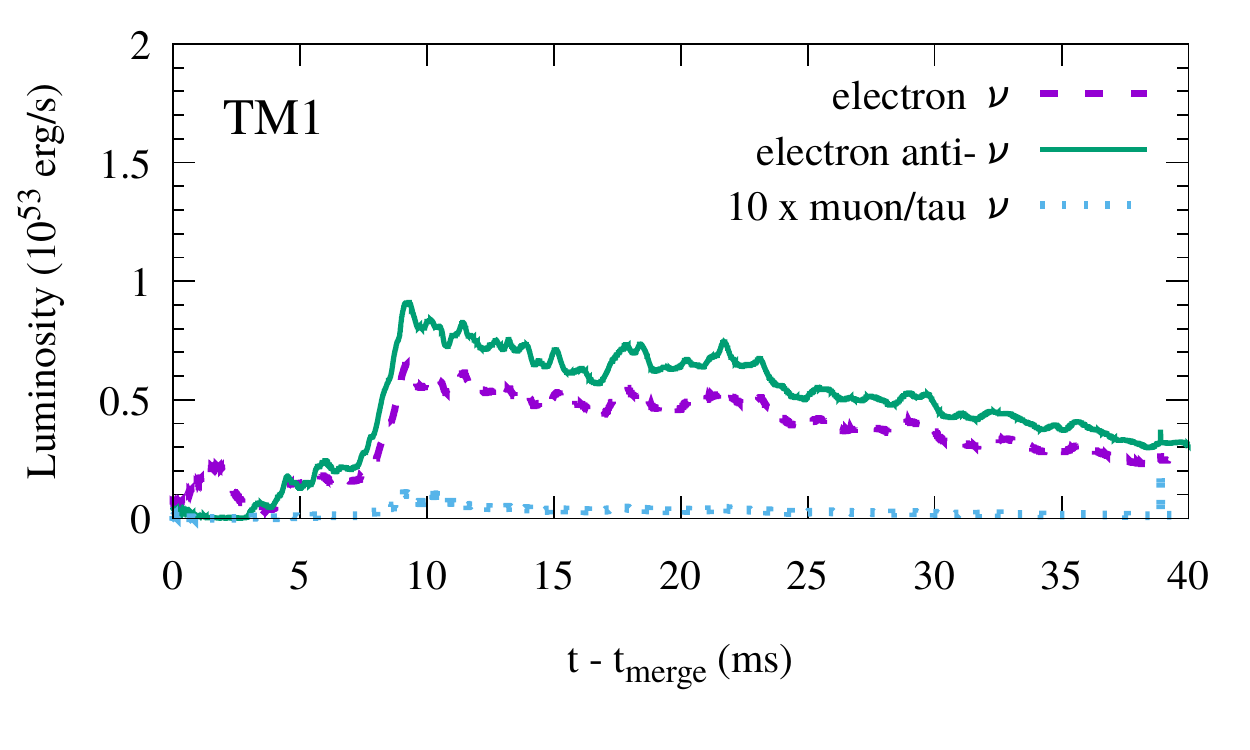}
 \end{tabular}
 \caption{Neutrino luminosity curves for SFHo (top), DD2 (middle), and
 TM1 (bottom). The purple-dashed, green-solid, and cyan-dotted curves
 correspond to the electron neutrino, electron antineutrino, and one of
 the heavy-lepton neutrinos, respectively. The luminosity curves for the
 heavy-lepton neutrinos are multiplied by a factor of 10 to make the
 plot visible.} \label{fig:lightcurve}
\end{figure}

Figure \ref{fig:lightcurve} shows the evolution of the neutrino
luminosity for all the models and flavors. All the models dominantly
emit electron neutrinos and antineutrinos with the peak luminosity of
$\sim 0.5$--\SI{2e53}{erg.s^{-1}} for each flavor at $\approx
\SI{10}{\milli\second}$ after the onset of merger. Because the
self-collision of the tidal tail occurs only after a single revolution
around the black hole (see also Ref.~\cite{kyutoku_iost2015}), the onset
of neutrino emission is delayed by $\approx$ \SI{10}{\milli\second} with
respect to the tidal disruption of the neutron star. The emission is
dominated by electron/positron captures onto free nucleons (see also
below for the discussion of the pair process). The luminosity is higher
for $\bar{\nu}_e$ by 20\%--30\% than for $\nu_e$ as in the case of
binary-neutron-star mergers
\cite{ruffert_jts1997,rosswog_liebendorfer2003,sekiguchi_kks2011,sekiguchi_kks2015,sekiguchi_kkst2016},
because neutrons are more abundant than protons in the accretion disk
formed from neutron stars. This hierarchy has also been found in
previous studies for the merger of black hole-neutron star binaries
\cite{janka_erf1999,deaton_dfookmss2013,foucart_etal2014}.

\begin{figure}
 \includegraphics[width=.95\linewidth]{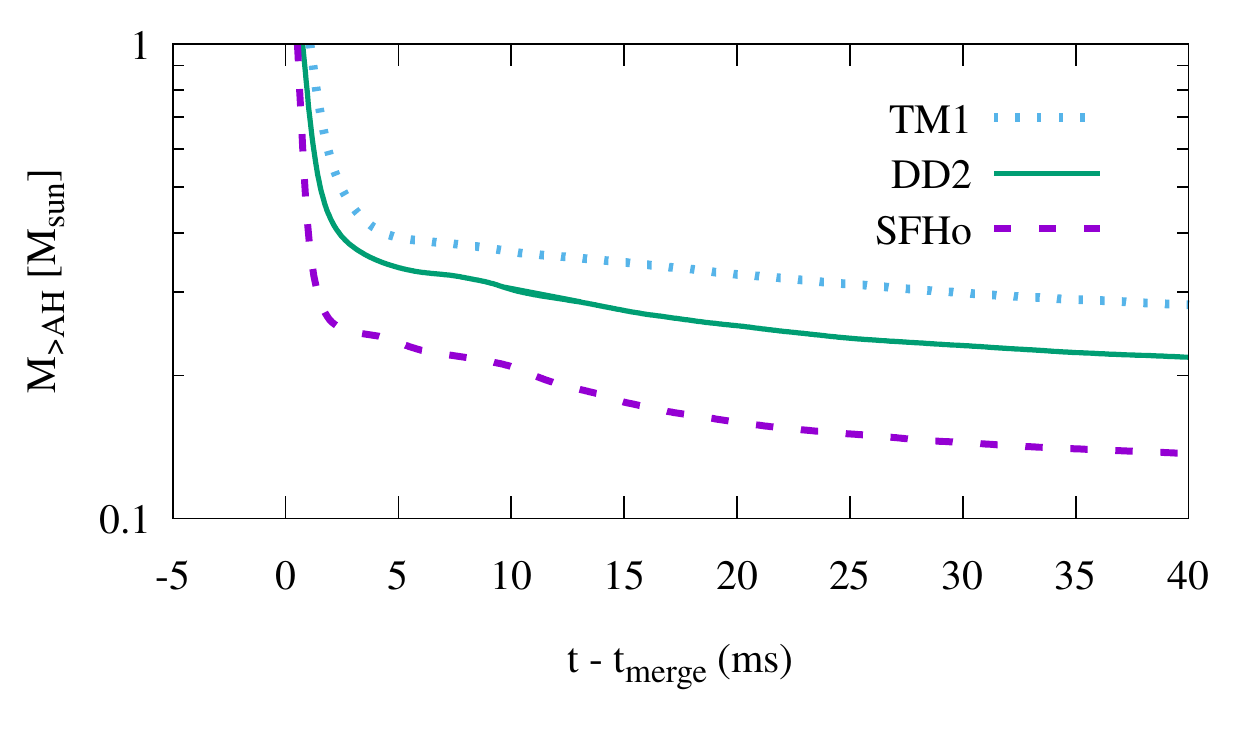} \caption{Time evolution
 of the mass remaining outside the apparent horizon for SFHo (purple
 dashed), DD2 (green solid), and TM1 (cyan dotted).} \label{fig:mdisk}
\end{figure}

The peak luminosity is higher for the model with a softer equation of
state and thus a smaller neutron-star radius such as SFHo. This trend is
consistent with the mergers of binary neutron stars
\cite{sekiguchi_kks2015,palenzuela_lnlcoa2015,sekiguchi_kkst2016,foucart_hdoorklps2016}.\footnote{This
trend seems to be consistent with previous simulations of black
hole-neutron star binary mergers \cite{foucart_etal2014}, whereas it is
not explicitly stated.} For black hole-neutron star binaries, this
dependence of the neutrino luminosity on the neutron-star radius
immediately means that the peak luminosity is anticorrelated with the
disk mass, because it is established by various previous work that the
disk mass is an increasing function of the neutron-star radius for given
binary parameters \cite{shibata_taniguchi2011}. Figure \ref{fig:mdisk}
shows the time evolution of the mass remaining outside the apparent
horizon. As expected, the larger the neutron-star radius, the larger the
mass outside the apparent horizon. Quantitatively, the values at
\SI{10}{\milli\second} after the onset of merger agree approximately
with those found in our purely hydrodynamic studies performed adopting
piecewise polytropic equations of state for comparable neutron-star
compactnesses \cite{kyutoku_ost2011} (see also
Ref.~\cite{foucart2012}). This agreement confirms the irrelevance of the
neutrino transport in the inspiral and merger phases.

The variation of the neutrino luminosity cannot be ascribed to the spin
of the remnant black hole, which is not different substantially among
the models. Specifically, the black-hole spins are 0.84--0.86 at
\SI{10}{\milli\second} after the onset of merger, which agree with those
obtained in our purely hydrodynamic studies \cite{kyutoku_ost2011} (see
also Ref.~\cite{pannarale2013}), and the accretion increases them by
0.01--0.02 in \SI{30}{\milli\second}. We find that the spin parameter is
the largest for SFHo and the smallest for TM1, but the difference is
very minor.

\begin{figure}
 \includegraphics[width=.95\linewidth]{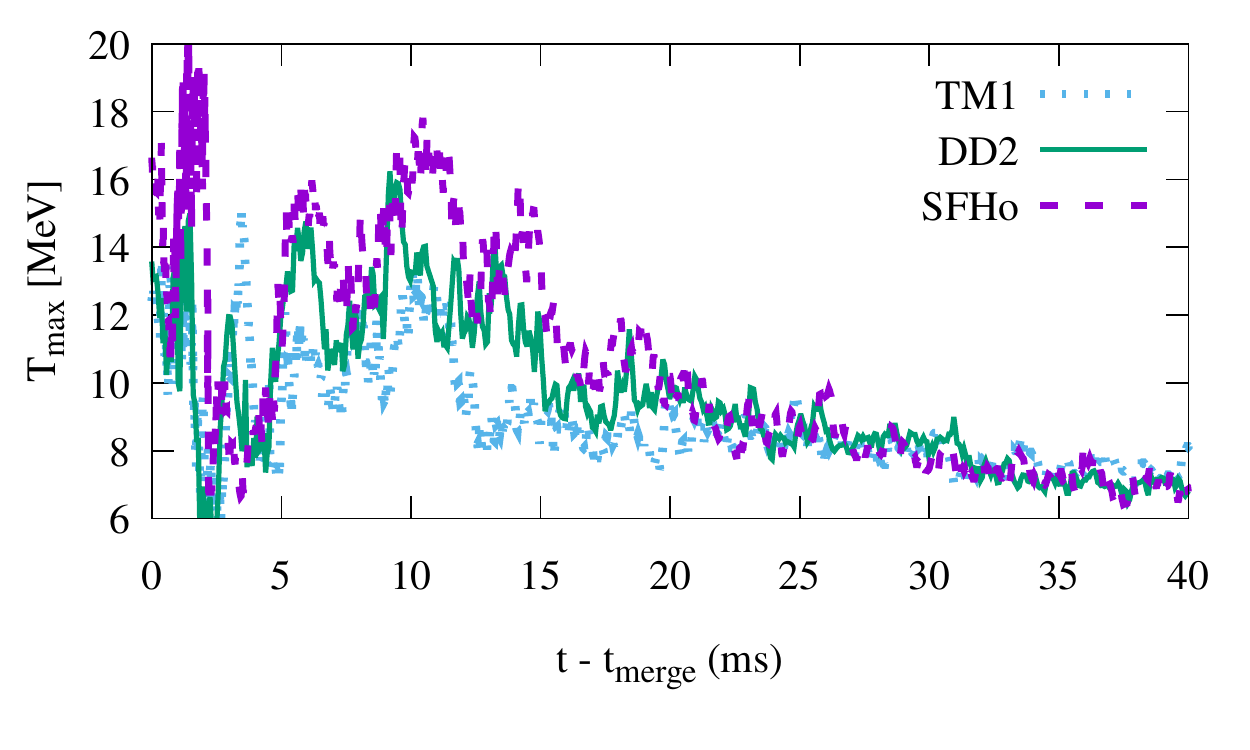} \caption{Time evolution
 of the maximum temperature in the accretion disk for SFHo (purple
 dashed), DD2 (green solid), and TM1 (cyan dotted).} \label{fig:tmax}
\end{figure}

The high neutrino luminosity for a small neutron-star radius is ascribed
to high temperature realized in a compact accretion disk. Figure
\ref{fig:tmax} compares the time evolution of the maximum temperature in
the accretion disk and shows that the temperature is higher for smaller
neutron stars around $t - t_\mathrm{merge} \approx
\SI{10}{\milli\second}$, i.e., the peak of the neutrino luminosity. For
neutron stars with a small radius such as SFHo, tidal disruption occurs
at an orbit very close to the innermost stable circular orbit of the
black hole. Accordingly, the self-collision of the tidal tail occurs
also near the innermost stable circular orbit. Reflecting high orbital
velocity of the close orbit, the energy liberation near the black hole
results in the high temperature. Because the emissivity of neutrinos by
electron/positron captures is sensitive to the temperature
\cite{sekiguchi2010,sekiguchi_kks2012}, the neutrino luminosity becomes
high for soft equations of state and small neutron-star radii.

\begin{figure}
 \includegraphics[width=.95\linewidth]{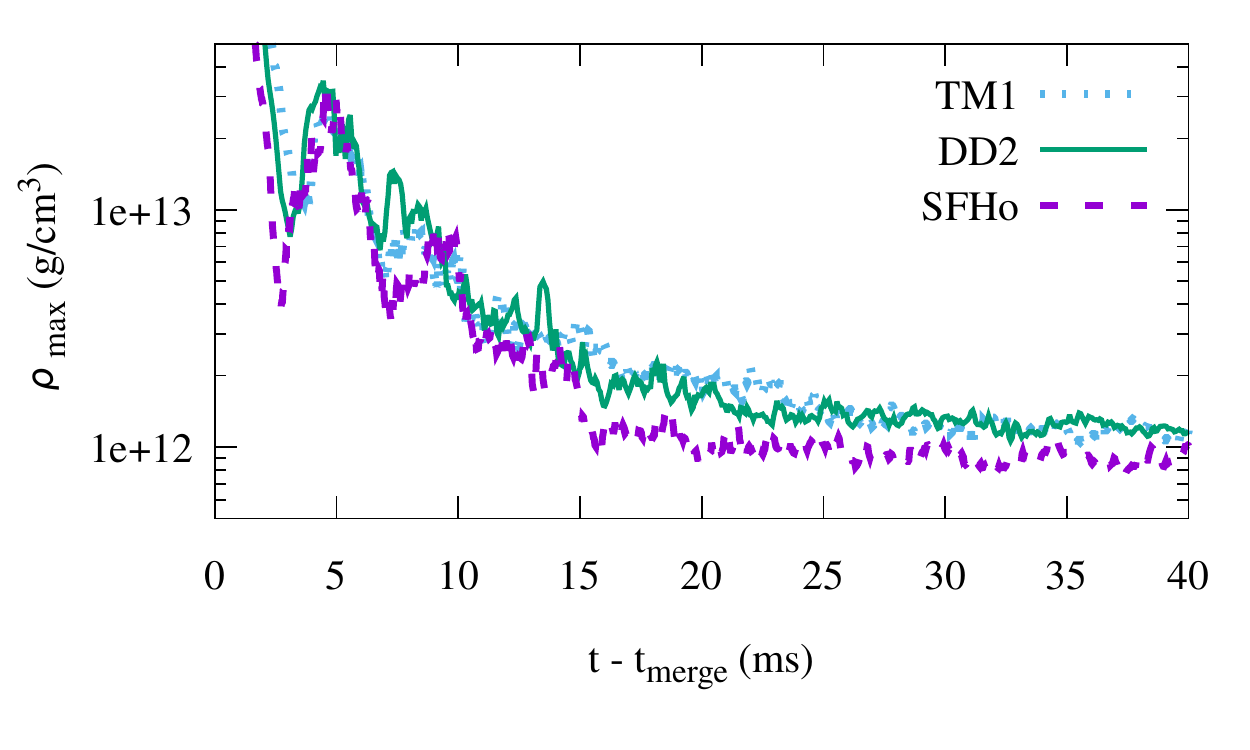} \caption{Time evolution
 of the maximum rest-mass density in the accretion disk for SFHo (purple
 dashed), DD2 (green solid), and TM1 (cyan dotted).} \label{fig:rhomax}
\end{figure}

The rest-mass density may have less impact on the neutrino
luminosity. Figure \ref{fig:rhomax} shows the time evolution of the
maximum rest-mass density in the accretion disk, which is higher for TM1
than for SFHo on average. The higher rest-mass density for a larger
neutron-star radius has already been found in our purely hydrodynamic
study as the correlation between the maximum rest-mass density and the
disk mass \cite{kyutoku_st2010,kyutoku_st2010e}. Although the high
rest-mass density of the disk for TM1 could be potentially advantageous
for neutrino emission, the high temperature of the disk for SFHo plays a
dominant role in determining the neutrino luminosity. Note that the high
rest-mass density also increases the optical depth and obscures the
inner hot portion to some extent, and the values of $Y_e$ in the outer
part of the disk are increased by neutrino irradiation.

We do not observe strong emission of heavy-lepton neutrinos from the
merger remnant of black hole-neutron star binaries. In fact, we are
required to multiply the luminosity by a factor of 10 to make the curve
for $\nu_x$ visible in Fig.~\ref{fig:lightcurve}. This weakness should
be compared with moderately strong emission from massive neutron stars
formed after binary-neutron-star mergers
\cite{sekiguchi_kks2011,sekiguchi_kks2011-2,sekiguchi_kks2015,sekiguchi_kkst2016},
for which heavy-lepton neutrinos are emitted via pair processes such as
the electron-positron pair annihilation. This difference reflects the
different temperature of merger remnants. While the massive neutron
stars formed after binary-neutron-star mergers are as hot as a few tens
of \si{\mega\electronvolt}, typical temperature of the accretion disk
formed after black hole-neutron star binary mergers is around
\SI{10}{\mega\electronvolt}. This is not sufficient to produce a copious
amount of heavy-lepton neutrinos via pair processes. Note that the
difference of the heavy-lepton neutrino luminosity between SFHo and TM1
is by a factor of 7--8 around the peak, whereas those of the electron
neutrino and antineutrino luminosity are at most by a factor of 2. This
reflects the steep dependence of the pair emission rate on the
temperature.

\begin{figure}
 \begin{tabular}{c}
  \includegraphics[width=.95\linewidth]{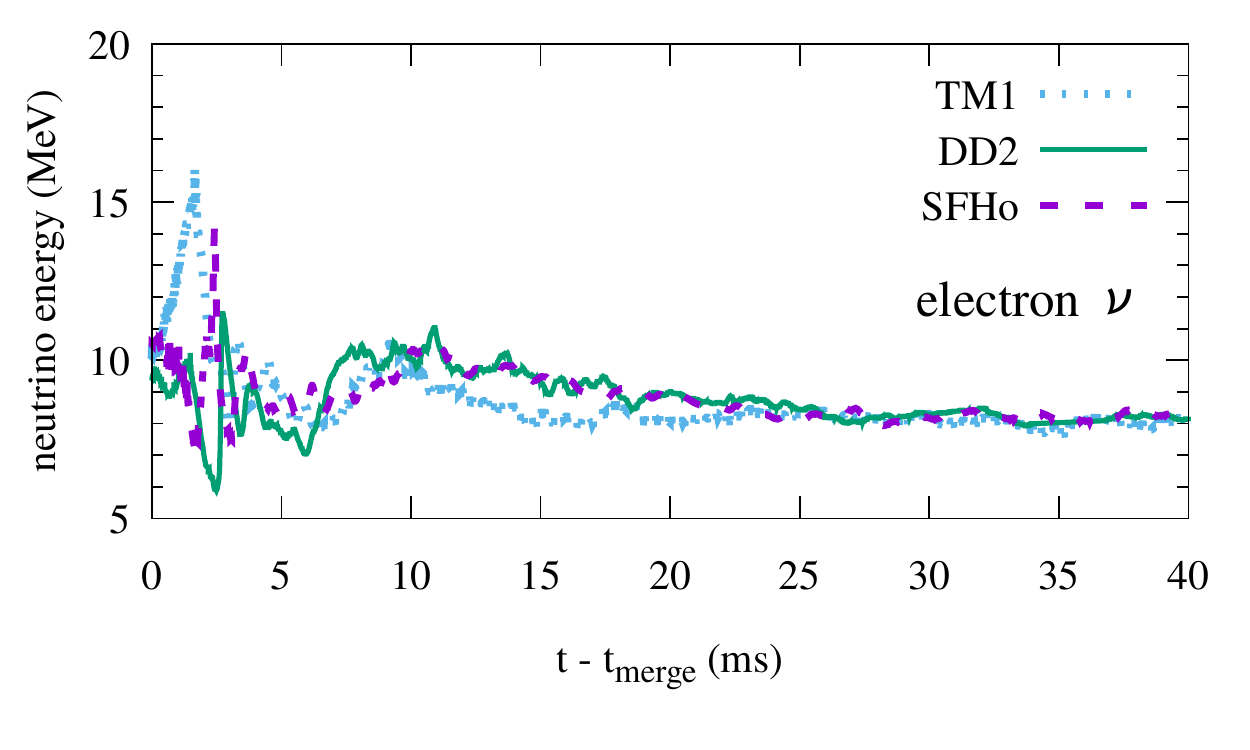} \\
  \includegraphics[width=.95\linewidth]{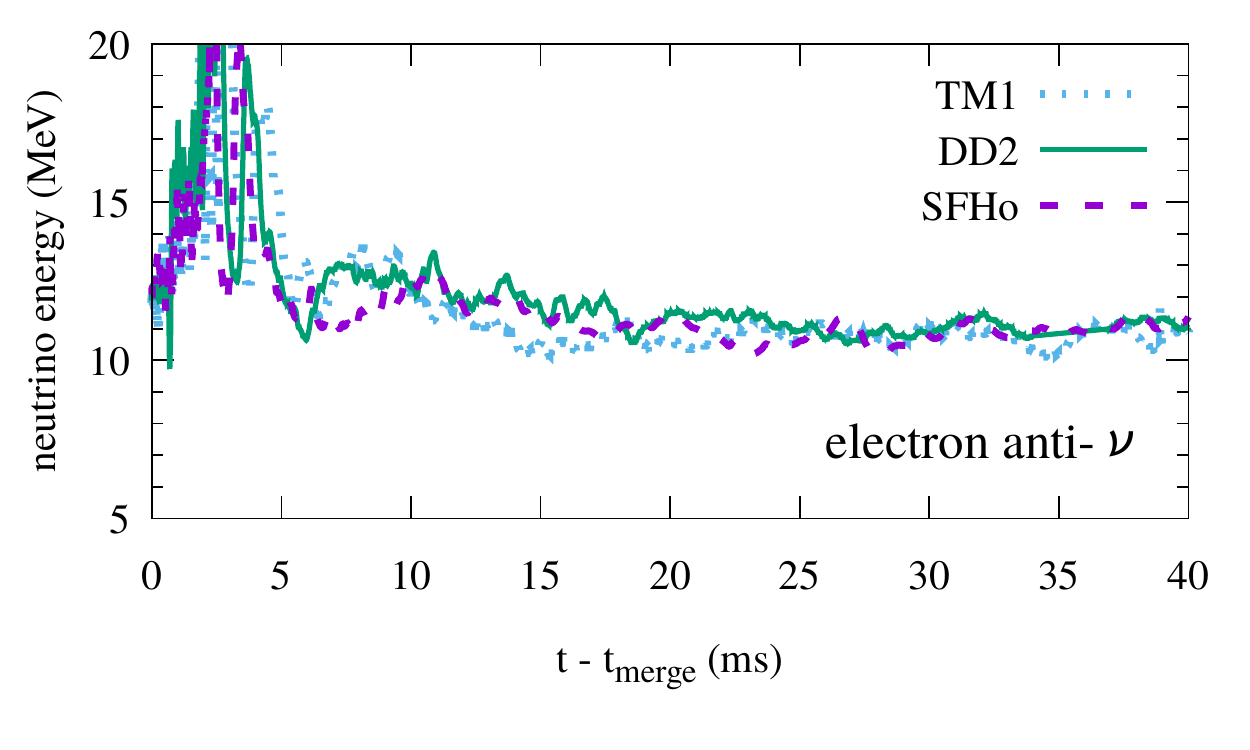} \\
  \includegraphics[width=.95\linewidth]{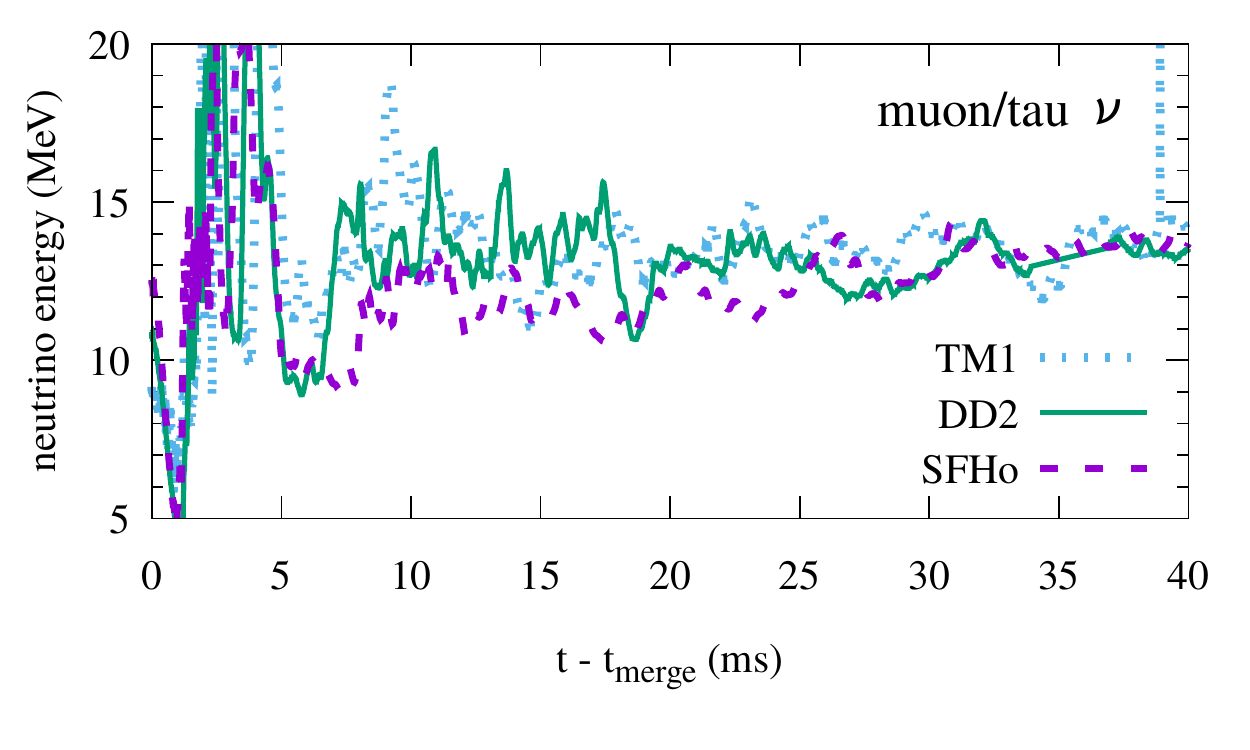}
 \end{tabular}
 \caption{Time evolution of the average neutrino energy for $\nu_e$
 (top), $\bar{\nu}_e$ (middle), and $\nu_x$ (bottom). The purple-dashed,
 green-solid, and cyan-dotted curves correspond to SFHo, DD2, and TM1,
 respectively. The average energy of $\bar{\nu}_e$ and $\nu_x$
 approaches \SI{30}{\mega\electronvolt} at $t - t_\mathrm{merge} \approx
 3$--\SI{4}{\milli\second}, whereas we restrict the range in this figure
 to focus on long-term behavior.} \label{fig:energy}
\end{figure}

Figure \ref{fig:energy} compares the time evolution of the average
neutrino energy among the models. The average energy of neutrinos is not
distinctively higher for a softer equation of state, say SFHo, at $t -
t_\mathrm{merge} \approx \SI{10}{\milli\second}$ for any flavor, unlike
the maximum temperature in the accretion disk. After $t -
t_\mathrm{merge} \approx \SI{20}{\milli\second}$, the average neutrino
energy for each flavor settles approximately to constant values common
among the models. This suggests that the temperature at the neutrino
sphere for a given flavor is similar at the late time irrespective of
the neutron-star radius. Specifically, the energy settles to $\approx
8$, 11, and \SI{13}{\mega\electronvolt} for $\nu_e$, $\bar{\nu}_e$, and
$\nu_x$, respectively. Note that, however, the average neutrino energy
in this work is estimated as the ratio of the energy emission rate
(luminosity) to the number emission rate computed within a gray leakage
scheme, and thus only semiquantitative. Precise estimation requires
simulations with a multienergy transfer scheme (see also
Ref.~\cite{foucart_orkps2016}).

The luminosity gradually decreases in time due to the accretion of hot
material by the remnant black hole. From Fig.~\ref{fig:lightcurve}, the
luminosity is expected to decrease by an order of magnitude from the
peak value in $\approx \SI{100}{\milli\second}$ after the onset of
merger, which approximately amounts to the accretion time scale of the
remnant due to hydrodynamic processes associated with nonaxisymmetric
disk structures
\cite{foucart_dkt2011,foucart_etal2014,kawaguchi_knost2015,kyutoku_iost2015}. Specifically,
the accretion time scale is estimated to be $\approx
130$--\SI{150}{\milli\second} at $\approx \SI{30}{\milli\second}$ after
the onset of merger, being consistent with our previous work when
measured at a similar epoch
\cite{kyutoku_st2010,kyutoku_st2010e,kyutoku_iost2015,kawaguchi_knost2015}. The
accretion time scale depends only weakly on the grid resolution
\cite{kawaguchi_knost2015}. An effective viscosity may be characterized
by $\alpha \sim 0.005$--0.01 in a standard prescription \cite{kato_fm}.
In reality, however, the temperature of the disk should be increased by
viscous heating due to magnetohydrodynamic processes in the accretion
phase. If the material is heated efficiently, the neutrino luminosity is
enhanced (see, e.g.,
Refs.~\cite{popham_wf1999,kohri_mineshige2002,shibata_st2007,fernandez_metzger2013,metzger_fernandez2014}),
and therefore our estimate should be regarded as a lower limit. We would
like to revisit this topic in the near future
\cite{shibata_ks2017,shibata_kiuchi2017,fujibayashi_sks2017}.

\subsection{Mass ejection} \label{sec:result_ej}

\begin{figure*}
 \begin{tabular}{c}
  \includegraphics[width=.95\linewidth]{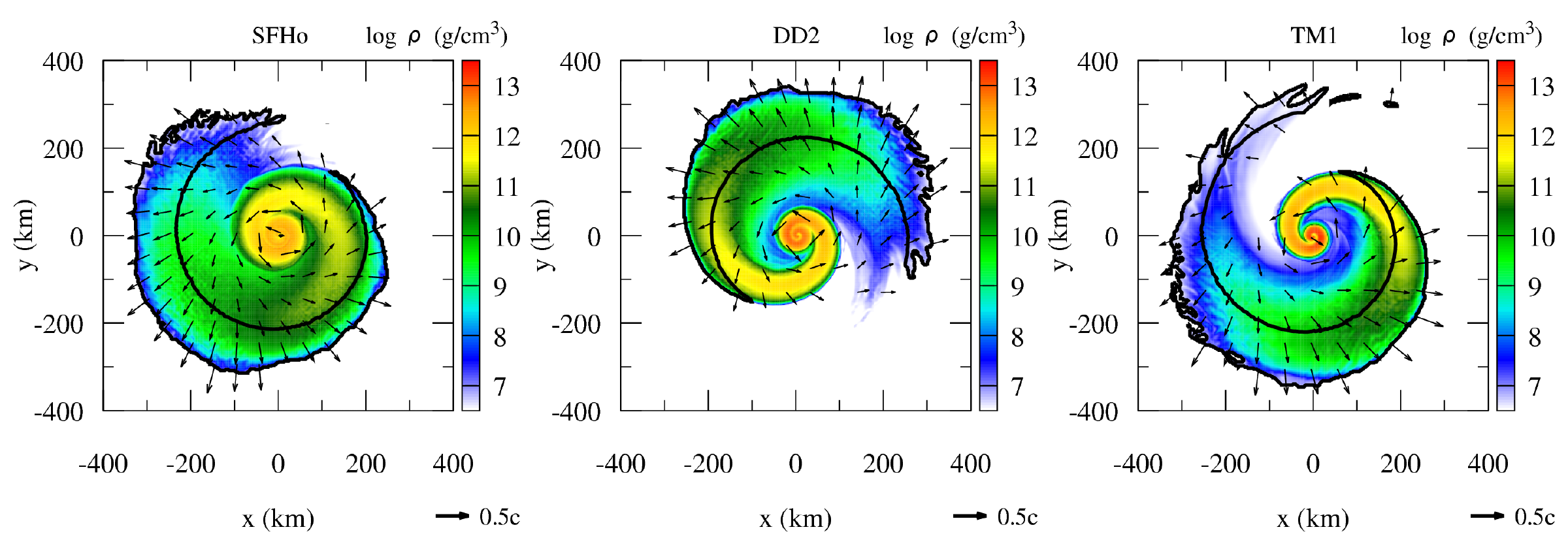} \\
  \includegraphics[width=.95\linewidth]{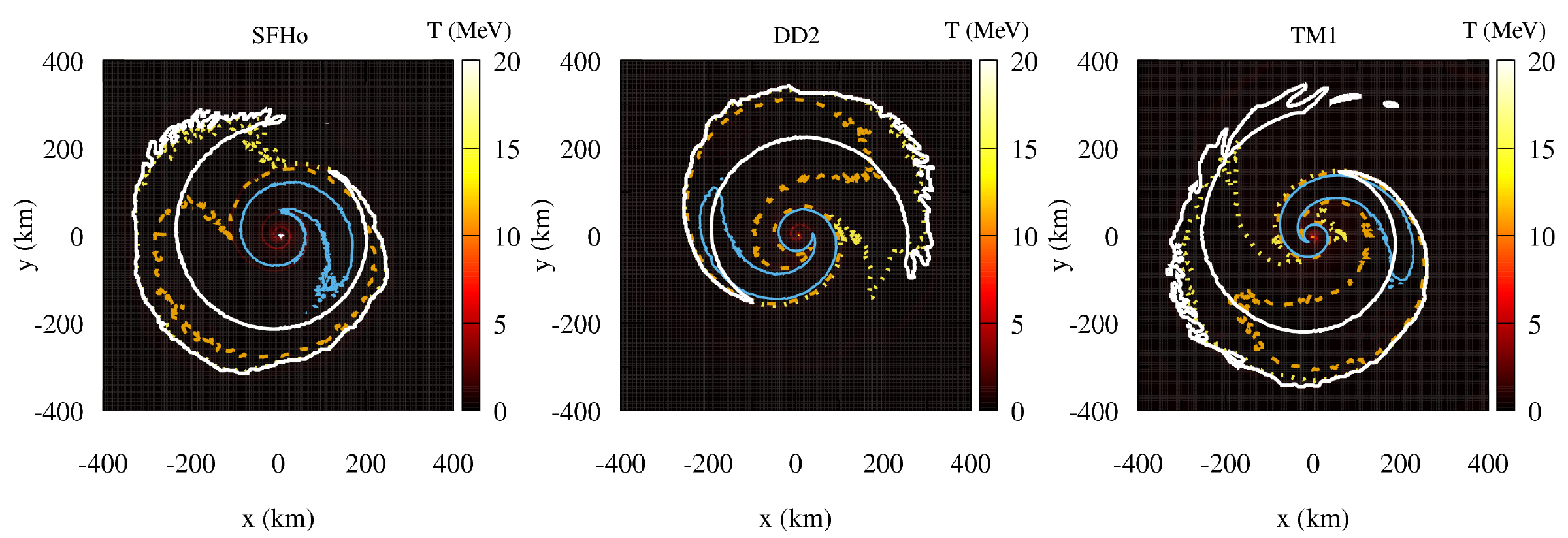} \\
  \includegraphics[width=.95\linewidth]{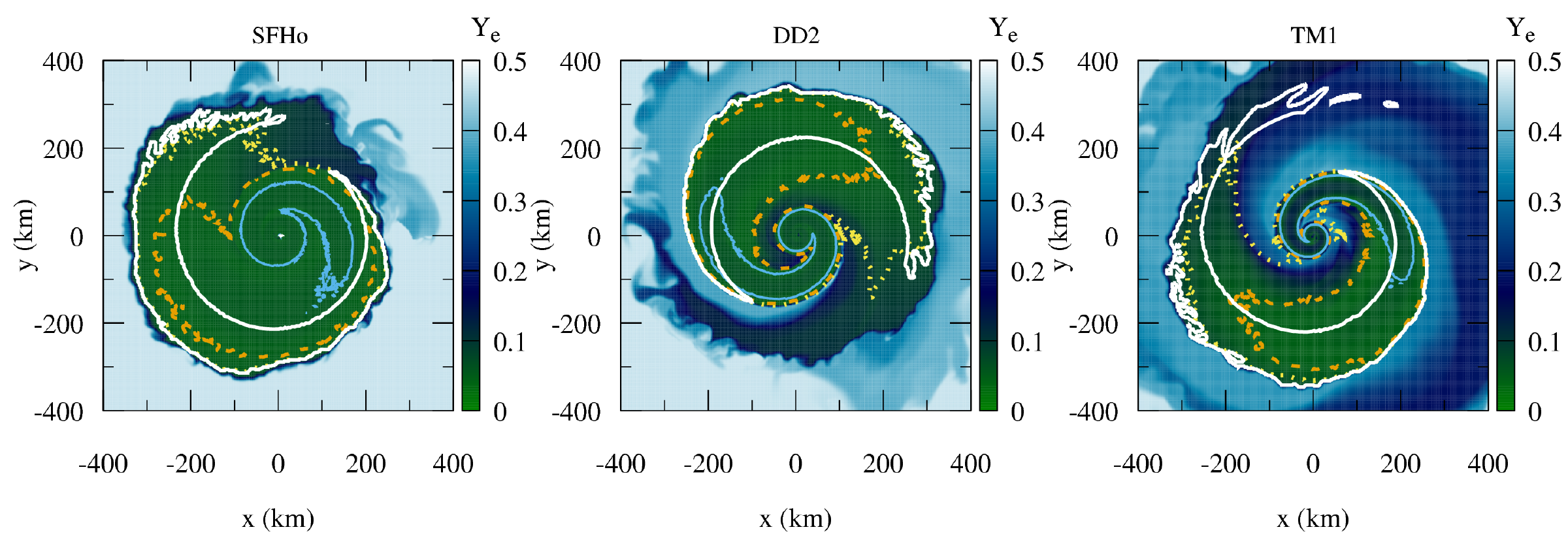}
  \end{tabular}
 \caption{Rest-mass density (top), temperature (middle), and electron
 fraction (bottom) profiles on the equatorial plane at
 \SI{3}{\milli\second} after the onset of merger, when the mass and
 electron fraction of the ejecta approximately settle to asymptotic
 values. Note the different spatial scale compared to
 Figs.~\ref{fig:snapdiskrho}, \ref{fig:snapdiskt} and
 \ref{fig:snapdiskye}. Black (top) and white (middle and bottom) curves
 indicate the unbound component identified by the conditions $u_t < -1$
 (shown only for $\rho \ge \SI{e6.5}{\gram\per\cubic\centi\meter}$ to
 match the top row), which are indistinguishable from $h u_t < -1$ on
 panels presented here \cite{kyutoku_iost2015}. The velocity vector $v^i
 \equiv u^i / u^t$ is overplotted on the rest-mass density profiles. We
 also show isodensity contours for $\rho = \num{e7}$ (yellow dotted),
 \num{e9} (orange dashed), and \SI{e11}{\gram\per\cubic\centi\meter}
 (light blue solid) on the temperature and electron-fraction profiles.}
 \label{fig:snapej}
\end{figure*}

The outer part of the tidally elongated neutron stars is ejected
dynamically at 1--\SI{2}{\milli\second} after the onset of merger
\cite{kyutoku_is2013,kyutoku_iost2015}. Figure \ref{fig:snapej} shows
the rest-mass density, temperature, and electron fraction of the ejecta
as well as the bound material on the equatorial plane at
\SI{3}{\milli\second} after the onset of merger. The amount of dynamical
ejecta in the polar region is tiny
\cite{kyutoku_is2013,kyutoku_iost2015}. The ejecta are anisotropic and
cold as found in previous work
\cite{kyutoku_is2013,kyutoku_iost2015}. In this work, we further found
that the electron fraction is as low as $Y_e \lesssim 0.1$ for the most
part of the dynamical ejecta, even though the neutrino transport is
solved in this work. In the following, we investigate the properties of
ejecta quantitatively.

\begin{figure}
 \includegraphics[width=.95\linewidth]{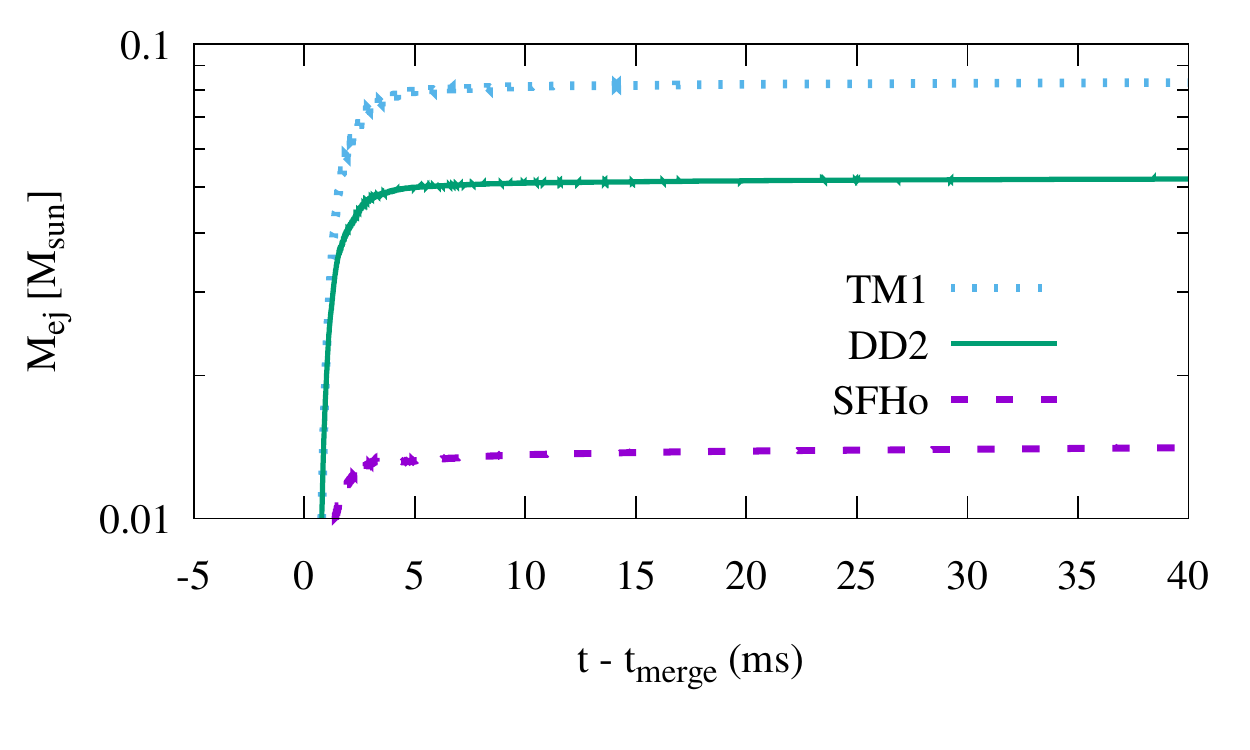} \caption{Time
 evolution of the ejecta mass identified by the condition $u_t<-1$ for
 SFHo (purple dashed), DD2 (green solid), and TM1 (cyan dotted).}
 \label{fig:mej}
\end{figure}

Figure \ref{fig:mej} shows the time evolution of the ejecta mass. All
the curves in the figure exhibit a steep rise associated with the
dynamical mass ejection in a few milliseconds after the onset of
merger. The values of the ejecta mass are $\approx 0.01 M_\odot$, $0.05
M_\odot$, and $0.08 M_\odot$ for SFHo, DD2, and TM1, respectively, at
\SI{10}{\milli\second} after the onset of merger. They are compatible
with the results of our previous purely hydrodynamic study performed
with piecewise polytropic equations of state for models with comparable
binary parameters \cite{kyutoku_iost2015}. We also find that our results
are consistent with a fitting
formula\footnote{http:\slash\slash{}www2.yukawa.kyoto-u.ac.jp\slash\~{}kyohei.kawaguchi\slash{}kn\_calc\slash{}main.html}
for dynamical ejecta calibrated against the results of simulations
performed employing piecewise polytropes
\cite{kawaguchi_kst2016}. Specifically, the fitting formula predicts
$0.009 M_\odot$, $0.04 M_\odot$, and $0.07 M_\odot$ for SFHo, DD2, and
TM1, respectively. This agreement implies that the fitting formula of
Ref.~\cite{kawaguchi_kst2016} is valid also for finite-temperature
equations of state.

\begin{figure}
 \includegraphics[width=.95\linewidth]{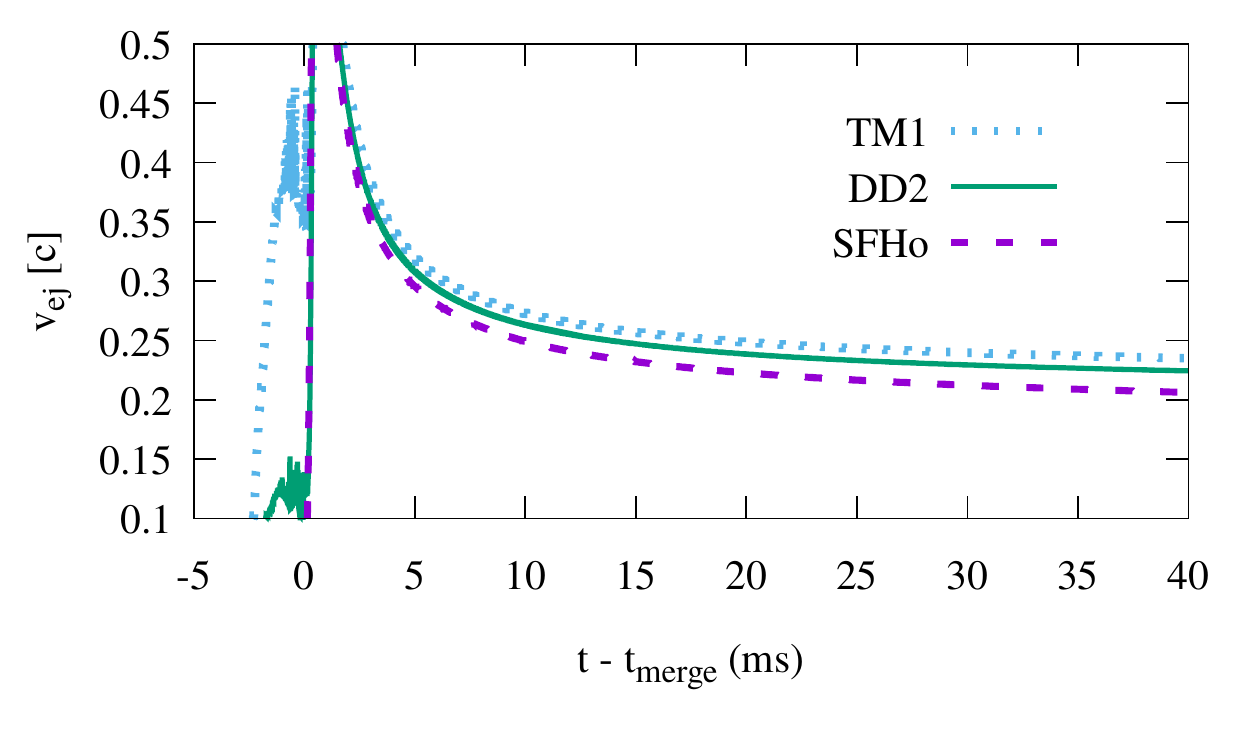} \caption{Time
 evolution of the ejecta velocity estimated in the same manner as
 described in Refs.~\cite{hotokezaka_kkosst2013,kyutoku_iost2015} for
 SFHo (purple dashed), DD2 (green solid), and TM1 (cyan dotted). The
 values before $t=t_\mathrm{merge}$ are numerical artifacts associated
 with the small ejecta mass. The values right after the onset of merger
 is $\approx 0.7c$, but we do not show such transitional values.}
 \label{fig:vej}
\end{figure}

Figure \ref{fig:vej} shows the time evolution of the ejecta velocity,
which is $0.25$--$0.3 c$ at \SI{10}{\milli\second} after the onset of
merger. This also agrees with our previous results and the prediction of
the fitting formula, $0.25 c$, for the dynamical ejecta
\cite{kawaguchi_kst2016}. However, we see that the ejecta velocity
asymptotes to $0.2$--$0.25 c$ on a long time scale as they escape from
the gravitational binding of the remnant black hole. The decrease by
$\sim 20\%$ is consistent with the estimate in
Ref.~\cite{kyutoku_iost2015}.

\begin{figure}
 \includegraphics[width=.95\linewidth]{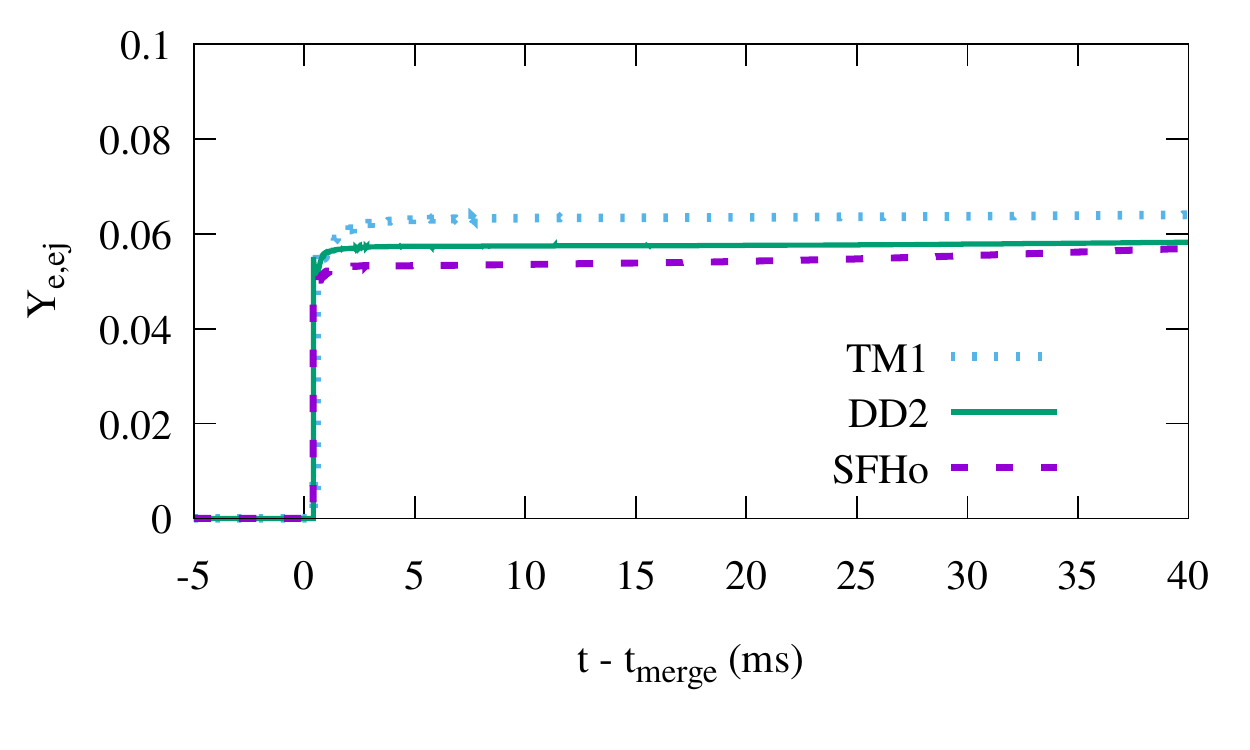} \caption{Time
 evolution of the mass-averaged electron fraction of the ejecta for SFHo
 (purple dashed), DD2 (green solid), and TM1 (cyan dotted).}
 \label{fig:yej}
\end{figure}

The electron fraction of the dynamical ejecta is lower than $Y_e = 0.1$
for most of the ejecta components. Figure \ref{fig:yej} shows the time
evolution of the mass-averaged electron fraction, $Y_{e,\mathrm{ej}}$,
which is identical to the electron fraction of the entire ejecta. The
values fall in the range between 0.05 and 0.07 for all the models. This
low electron fraction directly reflects the composition of neutron-star
matter in neutrinoless $\beta$-equilibrium at zero temperature. Stated
differently, neither the shock heating nor the neutrino irradiation (see
Fig.~\ref{fig:lightcurve}) has a significant impact on the dynamical
ejecta that leave the central region in a short time scale after the
onset of merger.

The averaged electron fraction, $Y_{e,\mathrm{ej}}$, is higher when the
ejecta mass is larger for the models considered in this study, and this
trend is opposite to that observed in the case of binary neutron stars
\cite{sekiguchi_kks2015,sekiguchi_kkst2016}. Whereas numerical errors
associated with the finite resolution prevent us from concluding this
correlation to be definitive (see Sec.~\ref{sec:result_res}), we argue
below that this can reasonably stem from the correlation between the
symmetry energy for nuclear matter and the neutron-star radius
\cite{lattimer_prakash2001}. When the symmetry energy is higher, the
pressure of nuclear matter at and above the saturation density tends to
be higher. High pressure at and above the saturation density is
empirically found to give a large neutron-star radius
\cite{lattimer_prakash2001}, and the large radius gives the large ejecta
mass for black hole-neutron star binaries. Thus, the symmetry energy can
be correlated with the ejecta mass via the tidal effect. At the same
time, the material prefers relatively proton-rich conditions
characterized by a higher value of $Y_e$ when the symmetry energy is
higher. Hence, the ejecta mass and averaged electron fraction can be
reasonably correlated. Note that the ejecta mass also depends on binary
parameters other than the neutron-star radius.

\begin{figure}
 \includegraphics[width=.95\linewidth]{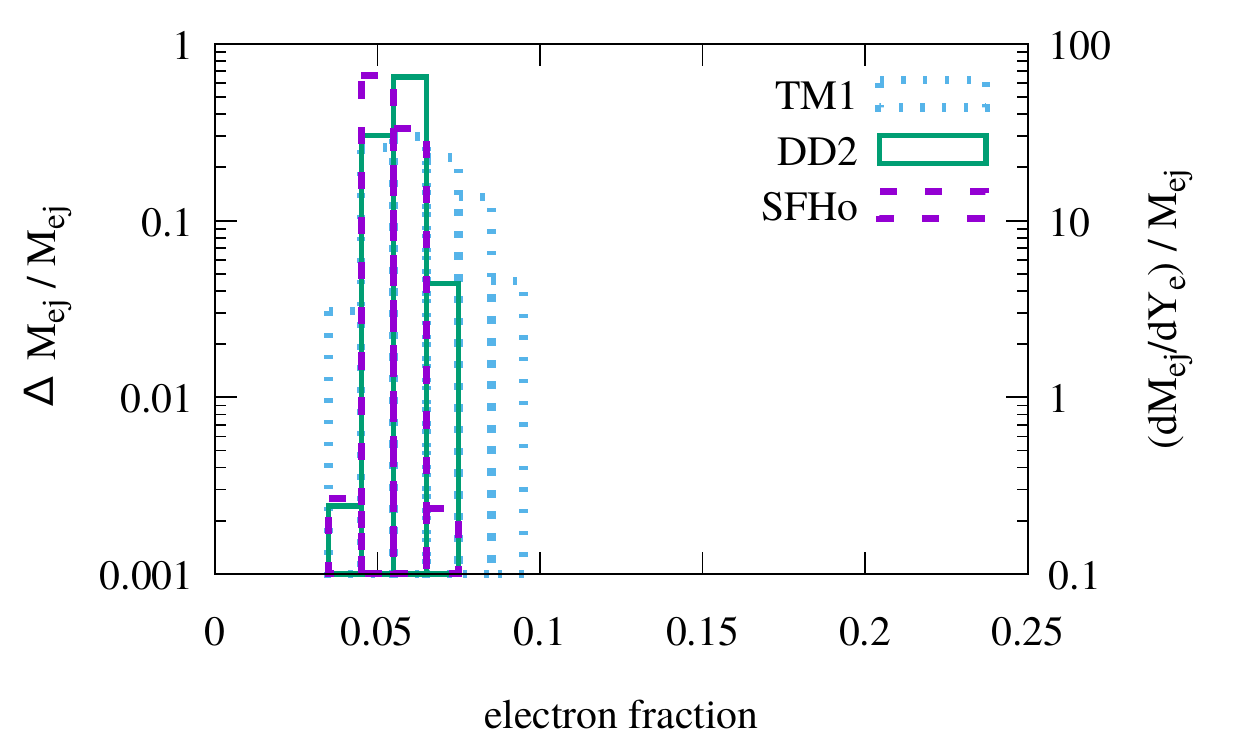} \caption{Distribution
 of the electron fraction in the unbound material at $\approx
 \SI{10}{\milli\second}$ after the onset of merger. The left axis shows
 the fraction of material in the bin of width $\Delta Y_e = 0.01$, and
 the right axis shows the differential distribution normalized by the
 total ejecta mass. The latter is independent of the bin width.}
 \label{fig:ydistrib}
\end{figure}

Figures \ref{fig:ydistrib} present the mass spectrum of the electron
fraction measured at $\approx \SI{10}{\milli\second}$ after the onset of
merger. These spectra do not change appreciably in time. The electron
fraction peaks sharply around the averaged value, $Y_{e,\mathrm{ej}}$,
shown in Fig.~\ref{fig:yej}. The peak shifts slightly towards the
high-$Y_e$ side as the equation of state becomes stiff in accordance
with Fig.~\ref{fig:yej}. We find that the specific entropy also takes a
low value of $\lesssim 10 k_\mathrm{B}^{-1}$ irrespective of the models
peaking around $5 k_\mathrm{B}^{-1}$. This is consistent with the
finding that the ejecta do not experience substantial shock heating. The
low electron fraction and entropy of the dynamical ejecta found in this
study agree with the results of previous simulations performed without
neutrino absorption
\cite{rosswog2005,rosswog_pn2013,just_bagj2015,foucart_dbdkhkps2017}.

We observe no substantial mass ejection from the remnant disk due to the
neutrino heating, i.e., neutrino-driven wind, in our simulations
performed until a few tens milliseconds after the onset of merger. This
is evident from the approximately constant values of the ejecta mass at
late times (see Fig.~\ref{fig:mej}). The insignificance of the purely
neutrino-driven wind has been pointed out in various simulations for the
remnant of black hole-neutron star binary mergers with different levels
of sophistication
\cite{fernandez_metzger2013,fernandez_kmq2015,just_bagj2015,foucart_etal2015}. Our
study confirms this fact by fully general-relativistic
radiation-hydrodynamics simulations starting from the inspiral phase for
the first time. We caution that our current simulations cannot evaluate
the amount of the viscously driven wind and the effect of neutrino
irradiation on it, because physical viscosity is not implemented.

\subsection{Convergence} \label{sec:result_res}

\begin{figure*}
 \begin{tabular}{cc}
  \includegraphics[width=.48\linewidth]{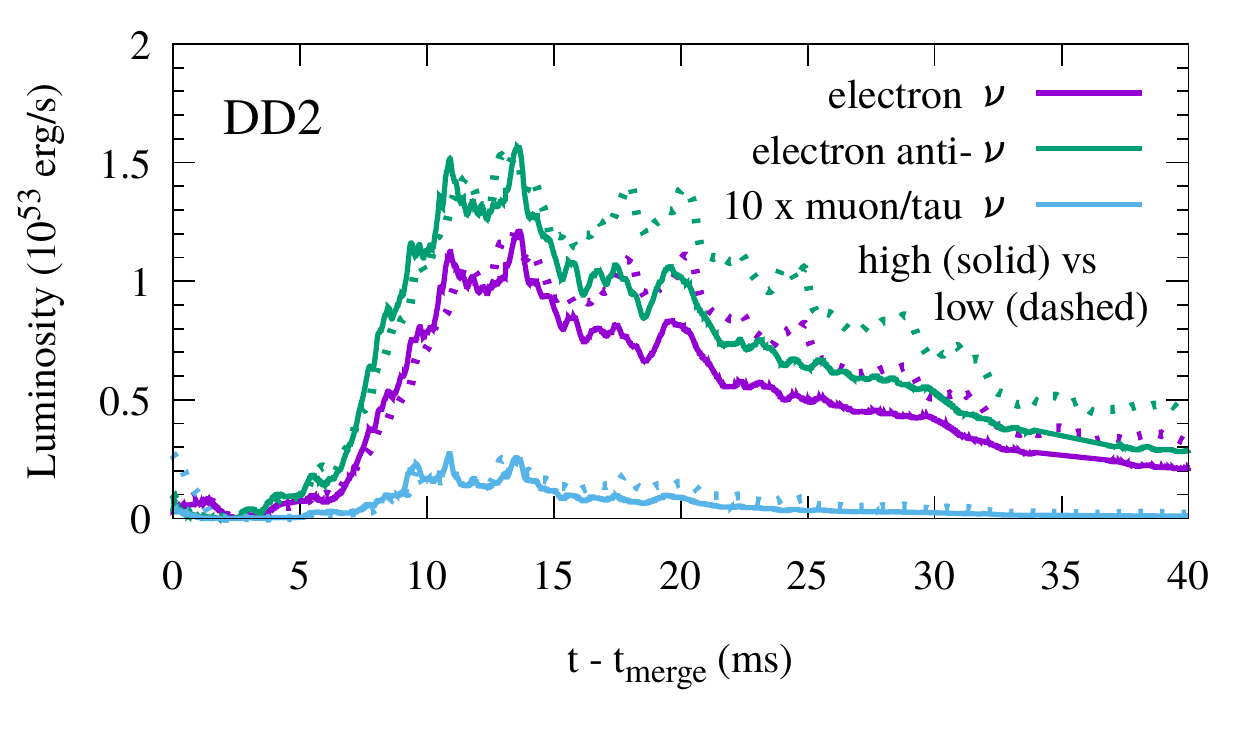} &
  \includegraphics[width=.48\linewidth]{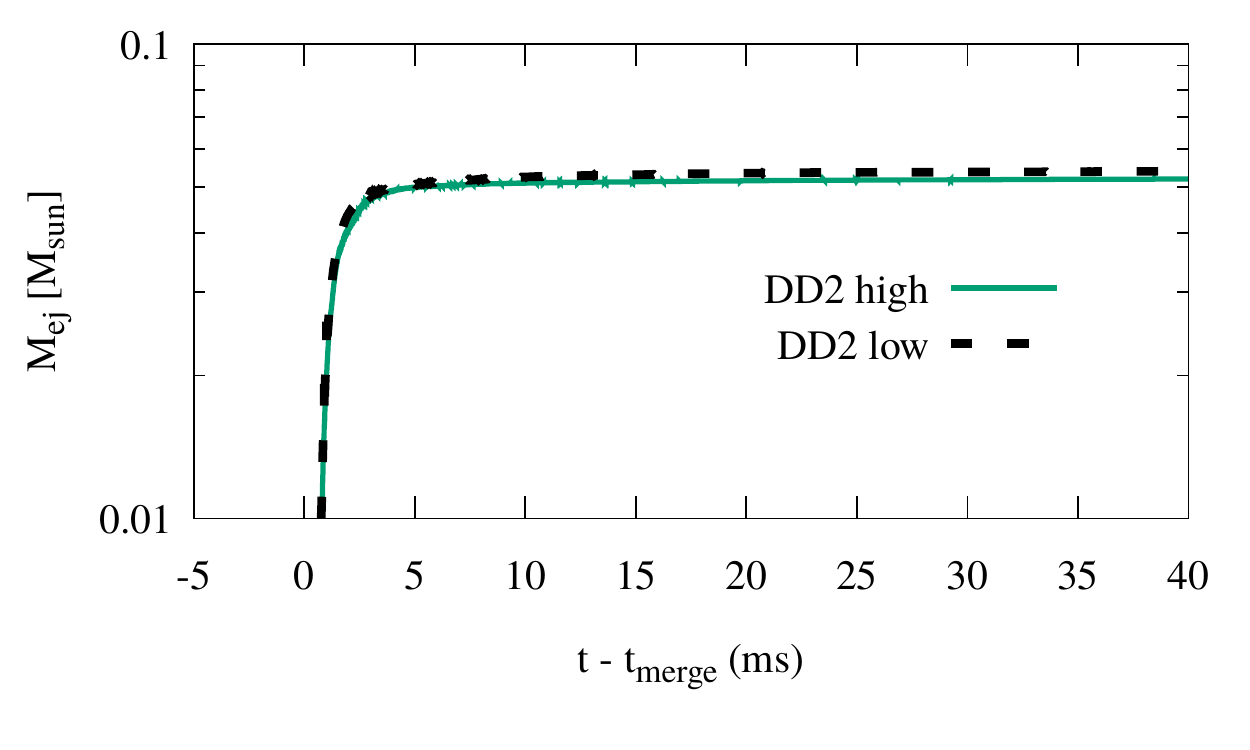} \\
  \includegraphics[width=.48\linewidth]{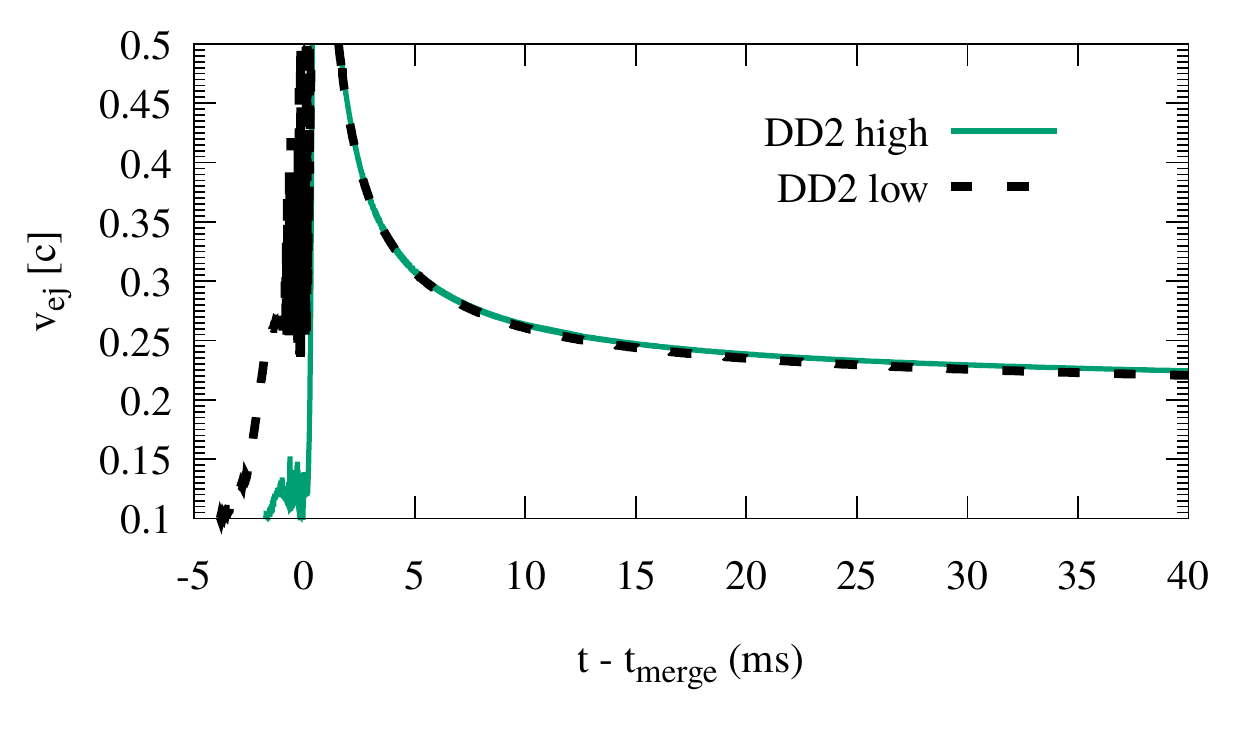} &
  \includegraphics[width=.48\linewidth]{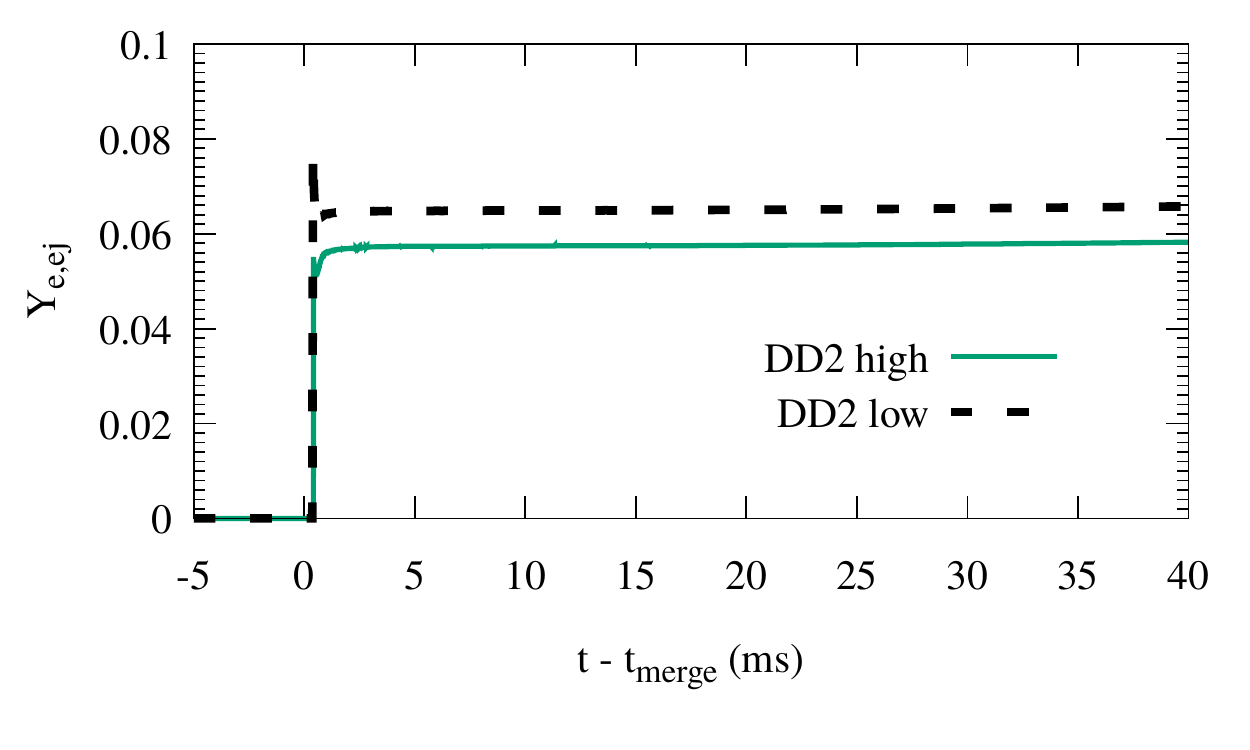}
 \end{tabular}
 \caption{Comparison of results for the DD2 model between high ($\Delta
 x = \SI{270}{\meter}$, solid curve) and low ($\Delta x =
 \SI{400}{\meter}$, dashed curve) resolutions. The former is the same as
 those have been shown in other figures. The top-left, top-right,
 bottom-left, and bottom-right panels show the neutrino luminosity for
 all the flavors, the ejecta mass, the ejecta velocity, and the averaged
 electron fraction of the ejecta, respectively. The luminosity curves
 for the heavy-lepton neutrinos are multiplied by a factor of 10 to make
 the plot visible in the top-left panel. The velocity right after the
 onset of merger is out of the vertical range and is not shown in the
 bottom-left panel as described in the caption of Fig.~\ref{fig:vej}.}
 \label{fig:res}
\end{figure*}

We check the convergence of our results by simulating the DD2 model with
a low resolution of $\Delta x = \SI{400}{\meter}$, which is coarser by
$\approx 50\%$ than our fiducial run. Figure \ref{fig:res} shows various
quantities derived by the high- (discussed so far) and low-resolution
simulations. By conservatively assuming first-order convergence, the
error in the high-resolution results will be twice the difference
between high- and low-resolution ones. While some physical quantities do
not converge very well with our current resolutions, we may safely
conclude that, e.g., the dynamical ejecta from black hole-neutron star
binaries have a low electron fraction of $Y_e \lesssim 0.1$ as discussed
in the following.

The top-left panel of Fig.~\ref{fig:res} compares the time evolution of
the neutrino luminosity. The peak luminosity agrees within a few \% for
the electron neutrino/antineutrino and 10\% for the heavy-lepton
neutrino. This agreement implies that our simulations appropriately
resolve the process of the disk formation. After the peak, the
luminosity is systematically higher by $\sim$ a few tens \% for the low
resolution irrespective of the flavors. This difference stems from
spurious heating at the low resolution due to enhanced numerical
dissipation in the remnant disk. Thus, the late-time luminosity may be
overestimated by a factor of up to $\sim 2$ for our fiducial runs under
adopted physical inputs. The realistic values should be set by physical
viscosity not modeled in this study.

Kinematic properties of the ejecta are approximately
convergent. Specifically, Fig.~\ref{fig:res} shows that the mass (top
right) and velocity (bottom left) agree within 5\% between two
resolutions. Thus, the error in the results of the high-resolution run
may be less than 10\%. Such good convergence is expected for the case in
which the ejecta mass is large \cite{kyutoku_iost2015} and may also
apply to the other models considered in this study.

The averaged electron fraction of the ejecta shown in the bottom-right
panel of Fig.~\ref{fig:res} is less convergent than the kinematic
properties, although the error does not seem crucial. The difference
between the high and low resolutions is $\approx 10\%$, and the error in
the results of the high-resolution run may be as large as $\approx
20\%$. We observe that the material with high values of $Y_e$ is
slightly more abundant for the low resolution than for the high
resolution, but the mass with $Y_e \ge 0.1$ is less than 0.1\% of the
ejecta mass and is much smaller than the numerical error. Thus, we may
safely conclude that the dynamical ejecta from black hole-neutron star
binaries have a low electron fraction of $Y_e \lesssim 0.1$.

\section{Result without neutrino absorption} \label{sec:nh}

\begin{figure}
 \includegraphics[width=.95\linewidth]{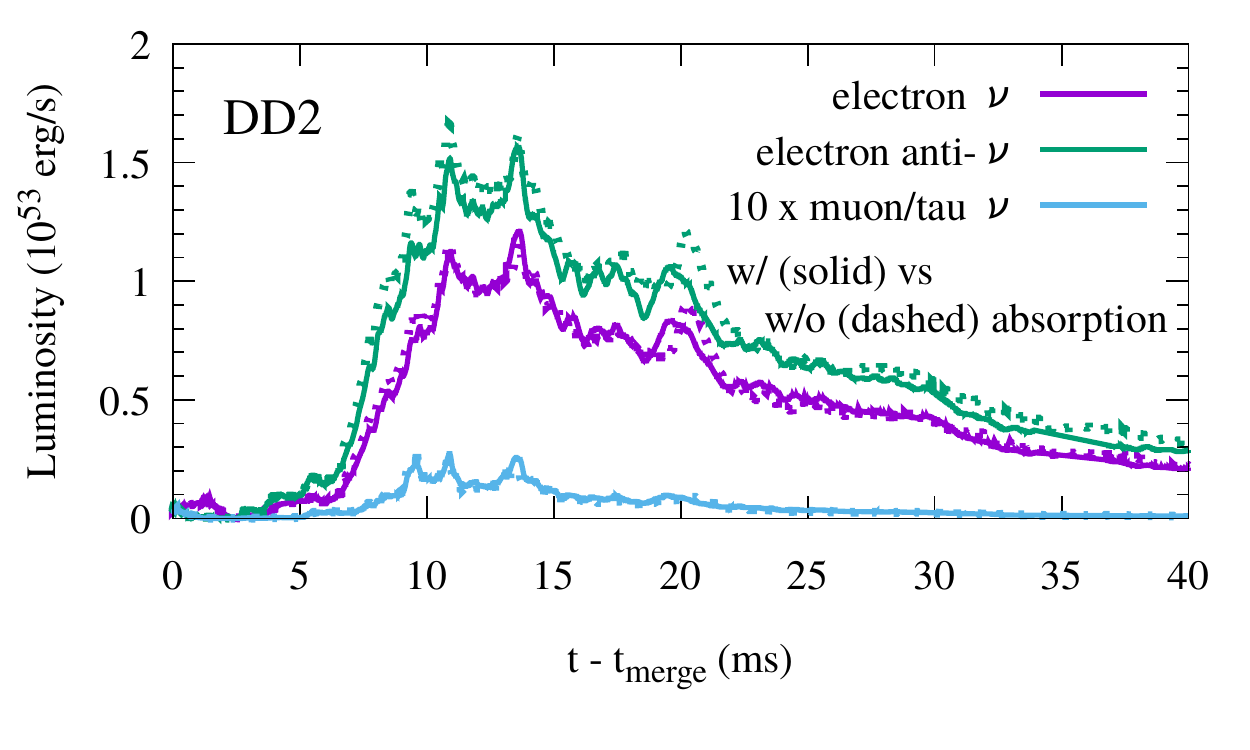} \caption{Neutrino
 luminosity curves of all the flavors for DD2. The color is the same as
 Fig.~\ref{fig:lightcurve}. The solid and dashed curves are results for
 simulations with and without neutrino absorption, respectively, and
 thus the former is the same as the middle panel of
 Fig.~\ref{fig:lightcurve}. The luminosity curves for the heavy-lepton
 neutrinos are multiplied by a factor of 10 to make the plot visible.}
 \label{fig:nhlightcurve}
\end{figure}

To single out the effect of neutrino transport, we perform a simulation
for the DD2 model without incorporating neutrino absorption. The
neutrino light curves are compared in Fig.~\ref{fig:nhlightcurve}. This
figure shows that the neutrino emission is not affected substantially by
the neutrino absorption onto the material. Note that the optical depth
is always taken into account for calculating the neutrino luminosity in
the leakage scheme irrespective of the neutrino transport. Close
inspection reveals that the luminosity becomes slightly low when the
absorption is taken into account as a natural outcome.

\begin{figure*}
 \begin{tabular}{ccc}
  \includegraphics[width=.48\linewidth]{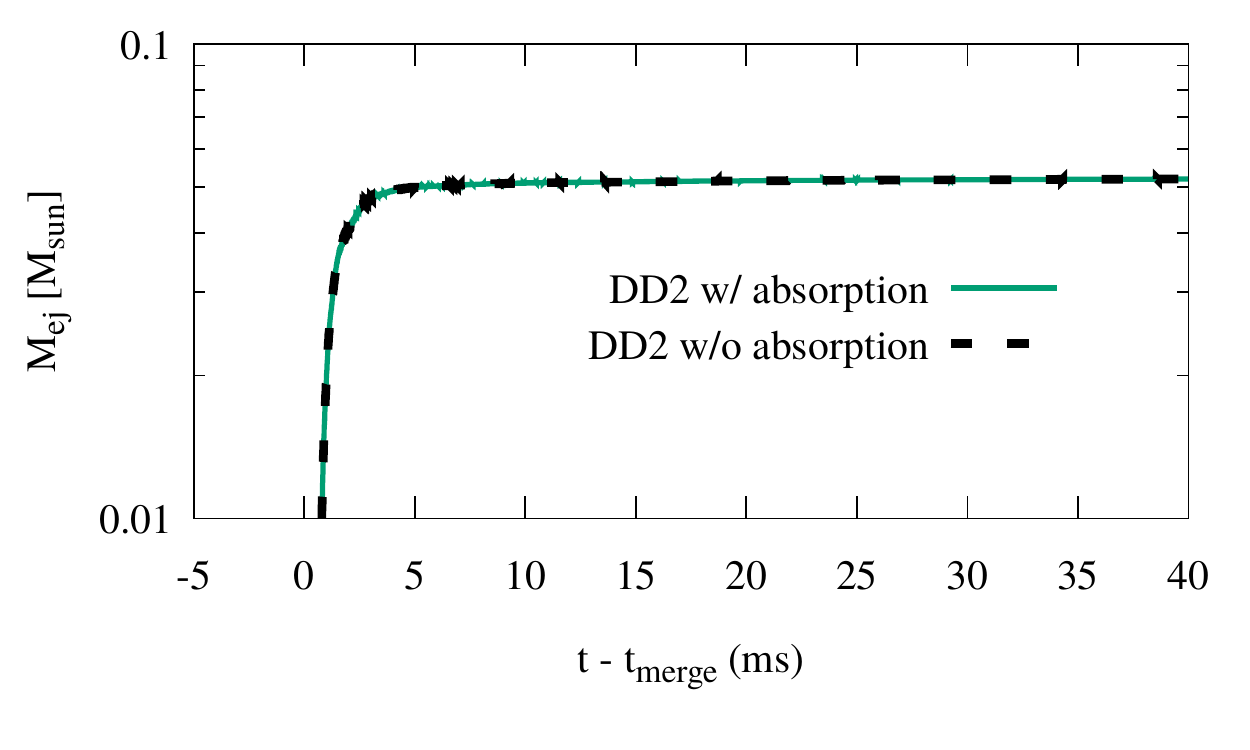} &
  \includegraphics[width=.48\linewidth]{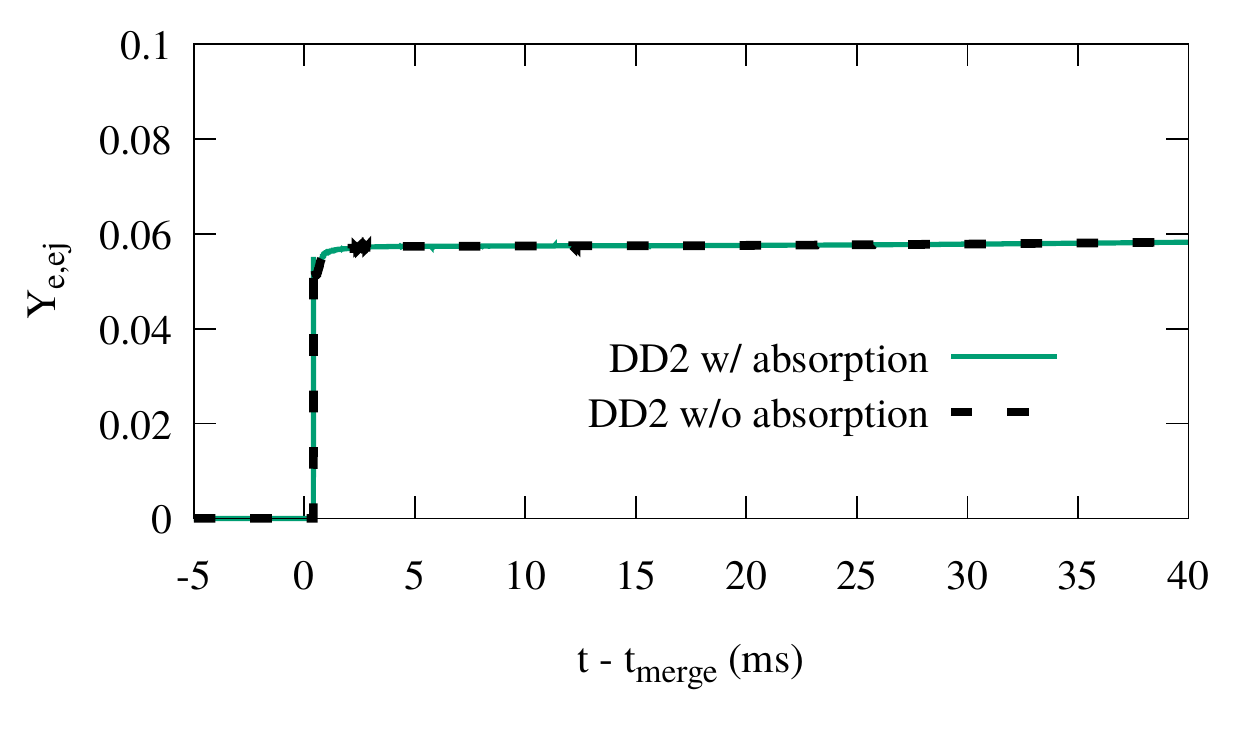}
 \end{tabular}
 \caption{Time evolution of the mass (left) and averaged electron
 fraction (right) of ejecta for the DD2 model with (solid) and without
 (dashed) neutrino absorption.} \label{fig:nhejecta}
\end{figure*}

The insignificance of the neutrino absorption in the mass ejection from
black hole-neutron star binaries is confirmed by comparing the results
of simulations with and without neutrino absorption. Figure
\ref{fig:nhejecta} shows the time evolution of the mass and averaged
electron fraction of the ejecta. Neither quantity changes appreciably as
a result of the neutrino absorption. The differences are much smaller
than the errors associated with the finite resolution discussed in
Sec.~\ref{sec:result_res}. This comparison clearly indicates that the
purely neutrino-driven wind is negligible and also that the neutrino
irradiation does not have a significant impact at least within our
approximate transport scheme for this model.

We also perform a simulation without neutrino absorption for the SFHo
model up to $t - t_\mathrm{merge} \approx \SI{20}{\milli\second}$ and
obtain the same conclusion. Because the ejecta mass is smaller for SFHo
than for DD2, this comparison gives stronger evidence of the
insignificance of the neutrino transport. Quantitatively, the averaged
electron fraction, $Y_{e,\mathrm{ej}}$, decreases only by $\lesssim
0.001$.

The ineffectiveness of neutrino absorption can be understood by
following analytic arguments. By approximating that the ejecta are
composed purely of neutrons, the time scale for neutrino absorption to
change the electron fraction of ejecta by (arbitrarily chosen) $\Delta
Y_{e,\mathrm{ej}} \approx 0.1$ may be estimated by [see Eq.~(3.3) of
Ref.~\cite{sekiguchi_kkst2016}]
\begin{align}
 t_\mathrm{abs} & \approx \Delta Y_{e,\mathrm{ej}} \left[ \frac{1}{4 \pi
 ( v_\mathrm{ej} t_\mathrm{exp} )^2} \frac{\sigma L_\nu}{\langle
 \epsilon_\nu \rangle} \right]^{-1} , \\
 & \approx \SI{100}{\milli\second} \times \left(
 \frac{t_\mathrm{exp}}{\SI{10}{\milli\second}} \right)^2 \left(
 \frac{L_\nu}{\SI{e53}{erg.s^{-1}}} \right)^{-1} ,
\end{align}
where $t_\mathrm{exp} \approx t - t_\mathrm{merge}$ is the time scale of
the expansion of the ejecta, $\sigma \approx
\SI{e-41}{\square\centi\meter}$ is the cross section for the capture of
neutrinos with average energy of $\langle \epsilon_\nu \rangle \approx
\SI{10}{\mega\electronvolt}$, and $L_\nu$ is the neutrino luminosity,
which takes a value of $\approx \SI{e53}{erg.s^{-1}}$ around the peak of
emission, $t_\mathrm{exp} \approx \SI{10}{\milli\second}$. The time
scale for the appreciable increase of internal energy is longer than
this, because the kinetic energy per nucleon, a few tens of
\si{\mega\electronvolt}, is higher than the average energy of
neutrinos. These time scales are always longer by at least an order of
magnitude than the expansion time scale of the ejecta,
$t_\mathrm{exp}$. In reality, $t_\mathrm{abs}$ is likely to be longer
than this estimate, because neutrinos have to catch up the ejecta, the
neutrino luminosity decreases in time, and the ejecta contain a small
fraction of protons. Thus, the neutrino absorption should be
ineffective, and our simulations confirm this expectation.

We note that our treatment of the neutrino absorption is approximate,
while the associated error is expected to be minor. Specifically, we
calculate the absorption rate in a local manner assuming that streaming
neutrinos obey the Fermi-Dirac distribution with the temperature of the
fluid. This prescription is expected to be valid in the optically thick
and also intermediate (gray) regions. However, the neutrino temperature
is underestimated in the cold and optically thin region, where high
temperatures in the emitting region should be appropriate. At the same
time, we assign finite chemical potential to streaming neutrinos by
requiring that the energy density agrees with the value obtained from
the time evolution. Mitigated by this finite chemical potential, the
absorption rate is underestimated by typically less than a factor of 2
even in the optically thin region. This amount of the error is unlikely
to affect our conclusion that the neutrino absorption is inefficient in
dynamical mass ejection from black hole-neutron star binaries. It is
also assuring that our results are consistent with those of a
postprocess study of neutrino irradiation for black hole-neutron star
binaries \cite{roberts_ldfflnop2017}, and the weakness of the purely
neutrino-driven wind from the hot remnant disk may be affected only very
weakly by this treatment of the temperature. Still, the assumption that
neutrinos obey the Fermi-Dirac distribution itself is not strictly valid
\cite{richers_kofo2015}, and multienergy transport simulations are
necessary to obtain reliable results (see also
Ref.~\cite{foucart_orkps2016}). We leave this task as our future study.

\section{Discussion} \label{sec:discussion}

\subsection{\textit{r}-process nucleosynthesis and macronova/kilonova}

The very neutron-rich dynamical ejecta from black hole-neutron star
binaries are promising as a site of the strong \textit{r}-process
nucleosynthesis, which produces the second and third peaks of the
abundance pattern. However, it is very unlikely that the dynamical
ejecta from black hole-neutron star binaries can produce elements below
the second peak, because the ejecta component with the electron fraction
appropriate for the first-peak production, say $Y_e \gtrsim 0.25$
\cite{wanajo_snkks2014,roberts_ldfflnop2017}, is very minor even under
the neutrino irradiation (see Fig.~\ref{fig:ydistrib}). Furthermore, the
purely neutrino-driven wind is negligible even if its electron fraction
could be high. Still, the remnants of black hole-neutron star binaries
could produce the first-peak elements if the viscously driven wind is
strong and only mildly neutron rich
\cite{just_bagj2015,wu_fmm2016}. Sufficient viscosity may be provided
via the magnetorotational instability in the accretion disk
\cite{kiuchi_skstw2015,siegel_metzger2017}, and we would like to study
the chemical composition of such wind components and associated
nucleosynthesis in the near future. Because the electron fraction of the
remnant disk is not necessarily as low as $Y_e = 0.1$, particularly if
the neutron star is compact like our SFHo model (see
Fig.~\ref{fig:snapdiskye}), strong \textit{r}-process nucleosynthesis
and lanthanide formation may be suppressed \cite{fernandez_fkldr2017}.

The dynamical ejecta from black hole-neutron star binaries will give
rise to a bright macronova/kilonova due to not only the large mass but
also various reasons. Because the lanthanides and actinides must be
synthesized enormously for the low electron fraction
\cite{tanaka_hotokezaka2013,kasen_bb2013}, the opacity should be similar
to that of the dynamical ejecta from binary neutron stars except for
high-$Y_e$ components around the polar direction
\cite{wanajo_snkks2014,sekiguchi_kks2015,sekiguchi_kkst2016} and will
not make a substantial difference in the emission. The luminosity during
the early, optically thick phase will be enhanced compared to the
spherical ejecta due to the anisotropic geometry, which increases the
temperature for a given heating source, and accordingly the color will
tend to be blue
\cite{kyutoku_is2013,tanaka_hkwkss2014,kyutoku_iost2015}.

Dependence of the heating efficiency on the electron fraction has
recently been discussed vigorously
\cite{hotokezaka_wtbtp2016,barnes_kwm2016}. The $r$-process
nucleosynthesis under a very low electron fraction produces transuranic
elements and even induces fission cycling, and thus the dynamical ejecta
from black hole-neutron star binaries should contain many heavy nuclides
that experience the fission and/or $\alpha$-decay. Decay heat liberated
in these channels is deposited more efficiently to the ejecta than that
in the $\beta$-decay, in which the energy can be taken away by escaping
neutrinos and $\gamma$ rays. While it is debated whether fissile
nuclides remain until the epoch relevant to the macronova/kilonova
\cite{hotokezaka_wtbtp2016,barnes_kwm2016}, decay heat of the
$\beta$-decay can also be thermalized efficiently in the anisotropic
ejecta due to the large optical depth to $\gamma$ rays associated with
the high density. Thus, the macronova/kilonova from black hole-neutron
star binaries will be brighter than that from binary neutron stars for
given values of the ejecta mass and kinetic energy.

\subsection{Short-hard gamma-ray burst}

The key to a successful short-hard gamma-ray burst is the launch of a
collimated ultrarelativistic jet. One plausible mechanism for the jet
launch is pair annihilation of neutrinos and antineutrinos in the polar
region of the black hole
\cite{goodman_dn1987,eichler_lps1989,mochkovitch_him1993}. In the
following, we discuss the implication of the results obtained by our
neutrino-radiation-hydrodynamics simulations focusing on this
mechanism. Our discussion may not apply to other mechanisms of the jet
launch, such as the magnetically driven model
\cite{paschalidis_rs2015,kiuchi_skstw2015}, whereas neutrino cooling and
heating are always essential to determine dynamics and properties of the
accretion flow \cite{kawanaka_pk2013}.

Our results suggest that a massive accretion disk is not necessarily
promising for driving an energetic jet from black hole-neutron star
binary mergers. Among the three models considered in this work, the peak
neutrino luminosity is higher when the equation of state is softer and
thus the neutron-star radius is smaller because of the higher
temperature in the disk. On another front, the large neutron-star radius
is advantageous for producing a massive accretion disk as is well known
\cite{kyutoku_st2010,kyutoku_st2010e,kyutoku_ost2011,foucart2012} and is
confirmed in this work. Thus, the large disk mass does not necessarily
give rise to high neutrino luminosity. Although our simulations do not
incorporate viscous heating that will affect late-time behavior, this
effect is not likely to modify the peak luminosity substantially,
because the temperature achieves an approximately virialized value at
the peak of emission due solely to the shock heating. Moreover, the disk
mass as an energy source decreases monotonically in time, and the
viscosity only accelerates the accretion. Therefore, the viscosity is
not likely to produce a later and stronger peak than that seen in this
study (but see also Refs.~\cite{lee_rp2005,setiawan_rj2006}). By
considering the quadratic dependence of the neutrino pair-annihilation
power on the neutrino luminosity
\cite{goodman_dn1987,eichler_lps1989,mochkovitch_him1993,fujibayashi_sks2017},
a large disk mass is not necessarily advantageous for launching a high
peak-luminosity jet via this process. It should be cautioned that the
peak neutrino luminosity may not be directly relevant to the total
energy of the jet, because it will be determined by late-time luminosity
that can be enhanced by viscosity. The annihilation efficiency can also
be enhanced if the viscosity changes the geometry of accretion flows
from the disk to a torus.

The accretion disks formed in our models of black hole-neutron star
binaries are not likely to sustain the jet for a duration longer than
$\approx \SI{100}{\milli\second}$. This is significantly shorter than
the typical duration of prompt emission for observed short-hard
gamma-ray bursts \cite{berger2014}. The viscosity neglected in this work
will not lengthen the emission time scale, because the accretion time
scale will only be decreased by the enhanced angular momentum
transport. Although our simulations explore a very limited region of
possible parameters, it will not be easy to obtain a sizable disk with
the accretion time scale much longer than
$\SI{100}{\milli\second}$. Thus, it will be challenging for black
hole-neutron star binaries to explain the duration of typical prompt
emission as well as extended emission \cite{kagawa_etal2015}.

The merger remnants of our current simulations are also lacking
plausible mechanisms to collimate the jet, while the viscosity is
essential to draw a conclusion on this topic. In the case of
binary-neutron-star mergers, the dynamical ejecta as well as the disk
wind will surround the polar region of the central black hole, and
indeed it has been shown that they can collimate a hypothetical jet via
hydrodynamic interaction depending on the models
\cite{nagakura_hssi2014,murguiabertheir_mrdl2014,duffell_qm2015,just_ojbs2016,murguiabertheir_rmdrrtpl2017}.
By contrast, our simulations indicate that neither the dynamical ejecta
nor the purely neutrino-driven wind supply substantial material to the
polar region in black hole-neutron star binary mergers. This fact does
not mean that the hydrodynamic collimation cannot work in black
hole-neutron star binaries, because the viscously driven wind together
with the neutrino irradiation could be responsible for feeding the polar
region. Note that the authors are not aware of simulations for jet
propagation starting from geometrically thin initial disk configurations
in this context. We would like to revisit this topic in the near future
with viscous-hydrodynamics simulations
\cite{shibata_ks2017,shibata_kiuchi2017}.

\section{Summary} \label{sec:summary}

We performed a series of neutrino-radiation-hydrodynamics simulations in
full general relativity for black hole-neutron star binary mergers. We
adopt an approximate but self-consistent neutrino transport scheme
including both emission (cooling) and absorption (heating). The neutron
star is modeled by three finite-temperature equations of state (SFHo,
DD2, and TM1), whereas we focus only on the cases that the neutron-star
mass ($1.35 M_\odot$), black-hole mass ($5.4 M_\odot$), and black-hole
spin ($0.75$ in terms of the dimensionless parameter and aligned with
the orbital angular momentum) are fixed.

We find that the mergers of our black hole-neutron star binary models
result in the formation of massive remnant disks with 0.2--$0.4M_\odot$
that emit a copious amount of the electron neutrinos and antineutrinos
with the peak luminosity $\sim 0.5$--\SI{2e53}{erg.s^{-1}} for each
flavor. Electron antineutrinos are brighter by 20\%--30\% than electron
neutrinos due to the predominant positron capture on neutrons in the
accretion disk. The emission of heavy-lepton neutrinos is quite minor
\cite{foucart_etal2014}, because the temperature in the remnant disk is
not high enough for the pair processes to become efficient. The peak
luminosity is higher when the neutron-star radius is smaller among the
three models considered in this study due to the higher temperature in
the accretion disk.

Properties of dynamical ejecta from black hole-neutron star binaries are
basically the same as those estimated in previous purely hydrodynamic
simulations \cite{kyutoku_iost2015}. In this work, we confirm the
expectation that the electron fraction of the dynamical ejecta is very
low keeping the initial composition of the cold neutron star. We show,
for the first time by merger simulations, that the neutrino irradiation
does not affect significantly the properties of the dynamical ejecta
such as the electron fraction by comparing the results obtained with and
without neutrino absorption. The reason is that the dynamical ejecta
escape to a distant region too rapidly to be irradiated by neutrinos
from the remnant disk. As a result, the extreme neutron richness of the
neutron-star material is approximately preserved during the dynamical
mass ejection irrespective of the neutrino transport. This fact
indicates that dynamical mass ejection from black hole-neutron star
binaries can be studied accurately without implementing a detailed
neutrino transport scheme unlike that from binary neutron stars
\cite{sekiguchi_kks2015,sekiguchi_kkst2016}. Our results also suggest
that the nucleosynthesis outcome will be dominated by heavy
\textit{r}-process nuclei around second and third peaks.

We also find that the remnant disks do not appreciably launch
neutrino-driven winds. This is consistent with previous studies of
accretion disks with different levels of sophistication
\cite{fernandez_metzger2013,fernandez_kmq2015,just_bagj2015,foucart_etal2015}. However,
it is premature to conclude that the neutrino transport does not play
any role in the mass ejection from black hole-neutron star binaries. If
magnetic fields and associated viscosity play an important role in
ejecting substantial material from the remnant \cite{kiuchi_skstw2015},
the neutrino interaction can be important for determining the properties
of ejecta. We leave such multiphysics simulations as our future task.

\begin{acknowledgments}
 We thank Francois Foucart and Norita Kawanaka for valuable
 discussions. Numerical simulations are performed on the supercomputer K
 at AICS (Project No. hp170313), Cray XC30 at CfCA of National
 Astronomical Observatory of Japan, and XC30 at Yukawa Institute for
 Theoretical Physics of Kyoto University. This work is supported by
 Japanese Society for Promotion of Science (JSPS) KAKENHI Grant-in-Aid
 for Scientific Research (No.~JP26400267, No.~JP15H00782,
 No.~JP15K05077, No.~JP16H02183, No.~JP16H06341, No.~JP16H06342,
 No.~JP16K17706, No.~JP17H01131, No.~JP17K05447, No.~JP17H06361, and
 No.~JP17H06363) and a post-K computer project (priority issue No.~9) of
 Japanese Ministry of Education, Culture, Sports, Science and Technology
 (MEXT).
\end{acknowledgments}


\begin{thebibliography}{119}%
\makeatletter
\providecommand \@ifxundefined [1]{%
 \@ifx{#1\undefined}
}%
\providecommand \@ifnum [1]{%
 \ifnum #1\expandafter \@firstoftwo
 \else \expandafter \@secondoftwo
 \fi
}%
\providecommand \@ifx [1]{%
 \ifx #1\expandafter \@firstoftwo
 \else \expandafter \@secondoftwo
 \fi
}%
\providecommand \natexlab [1]{#1}%
\providecommand \enquote  [1]{``#1''}%
\providecommand \bibnamefont  [1]{#1}%
\providecommand \bibfnamefont [1]{#1}%
\providecommand \citenamefont [1]{#1}%
\providecommand \href@noop [0]{\@secondoftwo}%
\providecommand \href [0]{\begingroup \@sanitize@url \@href}%
\providecommand \@href[1]{\@@startlink{#1}\@@href}%
\providecommand \@@href[1]{\endgroup#1\@@endlink}%
\providecommand \@sanitize@url [0]{\catcode `\\12\catcode `\$12\catcode
  `\&12\catcode `\#12\catcode `\^12\catcode `\_12\catcode `\%12\relax}%
\providecommand \@@startlink[1]{}%
\providecommand \@@endlink[0]{}%
\providecommand \url  [0]{\begingroup\@sanitize@url \@url }%
\providecommand \@url [1]{\endgroup\@href {#1}{\urlprefix }}%
\providecommand \urlprefix  [0]{URL }%
\providecommand \Eprint [0]{\href }%
\providecommand \doibase [0]{http://dx.doi.org/}%
\providecommand \selectlanguage [0]{\@gobble}%
\providecommand \bibinfo  [0]{\@secondoftwo}%
\providecommand \bibfield  [0]{\@secondoftwo}%
\providecommand \translation [1]{[#1]}%
\providecommand \BibitemOpen [0]{}%
\providecommand \bibitemStop [0]{}%
\providecommand \bibitemNoStop [0]{.\EOS\space}%
\providecommand \EOS [0]{\spacefactor3000\relax}%
\providecommand \BibitemShut  [1]{\csname bibitem#1\endcsname}%
\let\auto@bib@innerbib\@empty
\bibitem [{\citenamefont {Rosswog}(2005)}]{rosswog2005}%
  \BibitemOpen
  \bibfield  {author} {\bibinfo {author} {\bibfnamefont {S.}~\bibnamefont
  {Rosswog}},\ }\href {\doibase 10.1086/497062} {\bibfield  {journal} {\bibinfo
   {journal} {Astrophys. J.}\ }\textbf {\bibinfo {volume} {634}},\ \bibinfo
  {pages} {1202} (\bibinfo {year} {2005})}\BibitemShut {NoStop}%
\bibitem [{\citenamefont {Rantsiou}\ \emph {et~al.}(2008)\citenamefont
  {Rantsiou}, \citenamefont {Kobayashi}, \citenamefont {Laguna},\ and\
  \citenamefont {Rasio}}]{rantsiou_klr2008}%
  \BibitemOpen
  \bibfield  {author} {\bibinfo {author} {\bibfnamefont {E.}~\bibnamefont
  {Rantsiou}}, \bibinfo {author} {\bibfnamefont {S.}~\bibnamefont {Kobayashi}},
  \bibinfo {author} {\bibfnamefont {P.}~\bibnamefont {Laguna}}, \ and\ \bibinfo
  {author} {\bibfnamefont {F.~A.}\ \bibnamefont {Rasio}},\ }\href {\doibase
  10.1086/587858} {\bibfield  {journal} {\bibinfo  {journal} {Astrophys. J.}\
  }\textbf {\bibinfo {volume} {680}},\ \bibinfo {pages} {1326} (\bibinfo {year}
  {2008})}\BibitemShut {NoStop}%
\bibitem [{\citenamefont {Kyutoku}\ \emph
  {et~al.}(2011{\natexlab{a}})\citenamefont {Kyutoku}, \citenamefont {Okawa},
  \citenamefont {Shibata},\ and\ \citenamefont {Taniguchi}}]{kyutoku_ost2011}%
  \BibitemOpen
  \bibfield  {author} {\bibinfo {author} {\bibfnamefont {K.}~\bibnamefont
  {Kyutoku}}, \bibinfo {author} {\bibfnamefont {H.}~\bibnamefont {Okawa}},
  \bibinfo {author} {\bibfnamefont {M.}~\bibnamefont {Shibata}}, \ and\
  \bibinfo {author} {\bibfnamefont {K.}~\bibnamefont {Taniguchi}},\ }\href
  {\doibase 10.1103/PhysRevD.84.064018} {\bibfield  {journal} {\bibinfo
  {journal} {Phys. Rev. D}\ }\textbf {\bibinfo {volume} {84}},\ \bibinfo
  {pages} {064018} (\bibinfo {year} {2011}{\natexlab{a}})}\BibitemShut
  {NoStop}%
\bibitem [{\citenamefont {Rosswog}\ \emph {et~al.}(2013)\citenamefont
  {Rosswog}, \citenamefont {Piran},\ and\ \citenamefont
  {Nakar}}]{rosswog_pn2013}%
  \BibitemOpen
  \bibfield  {author} {\bibinfo {author} {\bibfnamefont {S.}~\bibnamefont
  {Rosswog}}, \bibinfo {author} {\bibfnamefont {T.}~\bibnamefont {Piran}}, \
  and\ \bibinfo {author} {\bibfnamefont {E.}~\bibnamefont {Nakar}},\ }\href
  {\doibase 10.1093/mnras/sts708} {\bibfield  {journal} {\bibinfo  {journal}
  {Mon. Not. R. Astron. Soc.}\ }\textbf {\bibinfo {volume} {430}},\ \bibinfo
  {pages} {2585} (\bibinfo {year} {2013})}\BibitemShut {NoStop}%
\bibitem [{\citenamefont {Foucart}\ \emph {et~al.}(2013)\citenamefont
  {Foucart}, \citenamefont {Deaton}, \citenamefont {Duez}, \citenamefont
  {Kidder}, \citenamefont {MacDonald}, \citenamefont {Ott}, \citenamefont
  {Pfeiffer}, \citenamefont {Scheel}, \citenamefont {Szilagyi},\ and\
  \citenamefont {Teukolsky}}]{foucart_ddkmopsst2013}%
  \BibitemOpen
  \bibfield  {author} {\bibinfo {author} {\bibfnamefont {F.}~\bibnamefont
  {Foucart}}, \bibinfo {author} {\bibfnamefont {M.~B.}\ \bibnamefont {Deaton}},
  \bibinfo {author} {\bibfnamefont {M.~D.}\ \bibnamefont {Duez}}, \bibinfo
  {author} {\bibfnamefont {L.~E.}\ \bibnamefont {Kidder}}, \bibinfo {author}
  {\bibfnamefont {I.}~\bibnamefont {MacDonald}}, \bibinfo {author}
  {\bibfnamefont {C.~D.}\ \bibnamefont {Ott}}, \bibinfo {author} {\bibfnamefont
  {H.~P.}\ \bibnamefont {Pfeiffer}}, \bibinfo {author} {\bibfnamefont {M.~A.}\
  \bibnamefont {Scheel}}, \bibinfo {author} {\bibfnamefont {B.}~\bibnamefont
  {Szilagyi}}, \ and\ \bibinfo {author} {\bibfnamefont {S.~A.}\ \bibnamefont
  {Teukolsky}},\ }\href {\doibase 10.1103/PhysRevD.87.084006} {\bibfield
  {journal} {\bibinfo  {journal} {Phys. Rev. D}\ }\textbf {\bibinfo {volume}
  {87}},\ \bibinfo {pages} {084006} (\bibinfo {year} {2013})}\BibitemShut
  {NoStop}%
\bibitem [{\citenamefont {Lovelace}\ \emph {et~al.}(2013)\citenamefont
  {Lovelace}, \citenamefont {Duez}, \citenamefont {Foucart}, \citenamefont
  {Kidder}, \citenamefont {Pfeiffer}, \citenamefont {Scheel},\ and\
  \citenamefont {Szil{\'a}gyi}}]{lovelace_dfkpss2013}%
  \BibitemOpen
  \bibfield  {author} {\bibinfo {author} {\bibfnamefont {G.}~\bibnamefont
  {Lovelace}}, \bibinfo {author} {\bibfnamefont {M.~D.}\ \bibnamefont {Duez}},
  \bibinfo {author} {\bibfnamefont {F.}~\bibnamefont {Foucart}}, \bibinfo
  {author} {\bibfnamefont {L.~E.}\ \bibnamefont {Kidder}}, \bibinfo {author}
  {\bibfnamefont {H.~P.}\ \bibnamefont {Pfeiffer}}, \bibinfo {author}
  {\bibfnamefont {M.~A.}\ \bibnamefont {Scheel}}, \ and\ \bibinfo {author}
  {\bibfnamefont {B.}~\bibnamefont {Szil{\'a}gyi}},\ }\href {\doibase
  10.1088/0264-9381/30/13/135004} {\bibfield  {journal} {\bibinfo  {journal}
  {Classical Quantum Gravity}\ }\textbf {\bibinfo {volume} {30}},\ \bibinfo
  {pages} {135004} (\bibinfo {year} {2013})}\BibitemShut {NoStop}%
\bibitem [{\citenamefont {Kyutoku}\ \emph {et~al.}(2013)\citenamefont
  {Kyutoku}, \citenamefont {Ioka},\ and\ \citenamefont
  {Shibata}}]{kyutoku_is2013}%
  \BibitemOpen
  \bibfield  {author} {\bibinfo {author} {\bibfnamefont {K.}~\bibnamefont
  {Kyutoku}}, \bibinfo {author} {\bibfnamefont {K.}~\bibnamefont {Ioka}}, \
  and\ \bibinfo {author} {\bibfnamefont {M.}~\bibnamefont {Shibata}},\ }\href
  {\doibase 10.1103/PhysRevD.88.041503} {\bibfield  {journal} {\bibinfo
  {journal} {Phys. Rev. D}\ }\textbf {\bibinfo {volume} {88}},\ \bibinfo
  {pages} {041503} (\bibinfo {year} {2013})}\BibitemShut {NoStop}%
\bibitem [{\citenamefont {Deaton}\ \emph {et~al.}(2013)\citenamefont {Deaton},
  \citenamefont {Duez}, \citenamefont {Foucart}, \citenamefont {O'Connor},
  \citenamefont {Ott}, \citenamefont {Kidder}, \citenamefont {Muhlberger},
  \citenamefont {Scheel},\ and\ \citenamefont
  {Szilagyi}}]{deaton_dfookmss2013}%
  \BibitemOpen
  \bibfield  {author} {\bibinfo {author} {\bibfnamefont {M.~B.}\ \bibnamefont
  {Deaton}}, \bibinfo {author} {\bibfnamefont {M.~D.}\ \bibnamefont {Duez}},
  \bibinfo {author} {\bibfnamefont {F.}~\bibnamefont {Foucart}}, \bibinfo
  {author} {\bibfnamefont {E.}~\bibnamefont {O'Connor}}, \bibinfo {author}
  {\bibfnamefont {C.~D.}\ \bibnamefont {Ott}}, \bibinfo {author} {\bibfnamefont
  {L.~E.}\ \bibnamefont {Kidder}}, \bibinfo {author} {\bibfnamefont {C.~D.}\
  \bibnamefont {Muhlberger}}, \bibinfo {author} {\bibfnamefont {M.~A.}\
  \bibnamefont {Scheel}}, \ and\ \bibinfo {author} {\bibfnamefont
  {B.}~\bibnamefont {Szilagyi}},\ }\href {\doibase 10.1088/0004-637X/776/1/47}
  {\bibfield  {journal} {\bibinfo  {journal} {Astrophys. J.}\ }\textbf
  {\bibinfo {volume} {776}},\ \bibinfo {pages} {47} (\bibinfo {year}
  {2013})}\BibitemShut {NoStop}%
\bibitem [{\citenamefont {Foucart}\ \emph {et~al.}(2014)\citenamefont
  {Foucart}, \citenamefont {Deaton}, \citenamefont {Duez}, \citenamefont
  {O'Connor}, \citenamefont {Ott}, \citenamefont {Haas}, \citenamefont
  {Kidder}, \citenamefont {Pfeiffer}, \citenamefont {Scheel},\ and\
  \citenamefont {Szilagyi}}]{foucart_etal2014}%
  \BibitemOpen
  \bibfield  {author} {\bibinfo {author} {\bibfnamefont {F.}~\bibnamefont
  {Foucart}}, \bibinfo {author} {\bibfnamefont {M.~B.}\ \bibnamefont {Deaton}},
  \bibinfo {author} {\bibfnamefont {M.~D.}\ \bibnamefont {Duez}}, \bibinfo
  {author} {\bibfnamefont {E.}~\bibnamefont {O'Connor}}, \bibinfo {author}
  {\bibfnamefont {C.~D.}\ \bibnamefont {Ott}}, \bibinfo {author} {\bibfnamefont
  {R.}~\bibnamefont {Haas}}, \bibinfo {author} {\bibfnamefont {L.~E.}\
  \bibnamefont {Kidder}}, \bibinfo {author} {\bibfnamefont {H.~P.}\
  \bibnamefont {Pfeiffer}}, \bibinfo {author} {\bibfnamefont {M.~A.}\
  \bibnamefont {Scheel}}, \ and\ \bibinfo {author} {\bibfnamefont
  {B.}~\bibnamefont {Szilagyi}},\ }\href {\doibase 10.1103/PhysRevD.90.024026}
  {\bibfield  {journal} {\bibinfo  {journal} {Phys. Rev. D}\ }\textbf {\bibinfo
  {volume} {90}},\ \bibinfo {pages} {024026} (\bibinfo {year}
  {2014})}\BibitemShut {NoStop}%
\bibitem [{\citenamefont {Kawaguchi}\ \emph {et~al.}(2015)\citenamefont
  {Kawaguchi}, \citenamefont {Kyutoku}, \citenamefont {Nakano}, \citenamefont
  {Okawa}, \citenamefont {Shibata},\ and\ \citenamefont
  {Taniguchi}}]{kawaguchi_knost2015}%
  \BibitemOpen
  \bibfield  {author} {\bibinfo {author} {\bibfnamefont {K.}~\bibnamefont
  {Kawaguchi}}, \bibinfo {author} {\bibfnamefont {K.}~\bibnamefont {Kyutoku}},
  \bibinfo {author} {\bibfnamefont {H.}~\bibnamefont {Nakano}}, \bibinfo
  {author} {\bibfnamefont {H.}~\bibnamefont {Okawa}}, \bibinfo {author}
  {\bibfnamefont {M.}~\bibnamefont {Shibata}}, \ and\ \bibinfo {author}
  {\bibfnamefont {K.}~\bibnamefont {Taniguchi}},\ }\href {\doibase
  10.1103/PhysRevD.92.024014} {\bibfield  {journal} {\bibinfo  {journal} {Phys.
  Rev. D}\ }\textbf {\bibinfo {volume} {92}},\ \bibinfo {pages} {024014}
  (\bibinfo {year} {2015})}\BibitemShut {NoStop}%
\bibitem [{\citenamefont {Kyutoku}\ \emph {et~al.}(2015)\citenamefont
  {Kyutoku}, \citenamefont {Ioka}, \citenamefont {Okawa}, \citenamefont
  {Shibata},\ and\ \citenamefont {Taniguchi}}]{kyutoku_iost2015}%
  \BibitemOpen
  \bibfield  {author} {\bibinfo {author} {\bibfnamefont {K.}~\bibnamefont
  {Kyutoku}}, \bibinfo {author} {\bibfnamefont {K.}~\bibnamefont {Ioka}},
  \bibinfo {author} {\bibfnamefont {H.}~\bibnamefont {Okawa}}, \bibinfo
  {author} {\bibfnamefont {M.}~\bibnamefont {Shibata}}, \ and\ \bibinfo
  {author} {\bibfnamefont {K.}~\bibnamefont {Taniguchi}},\ }\href {\doibase
  10.1103/PhysRevD.92.044028} {\bibfield  {journal} {\bibinfo  {journal} {Phys.
  Rev. D}\ }\textbf {\bibinfo {volume} {92}},\ \bibinfo {pages} {044028}
  (\bibinfo {year} {2015})}\BibitemShut {NoStop}%
\bibitem [{\citenamefont {Kiuchi}\ \emph {et~al.}(2015)\citenamefont {Kiuchi},
  \citenamefont {Sekiguchi}, \citenamefont {Kyutoku}, \citenamefont {Shibata},
  \citenamefont {Taniguchi},\ and\ \citenamefont {Wada}}]{kiuchi_skstw2015}%
  \BibitemOpen
  \bibfield  {author} {\bibinfo {author} {\bibfnamefont {K.}~\bibnamefont
  {Kiuchi}}, \bibinfo {author} {\bibfnamefont {Y.}~\bibnamefont {Sekiguchi}},
  \bibinfo {author} {\bibfnamefont {K.}~\bibnamefont {Kyutoku}}, \bibinfo
  {author} {\bibfnamefont {M.}~\bibnamefont {Shibata}}, \bibinfo {author}
  {\bibfnamefont {K.}~\bibnamefont {Taniguchi}}, \ and\ \bibinfo {author}
  {\bibfnamefont {T.}~\bibnamefont {Wada}},\ }\href {\doibase
  10.1103/PhysRevD.92.064034} {\bibfield  {journal} {\bibinfo  {journal} {Phys.
  Rev. D}\ }\textbf {\bibinfo {volume} {92}},\ \bibinfo {pages} {064034}
  (\bibinfo {year} {2015})}\BibitemShut {NoStop}%
\bibitem [{\citenamefont {Just}\ \emph {et~al.}(2015)\citenamefont {Just},
  \citenamefont {Bauswein}, \citenamefont {Pulpillo}, \citenamefont {Goriely},\
  and\ \citenamefont {Janka}}]{just_bagj2015}%
  \BibitemOpen
  \bibfield  {author} {\bibinfo {author} {\bibfnamefont {O.}~\bibnamefont
  {Just}}, \bibinfo {author} {\bibfnamefont {A.}~\bibnamefont {Bauswein}},
  \bibinfo {author} {\bibfnamefont {R.~A.}\ \bibnamefont {Pulpillo}}, \bibinfo
  {author} {\bibfnamefont {S.}~\bibnamefont {Goriely}}, \ and\ \bibinfo
  {author} {\bibfnamefont {H.-T.}\ \bibnamefont {Janka}},\ }\href {\doibase
  10.1093/mnras/stv009} {\bibfield  {journal} {\bibinfo  {journal} {Mon. Not.
  R. Astron. Soc.}\ }\textbf {\bibinfo {volume} {448}},\ \bibinfo {pages} {541}
  (\bibinfo {year} {2015})}\BibitemShut {NoStop}%
\bibitem [{\citenamefont {Foucart}\ \emph {et~al.}(2017)\citenamefont
  {Foucart}, \citenamefont {Desai}, \citenamefont {Brege}, \citenamefont
  {Duez}, \citenamefont {Kasen}, \citenamefont {Hemberger}, \citenamefont
  {Kidder}, \citenamefont {Pfeiffer},\ and\ \citenamefont
  {Scheel}}]{foucart_dbdkhkps2017}%
  \BibitemOpen
  \bibfield  {author} {\bibinfo {author} {\bibfnamefont {F.}~\bibnamefont
  {Foucart}}, \bibinfo {author} {\bibfnamefont {D.}~\bibnamefont {Desai}},
  \bibinfo {author} {\bibfnamefont {W.}~\bibnamefont {Brege}}, \bibinfo
  {author} {\bibfnamefont {M.~D.}\ \bibnamefont {Duez}}, \bibinfo {author}
  {\bibfnamefont {D.}~\bibnamefont {Kasen}}, \bibinfo {author} {\bibfnamefont
  {D.~A.}\ \bibnamefont {Hemberger}}, \bibinfo {author} {\bibfnamefont {L.~E.}\
  \bibnamefont {Kidder}}, \bibinfo {author} {\bibfnamefont {H.~P.}\
  \bibnamefont {Pfeiffer}}, \ and\ \bibinfo {author} {\bibfnamefont {M.~A.}\
  \bibnamefont {Scheel}},\ }\href {\doibase 10.1088/1361-6382/aa573b}
  {\bibfield  {journal} {\bibinfo  {journal} {Classical Quantum Gravity}\
  }\textbf {\bibinfo {volume} {34}},\ \bibinfo {pages} {044002} (\bibinfo
  {year} {2017})}\BibitemShut {NoStop}%
\bibitem [{\citenamefont {Hotokezaka}\ \emph
  {et~al.}(2013{\natexlab{a}})\citenamefont {Hotokezaka}, \citenamefont
  {Kiuchi}, \citenamefont {Kyutoku}, \citenamefont {Okawa}, \citenamefont
  {Sekiguchi}, \citenamefont {Shibata},\ and\ \citenamefont
  {Taniguchi}}]{hotokezaka_kkosst2013}%
  \BibitemOpen
  \bibfield  {author} {\bibinfo {author} {\bibfnamefont {K.}~\bibnamefont
  {Hotokezaka}}, \bibinfo {author} {\bibfnamefont {K.}~\bibnamefont {Kiuchi}},
  \bibinfo {author} {\bibfnamefont {K.}~\bibnamefont {Kyutoku}}, \bibinfo
  {author} {\bibfnamefont {H.}~\bibnamefont {Okawa}}, \bibinfo {author}
  {\bibfnamefont {Y.-i.}\ \bibnamefont {Sekiguchi}}, \bibinfo {author}
  {\bibfnamefont {M.}~\bibnamefont {Shibata}}, \ and\ \bibinfo {author}
  {\bibfnamefont {K.}~\bibnamefont {Taniguchi}},\ }\href {\doibase
  10.1103/PhysRevD.87.024001} {\bibfield  {journal} {\bibinfo  {journal} {Phys.
  Rev. D}\ }\textbf {\bibinfo {volume} {87}},\ \bibinfo {pages} {024001}
  (\bibinfo {year} {2013}{\natexlab{a}})}\BibitemShut {NoStop}%
\bibitem [{\citenamefont {Bauswein}\ \emph {et~al.}(2013)\citenamefont
  {Bauswein}, \citenamefont {Goriely},\ and\ \citenamefont
  {Janka}}]{bauswein_gj2013}%
  \BibitemOpen
  \bibfield  {author} {\bibinfo {author} {\bibfnamefont {A.}~\bibnamefont
  {Bauswein}}, \bibinfo {author} {\bibfnamefont {S.}~\bibnamefont {Goriely}}, \
  and\ \bibinfo {author} {\bibfnamefont {H.-T.}\ \bibnamefont {Janka}},\ }\href
  {\doibase 10.1088/0004-637X/773/1/78} {\bibfield  {journal} {\bibinfo
  {journal} {Astrophys. J.}\ }\textbf {\bibinfo {volume} {773}},\ \bibinfo
  {pages} {78} (\bibinfo {year} {2013})}\BibitemShut {NoStop}%
\bibitem [{\citenamefont {Sekiguchi}\ \emph {et~al.}(2015)\citenamefont
  {Sekiguchi}, \citenamefont {Kiuchi}, \citenamefont {Kyutoku},\ and\
  \citenamefont {Shibata}}]{sekiguchi_kks2015}%
  \BibitemOpen
  \bibfield  {author} {\bibinfo {author} {\bibfnamefont {Y.}~\bibnamefont
  {Sekiguchi}}, \bibinfo {author} {\bibfnamefont {K.}~\bibnamefont {Kiuchi}},
  \bibinfo {author} {\bibfnamefont {K.}~\bibnamefont {Kyutoku}}, \ and\
  \bibinfo {author} {\bibfnamefont {M.}~\bibnamefont {Shibata}},\ }\href
  {\doibase 10.1103/PhysRevD.91.064059} {\bibfield  {journal} {\bibinfo
  {journal} {Phys. Rev. D}\ }\textbf {\bibinfo {volume} {91}},\ \bibinfo
  {pages} {064059} (\bibinfo {year} {2015})}\BibitemShut {NoStop}%
\bibitem [{\citenamefont {Palenzuela}\ \emph {et~al.}(2015)\citenamefont
  {Palenzuela}, \citenamefont {Liebling}, \citenamefont {Neilsen},
  \citenamefont {Lehner}, \citenamefont {Caballero}, \citenamefont {O'Connor},\
  and\ \citenamefont {Anderson}}]{palenzuela_lnlcoa2015}%
  \BibitemOpen
  \bibfield  {author} {\bibinfo {author} {\bibfnamefont {C.}~\bibnamefont
  {Palenzuela}}, \bibinfo {author} {\bibfnamefont {S.~L.}\ \bibnamefont
  {Liebling}}, \bibinfo {author} {\bibfnamefont {D.}~\bibnamefont {Neilsen}},
  \bibinfo {author} {\bibfnamefont {L.}~\bibnamefont {Lehner}}, \bibinfo
  {author} {\bibfnamefont {O.~L.}\ \bibnamefont {Caballero}}, \bibinfo {author}
  {\bibfnamefont {E.}~\bibnamefont {O'Connor}}, \ and\ \bibinfo {author}
  {\bibfnamefont {M.}~\bibnamefont {Anderson}},\ }\href {\doibase
  10.1103/PhysRevD.92.044045} {\bibfield  {journal} {\bibinfo  {journal} {Phys.
  Rev. D}\ }\textbf {\bibinfo {volume} {92}},\ \bibinfo {pages} {044045}
  (\bibinfo {year} {2015})}\BibitemShut {NoStop}%
\bibitem [{\citenamefont {Sekiguchi}\ \emph {et~al.}(2016)\citenamefont
  {Sekiguchi}, \citenamefont {Kiuchi}, \citenamefont {Kyutoku}, \citenamefont
  {Shibata},\ and\ \citenamefont {Taniguchi}}]{sekiguchi_kkst2016}%
  \BibitemOpen
  \bibfield  {author} {\bibinfo {author} {\bibfnamefont {Y.}~\bibnamefont
  {Sekiguchi}}, \bibinfo {author} {\bibfnamefont {K.}~\bibnamefont {Kiuchi}},
  \bibinfo {author} {\bibfnamefont {K.}~\bibnamefont {Kyutoku}}, \bibinfo
  {author} {\bibfnamefont {M.}~\bibnamefont {Shibata}}, \ and\ \bibinfo
  {author} {\bibfnamefont {K.}~\bibnamefont {Taniguchi}},\ }\href {\doibase
  10.1103/PhysRevD.93.124046} {\bibfield  {journal} {\bibinfo  {journal} {Phys.
  Rev. D}\ }\textbf {\bibinfo {volume} {93}},\ \bibinfo {pages} {124046}
  (\bibinfo {year} {2016})}\BibitemShut {NoStop}%
\bibitem [{\citenamefont {Radice}\ \emph {et~al.}(2016)\citenamefont {Radice},
  \citenamefont {Galeazzi}, \citenamefont {Lippuner}, \citenamefont {Roberts},
  \citenamefont {Ott},\ and\ \citenamefont {Rezzolla}}]{radice_glror2016}%
  \BibitemOpen
  \bibfield  {author} {\bibinfo {author} {\bibfnamefont {D.}~\bibnamefont
  {Radice}}, \bibinfo {author} {\bibfnamefont {F.}~\bibnamefont {Galeazzi}},
  \bibinfo {author} {\bibfnamefont {J.}~\bibnamefont {Lippuner}}, \bibinfo
  {author} {\bibfnamefont {L.~F.}\ \bibnamefont {Roberts}}, \bibinfo {author}
  {\bibfnamefont {C.~D.}\ \bibnamefont {Ott}}, \ and\ \bibinfo {author}
  {\bibfnamefont {L.}~\bibnamefont {Rezzolla}},\ }\href {\doibase
  10.1093/mnras/stw1227} {\bibfield  {journal} {\bibinfo  {journal} {Mon. Not.
  R. Astron. Soc.}\ }\textbf {\bibinfo {volume} {460}},\ \bibinfo {pages}
  {3255} (\bibinfo {year} {2016})}\BibitemShut {NoStop}%
\bibitem [{\citenamefont {Foucart}\ \emph
  {et~al.}(2016{\natexlab{a}})\citenamefont {Foucart}, \citenamefont
  {O'Connor}, \citenamefont {Roberts}, \citenamefont {Kidder}, \citenamefont
  {Pfeiffer},\ and\ \citenamefont {Scheel}}]{foucart_orkps2016}%
  \BibitemOpen
  \bibfield  {author} {\bibinfo {author} {\bibfnamefont {F.}~\bibnamefont
  {Foucart}}, \bibinfo {author} {\bibfnamefont {E.}~\bibnamefont {O'Connor}},
  \bibinfo {author} {\bibfnamefont {L.}~\bibnamefont {Roberts}}, \bibinfo
  {author} {\bibfnamefont {L.~E.}\ \bibnamefont {Kidder}}, \bibinfo {author}
  {\bibfnamefont {H.~P.}\ \bibnamefont {Pfeiffer}}, \ and\ \bibinfo {author}
  {\bibfnamefont {M.~A.}\ \bibnamefont {Scheel}},\ }\href {\doibase
  10.1103/PhysRevD.94.123016} {\bibfield  {journal} {\bibinfo  {journal} {Phys.
  Rev. D}\ }\textbf {\bibinfo {volume} {94}},\ \bibinfo {pages} {123016}
  (\bibinfo {year} {2016}{\natexlab{a}})}\BibitemShut {NoStop}%
\bibitem [{\citenamefont {Lattimer}\ and\ \citenamefont
  {Schramm}(1974)}]{lattimer_schramm1974}%
  \BibitemOpen
  \bibfield  {author} {\bibinfo {author} {\bibfnamefont {J.~M.}\ \bibnamefont
  {Lattimer}}\ and\ \bibinfo {author} {\bibfnamefont {D.~N.}\ \bibnamefont
  {Schramm}},\ }\href {\doibase 10.1086/181612} {\bibfield  {journal} {\bibinfo
   {journal} {Astrophys. J.}\ }\textbf {\bibinfo {volume} {192}},\ \bibinfo
  {pages} {L145} (\bibinfo {year} {1974})}\BibitemShut {NoStop}%
\bibitem [{\citenamefont {Li}\ and\ \citenamefont
  {Paczy{\'n}ski}(1998)}]{li_paczynski1998}%
  \BibitemOpen
  \bibfield  {author} {\bibinfo {author} {\bibfnamefont {L.-X.}\ \bibnamefont
  {Li}}\ and\ \bibinfo {author} {\bibfnamefont {B.}~\bibnamefont
  {Paczy{\'n}ski}},\ }\href {\doibase 10.1086/311680} {\bibfield  {journal}
  {\bibinfo  {journal} {Astrophys. J.}\ }\textbf {\bibinfo {volume} {507}},\
  \bibinfo {pages} {L59} (\bibinfo {year} {1998})}\BibitemShut {NoStop}%
\bibitem [{\citenamefont {Kulkarni}(2005)}]{kulkarni2005}%
  \BibitemOpen
  \bibfield  {author} {\bibinfo {author} {\bibfnamefont {S.~R.}\ \bibnamefont
  {Kulkarni}},\ }\href@noop {} {\  (\bibinfo {year} {2005})},\ \Eprint
  {http://arxiv.org/abs/arXiv:astro-ph/0510256} {arXiv:astro-ph/0510256}
  \BibitemShut {NoStop}%
\bibitem [{\citenamefont {Metzger}\ \emph {et~al.}(2010)\citenamefont
  {Metzger}, \citenamefont {Mart{\'i}nez-Pinedo}, \citenamefont {Darbha},
  \citenamefont {Quataert}, \citenamefont {Arcones}, \citenamefont {Kasen},
  \citenamefont {Thomas}, \citenamefont {Nugent}, \citenamefont {Panov},\ and\
  \citenamefont {Zinner}}]{metzger_mdqaktnpz2010}%
  \BibitemOpen
  \bibfield  {author} {\bibinfo {author} {\bibfnamefont {B.~D.}\ \bibnamefont
  {Metzger}}, \bibinfo {author} {\bibfnamefont {G.}~\bibnamefont
  {Mart{\'i}nez-Pinedo}}, \bibinfo {author} {\bibfnamefont {S.}~\bibnamefont
  {Darbha}}, \bibinfo {author} {\bibfnamefont {E.}~\bibnamefont {Quataert}},
  \bibinfo {author} {\bibfnamefont {A.}~\bibnamefont {Arcones}}, \bibinfo
  {author} {\bibfnamefont {D.}~\bibnamefont {Kasen}}, \bibinfo {author}
  {\bibfnamefont {R.}~\bibnamefont {Thomas}}, \bibinfo {author} {\bibfnamefont
  {P.}~\bibnamefont {Nugent}}, \bibinfo {author} {\bibfnamefont {I.~V.}\
  \bibnamefont {Panov}}, \ and\ \bibinfo {author} {\bibfnamefont {N.~T.}\
  \bibnamefont {Zinner}},\ }\href {\doibase 10.1111/j.1365-2966.2010.16864.x}
  {\bibfield  {journal} {\bibinfo  {journal} {Mon. Not. R. Astron. Soc.}\
  }\textbf {\bibinfo {volume} {406}},\ \bibinfo {pages} {2650} (\bibinfo {year}
  {2010})}\BibitemShut {NoStop}%
\bibitem [{\citenamefont {Tanaka}\ \emph {et~al.}(2014)\citenamefont {Tanaka},
  \citenamefont {Hotokezaka}, \citenamefont {Kyutoku}, \citenamefont {Wanajo},
  \citenamefont {Kiuchi}, \citenamefont {Sekiguchi},\ and\ \citenamefont
  {Shibata}}]{tanaka_hkwkss2014}%
  \BibitemOpen
  \bibfield  {author} {\bibinfo {author} {\bibfnamefont {M.}~\bibnamefont
  {Tanaka}}, \bibinfo {author} {\bibfnamefont {K.}~\bibnamefont {Hotokezaka}},
  \bibinfo {author} {\bibfnamefont {K.}~\bibnamefont {Kyutoku}}, \bibinfo
  {author} {\bibfnamefont {S.}~\bibnamefont {Wanajo}}, \bibinfo {author}
  {\bibfnamefont {K.}~\bibnamefont {Kiuchi}}, \bibinfo {author} {\bibfnamefont
  {Y.}~\bibnamefont {Sekiguchi}}, \ and\ \bibinfo {author} {\bibfnamefont
  {M.}~\bibnamefont {Shibata}},\ }\href {\doibase 10.1088/0004-637X/780/1/31}
  {\bibfield  {journal} {\bibinfo  {journal} {Astrophys. J.}\ }\textbf
  {\bibinfo {volume} {780}},\ \bibinfo {pages} {31} (\bibinfo {year}
  {2014})}\BibitemShut {NoStop}%
\bibitem [{\citenamefont {Kawaguchi}\ \emph {et~al.}(2016)\citenamefont
  {Kawaguchi}, \citenamefont {Kyutoku}, \citenamefont {Shibata},\ and\
  \citenamefont {Tanaka}}]{kawaguchi_kst2016}%
  \BibitemOpen
  \bibfield  {author} {\bibinfo {author} {\bibfnamefont {K.}~\bibnamefont
  {Kawaguchi}}, \bibinfo {author} {\bibfnamefont {K.}~\bibnamefont {Kyutoku}},
  \bibinfo {author} {\bibfnamefont {M.}~\bibnamefont {Shibata}}, \ and\
  \bibinfo {author} {\bibfnamefont {M.}~\bibnamefont {Tanaka}},\ }\href
  {\doibase 10.3847/0004-637X/825/1/52} {\bibfield  {journal} {\bibinfo
  {journal} {Astrophys. J.}\ }\textbf {\bibinfo {volume} {825}},\ \bibinfo
  {pages} {52} (\bibinfo {year} {2016})}\BibitemShut {NoStop}%
\bibitem [{\citenamefont {{Abbott}}\ \emph
  {et~al.}(2016{\natexlab{a}})\citenamefont {{Abbott}}, \citenamefont
  {{Abbott}}, \citenamefont {{Abbott}}, \citenamefont {{Abernathy}},
  \citenamefont {{Acernese}}, \citenamefont {{Ackley}}, \citenamefont
  {{Adams}}, \citenamefont {{Adams}}, \citenamefont {{Addesso}}, \citenamefont
  {{Adhikari}},\ and\ \citenamefont {et~al.}}]{ligovirgo2016}%
  \BibitemOpen
  \bibfield  {author} {\bibinfo {author} {\bibfnamefont {B.~P.}\ \bibnamefont
  {{Abbott}}}, \bibinfo {author} {\bibfnamefont {R.}~\bibnamefont {{Abbott}}},
  \bibinfo {author} {\bibfnamefont {T.~D.}\ \bibnamefont {{Abbott}}}, \bibinfo
  {author} {\bibfnamefont {M.~R.}\ \bibnamefont {{Abernathy}}}, \bibinfo
  {author} {\bibfnamefont {F.}~\bibnamefont {{Acernese}}}, \bibinfo {author}
  {\bibfnamefont {K.}~\bibnamefont {{Ackley}}}, \bibinfo {author}
  {\bibfnamefont {C.}~\bibnamefont {{Adams}}}, \bibinfo {author} {\bibfnamefont
  {T.}~\bibnamefont {{Adams}}}, \bibinfo {author} {\bibfnamefont
  {P.}~\bibnamefont {{Addesso}}}, \bibinfo {author} {\bibfnamefont {R.~X.}\
  \bibnamefont {{Adhikari}}}, \ and\ \bibinfo {author} {\bibnamefont
  {et~al.}},\ }\href {\doibase 10.1103/PhysRevLett.116.061102} {\bibfield
  {journal} {\bibinfo  {journal} {Phys. Rev. Lett.}\ }\textbf {\bibinfo
  {volume} {116}},\ \bibinfo {pages} {061102} (\bibinfo {year}
  {2016}{\natexlab{a}})}\BibitemShut {NoStop}%
\bibitem [{\citenamefont {{Abbott}}\ \emph
  {et~al.}(2016{\natexlab{b}})\citenamefont {{Abbott}}, \citenamefont
  {{Abbott}}, \citenamefont {{Abbott}}, \citenamefont {{Abernathy}},
  \citenamefont {{Acernese}}, \citenamefont {{Ackley}}, \citenamefont
  {{Adams}}, \citenamefont {{Adams}}, \citenamefont {{Addesso}}, \citenamefont
  {{Adhikari}},\ and\ \citenamefont {et~al.}}]{ligovirgo2016-4}%
  \BibitemOpen
  \bibfield  {author} {\bibinfo {author} {\bibfnamefont {B.~P.}\ \bibnamefont
  {{Abbott}}}, \bibinfo {author} {\bibfnamefont {R.}~\bibnamefont {{Abbott}}},
  \bibinfo {author} {\bibfnamefont {T.~D.}\ \bibnamefont {{Abbott}}}, \bibinfo
  {author} {\bibfnamefont {M.~R.}\ \bibnamefont {{Abernathy}}}, \bibinfo
  {author} {\bibfnamefont {F.}~\bibnamefont {{Acernese}}}, \bibinfo {author}
  {\bibfnamefont {K.}~\bibnamefont {{Ackley}}}, \bibinfo {author}
  {\bibfnamefont {C.}~\bibnamefont {{Adams}}}, \bibinfo {author} {\bibfnamefont
  {T.}~\bibnamefont {{Adams}}}, \bibinfo {author} {\bibfnamefont
  {P.}~\bibnamefont {{Addesso}}}, \bibinfo {author} {\bibfnamefont {R.~X.}\
  \bibnamefont {{Adhikari}}}, \ and\ \bibinfo {author} {\bibnamefont
  {et~al.}},\ }\href {\doibase 10.1103/PhysRevLett.116.241103} {\bibfield
  {journal} {\bibinfo  {journal} {Phys. Rev. Lett}\ }\textbf {\bibinfo {volume}
  {116}},\ \bibinfo {pages} {241103} (\bibinfo {year}
  {2016}{\natexlab{b}})}\BibitemShut {NoStop}%
\bibitem [{\citenamefont {{Abbott}}\ \emph
  {et~al.}(2017{\natexlab{a}})\citenamefont {{Abbott}}, \citenamefont
  {{Abbott}}, \citenamefont {{Abbott}}, \citenamefont {{Acernese}},
  \citenamefont {{Ackley}}, \citenamefont {{Adams}}, \citenamefont {{Adams}},
  \citenamefont {{Addesso}}, \citenamefont {{Adhikari}}, \citenamefont
  {{Adya}},\ and\ \citenamefont {et~al.}}]{ligovirgo2017}%
  \BibitemOpen
  \bibfield  {author} {\bibinfo {author} {\bibfnamefont {B.~P.}\ \bibnamefont
  {{Abbott}}}, \bibinfo {author} {\bibfnamefont {R.}~\bibnamefont {{Abbott}}},
  \bibinfo {author} {\bibfnamefont {T.~D.}\ \bibnamefont {{Abbott}}}, \bibinfo
  {author} {\bibfnamefont {F.}~\bibnamefont {{Acernese}}}, \bibinfo {author}
  {\bibfnamefont {K.}~\bibnamefont {{Ackley}}}, \bibinfo {author}
  {\bibfnamefont {C.}~\bibnamefont {{Adams}}}, \bibinfo {author} {\bibfnamefont
  {T.}~\bibnamefont {{Adams}}}, \bibinfo {author} {\bibfnamefont
  {P.}~\bibnamefont {{Addesso}}}, \bibinfo {author} {\bibfnamefont {R.~X.}\
  \bibnamefont {{Adhikari}}}, \bibinfo {author} {\bibfnamefont {V.~B.}\
  \bibnamefont {{Adya}}}, \ and\ \bibinfo {author} {\bibnamefont {et~al.}},\
  }\href {\doibase 10.1103/PhysRevLett.113.221101} {\bibfield  {journal}
  {\bibinfo  {journal} {Phys. Rev. Lett.}\ }\textbf {\bibinfo {volume} {118}},\
  \bibinfo {pages} {221101} (\bibinfo {year} {2017}{\natexlab{a}})}\BibitemShut
  {NoStop}%
\bibitem [{\citenamefont {{Abbott}}\ \emph
  {et~al.}(2017{\natexlab{b}})\citenamefont {{Abbott}}, \citenamefont
  {{Abbott}}, \citenamefont {{Abbott}}, \citenamefont {{Acernese}},
  \citenamefont {{Ackley}}, \citenamefont {{Adams}}, \citenamefont {{Adams}},
  \citenamefont {{Addesso}}, \citenamefont {{Adhikari}}, \citenamefont
  {{Adya}},\ and\ \citenamefont {et~al.}}]{ligovirgo2017-2}%
  \BibitemOpen
  \bibfield  {author} {\bibinfo {author} {\bibfnamefont {B.~P.}\ \bibnamefont
  {{Abbott}}}, \bibinfo {author} {\bibfnamefont {R.}~\bibnamefont {{Abbott}}},
  \bibinfo {author} {\bibfnamefont {T.~D.}\ \bibnamefont {{Abbott}}}, \bibinfo
  {author} {\bibfnamefont {F.}~\bibnamefont {{Acernese}}}, \bibinfo {author}
  {\bibfnamefont {K.}~\bibnamefont {{Ackley}}}, \bibinfo {author}
  {\bibfnamefont {C.}~\bibnamefont {{Adams}}}, \bibinfo {author} {\bibfnamefont
  {T.}~\bibnamefont {{Adams}}}, \bibinfo {author} {\bibfnamefont
  {P.}~\bibnamefont {{Addesso}}}, \bibinfo {author} {\bibfnamefont {R.~X.}\
  \bibnamefont {{Adhikari}}}, \bibinfo {author} {\bibfnamefont {V.~B.}\
  \bibnamefont {{Adya}}}, \ and\ \bibinfo {author} {\bibnamefont {et~al.}},\
  }\href {\doibase 10.1103/PhysRevLett.119.141101} {\bibfield  {journal}
  {\bibinfo  {journal} {Phys. Rev. Lett.}\ }\textbf {\bibinfo {volume} {119}},\
  \bibinfo {pages} {141101} (\bibinfo {year} {2017}{\natexlab{b}})}\BibitemShut
  {NoStop}%
\bibitem [{\citenamefont {{Abbott}}\ \emph
  {et~al.}(2017{\natexlab{c}})\citenamefont {{Abbott}}, \citenamefont
  {{Abbott}}, \citenamefont {{Abbott}}, \citenamefont {{Acernese}},
  \citenamefont {{Ackley}}, \citenamefont {{Adams}}, \citenamefont {{Adams}},
  \citenamefont {{Addesso}}, \citenamefont {{Adhikari}}, \citenamefont
  {{Adya}},\ and\ \citenamefont {et~al.}}]{ligovirgo2017-4}%
  \BibitemOpen
  \bibfield  {author} {\bibinfo {author} {\bibfnamefont {B.~P.}\ \bibnamefont
  {{Abbott}}}, \bibinfo {author} {\bibfnamefont {R.}~\bibnamefont {{Abbott}}},
  \bibinfo {author} {\bibfnamefont {T.~D.}\ \bibnamefont {{Abbott}}}, \bibinfo
  {author} {\bibfnamefont {F.}~\bibnamefont {{Acernese}}}, \bibinfo {author}
  {\bibfnamefont {K.}~\bibnamefont {{Ackley}}}, \bibinfo {author}
  {\bibfnamefont {C.}~\bibnamefont {{Adams}}}, \bibinfo {author} {\bibfnamefont
  {T.}~\bibnamefont {{Adams}}}, \bibinfo {author} {\bibfnamefont
  {P.}~\bibnamefont {{Addesso}}}, \bibinfo {author} {\bibfnamefont {R.~X.}\
  \bibnamefont {{Adhikari}}}, \bibinfo {author} {\bibfnamefont {V.~B.}\
  \bibnamefont {{Adya}}}, \ and\ \bibinfo {author} {\bibnamefont {et~al.}},\
  }\href {\doibase 10.3847/2041-8213/aa9f0c} {\bibfield  {journal} {\bibinfo
  {journal} {Astrophys. J.}\ }\textbf {\bibinfo {volume} {851}},\ \bibinfo
  {pages} {L35} (\bibinfo {year} {2017}{\natexlab{c}})}\BibitemShut {NoStop}%
\bibitem [{\citenamefont {{Abbott}}\ \emph
  {et~al.}(2016{\natexlab{c}})\citenamefont {{Abbott}}, \citenamefont
  {{Abbott}}, \citenamefont {{Abbott}}, \citenamefont {{Abernathy}},
  \citenamefont {{Acernese}}, \citenamefont {{Ackley}}, \citenamefont
  {{Adams}}, \citenamefont {{Adams}}, \citenamefont {{Addesso}}, \citenamefont
  {{Adhikari}},\ and\ \citenamefont {et~al.}}]{ligovirgo2016-5}%
  \BibitemOpen
  \bibfield  {author} {\bibinfo {author} {\bibfnamefont {B.~P.}\ \bibnamefont
  {{Abbott}}}, \bibinfo {author} {\bibfnamefont {R.}~\bibnamefont {{Abbott}}},
  \bibinfo {author} {\bibfnamefont {T.~D.}\ \bibnamefont {{Abbott}}}, \bibinfo
  {author} {\bibfnamefont {M.~R.}\ \bibnamefont {{Abernathy}}}, \bibinfo
  {author} {\bibfnamefont {F.}~\bibnamefont {{Acernese}}}, \bibinfo {author}
  {\bibfnamefont {K.}~\bibnamefont {{Ackley}}}, \bibinfo {author}
  {\bibfnamefont {C.}~\bibnamefont {{Adams}}}, \bibinfo {author} {\bibfnamefont
  {T.}~\bibnamefont {{Adams}}}, \bibinfo {author} {\bibfnamefont
  {P.}~\bibnamefont {{Addesso}}}, \bibinfo {author} {\bibfnamefont {R.~X.}\
  \bibnamefont {{Adhikari}}}, \ and\ \bibinfo {author} {\bibnamefont
  {et~al.}},\ }\href {\doibase 10.1103/PhysRevX.6.041015} {\bibfield  {journal}
  {\bibinfo  {journal} {Phys. Rev. X}\ }\textbf {\bibinfo {volume} {6}},\
  \bibinfo {pages} {041015} (\bibinfo {year} {2016}{\natexlab{c}})}\BibitemShut
  {NoStop}%
\bibitem [{\citenamefont {Lee}\ and\ \citenamefont
  {Ramirez-Ruiz}(2007)}]{lee_ramirezruiz2007}%
  \BibitemOpen
  \bibfield  {author} {\bibinfo {author} {\bibfnamefont {W.~H.}\ \bibnamefont
  {Lee}}\ and\ \bibinfo {author} {\bibfnamefont {E.}~\bibnamefont
  {Ramirez-Ruiz}},\ }\href {\doibase 10.1088/1367-2630/9/1/017} {\bibfield
  {journal} {\bibinfo  {journal} {New J. Phys.}\ }\textbf {\bibinfo {volume}
  {9}},\ \bibinfo {pages} {17} (\bibinfo {year} {2007})}\BibitemShut {NoStop}%
\bibitem [{\citenamefont {Nakar}(2007)}]{nakar2007}%
  \BibitemOpen
  \bibfield  {author} {\bibinfo {author} {\bibfnamefont {E.}~\bibnamefont
  {Nakar}},\ }\href {\doibase 10.1016/j.physrep.2007.02.005} {\bibfield
  {journal} {\bibinfo  {journal} {Phys. Rep.}\ }\textbf {\bibinfo {volume}
  {442}},\ \bibinfo {pages} {166} (\bibinfo {year} {2007})}\BibitemShut
  {NoStop}%
\bibitem [{\citenamefont {Berger}(2014)}]{berger2014}%
  \BibitemOpen
  \bibfield  {author} {\bibinfo {author} {\bibfnamefont {E.}~\bibnamefont
  {Berger}},\ }\href {\doibase 10.1146/annurev-astro-081913-035926} {\bibfield
  {journal} {\bibinfo  {journal} {Annu. Rev. Astron. Astrophys.}\ }\textbf
  {\bibinfo {volume} {52}},\ \bibinfo {pages} {43} (\bibinfo {year}
  {2014})}\BibitemShut {NoStop}%
\bibitem [{\citenamefont {Tanvir}\ \emph {et~al.}(2013)\citenamefont {Tanvir},
  \citenamefont {Levan}, \citenamefont {Fruchter}, \citenamefont {Hjorth},
  \citenamefont {Hounsell}, \citenamefont {Wiersema},\ and\ \citenamefont
  {Tunnicliffe}}]{tanvir_lfhhwt2013}%
  \BibitemOpen
  \bibfield  {author} {\bibinfo {author} {\bibfnamefont {N.~R.}\ \bibnamefont
  {Tanvir}}, \bibinfo {author} {\bibfnamefont {A.~J.}\ \bibnamefont {Levan}},
  \bibinfo {author} {\bibfnamefont {A.~S.}\ \bibnamefont {Fruchter}}, \bibinfo
  {author} {\bibfnamefont {J.}~\bibnamefont {Hjorth}}, \bibinfo {author}
  {\bibfnamefont {R.}~\bibnamefont {Hounsell}}, \bibinfo {author}
  {\bibfnamefont {K.}~\bibnamefont {Wiersema}}, \ and\ \bibinfo {author}
  {\bibfnamefont {R.~L.}\ \bibnamefont {Tunnicliffe}},\ }\href {\doibase
  10.1038/nature12505} {\bibfield  {journal} {\bibinfo  {journal} {Nature
  (London)}\ }\textbf {\bibinfo {volume} {500}},\ \bibinfo {pages} {547}
  (\bibinfo {year} {2013})}\BibitemShut {NoStop}%
\bibitem [{\citenamefont {Berger}\ \emph {et~al.}(2013)\citenamefont {Berger},
  \citenamefont {Fong},\ and\ \citenamefont {Chornock}}]{berger_fc2013}%
  \BibitemOpen
  \bibfield  {author} {\bibinfo {author} {\bibfnamefont {E.}~\bibnamefont
  {Berger}}, \bibinfo {author} {\bibfnamefont {W.}~\bibnamefont {Fong}}, \ and\
  \bibinfo {author} {\bibfnamefont {R.}~\bibnamefont {Chornock}},\ }\href
  {\doibase 10.1088/2041-8205/774/2/L23} {\bibfield  {journal} {\bibinfo
  {journal} {Astrophys. J.}\ }\textbf {\bibinfo {volume} {774}},\ \bibinfo
  {pages} {L23} (\bibinfo {year} {2013})}\BibitemShut {NoStop}%
\bibitem [{\citenamefont {Hotokezaka}\ \emph
  {et~al.}(2013{\natexlab{b}})\citenamefont {Hotokezaka}, \citenamefont
  {Kyutoku}, \citenamefont {Tanaka}, \citenamefont {Kiuchi}, \citenamefont
  {Sekiguchi}, \citenamefont {Shiata},\ and\ \citenamefont
  {Wanajo}}]{hotokezaka_ktkssw2013}%
  \BibitemOpen
  \bibfield  {author} {\bibinfo {author} {\bibfnamefont {K.}~\bibnamefont
  {Hotokezaka}}, \bibinfo {author} {\bibfnamefont {K.}~\bibnamefont {Kyutoku}},
  \bibinfo {author} {\bibfnamefont {M.}~\bibnamefont {Tanaka}}, \bibinfo
  {author} {\bibfnamefont {K.}~\bibnamefont {Kiuchi}}, \bibinfo {author}
  {\bibfnamefont {Y.}~\bibnamefont {Sekiguchi}}, \bibinfo {author}
  {\bibfnamefont {M.}~\bibnamefont {Shiata}}, \ and\ \bibinfo {author}
  {\bibfnamefont {S.}~\bibnamefont {Wanajo}},\ }\href {\doibase
  10.1088/2041-8205/778/1/L16} {\bibfield  {journal} {\bibinfo  {journal}
  {Astrophys. J.}\ }\textbf {\bibinfo {volume} {778}},\ \bibinfo {pages} {L16}
  (\bibinfo {year} {2013}{\natexlab{b}})}\BibitemShut {NoStop}%
\bibitem [{\citenamefont {Nagakura}\ \emph {et~al.}(2014)\citenamefont
  {Nagakura}, \citenamefont {Hotokezaka}, \citenamefont {Sekiguchi},
  \citenamefont {Shibata},\ and\ \citenamefont {Ioka}}]{nagakura_hssi2014}%
  \BibitemOpen
  \bibfield  {author} {\bibinfo {author} {\bibfnamefont {H.}~\bibnamefont
  {Nagakura}}, \bibinfo {author} {\bibfnamefont {K.}~\bibnamefont
  {Hotokezaka}}, \bibinfo {author} {\bibfnamefont {Y.}~\bibnamefont
  {Sekiguchi}}, \bibinfo {author} {\bibfnamefont {M.}~\bibnamefont {Shibata}},
  \ and\ \bibinfo {author} {\bibfnamefont {K.}~\bibnamefont {Ioka}},\ }\href
  {\doibase 10.1088/2041-8205/784/2/L28} {\bibfield  {journal} {\bibinfo
  {journal} {Astrophys. J.}\ }\textbf {\bibinfo {volume} {784}},\ \bibinfo
  {pages} {L28} (\bibinfo {year} {2014})}\BibitemShut {NoStop}%
\bibitem [{\citenamefont {Murguia-Berthier}\ \emph {et~al.}(2014)\citenamefont
  {Murguia-Berthier}, \citenamefont {Montes}, \citenamefont {Ramirez-Ruiz},
  \citenamefont {{De Colle}},\ and\ \citenamefont
  {Lee}}]{murguiabertheir_mrdl2014}%
  \BibitemOpen
  \bibfield  {author} {\bibinfo {author} {\bibfnamefont {A.}~\bibnamefont
  {Murguia-Berthier}}, \bibinfo {author} {\bibfnamefont {G.}~\bibnamefont
  {Montes}}, \bibinfo {author} {\bibfnamefont {E.}~\bibnamefont
  {Ramirez-Ruiz}}, \bibinfo {author} {\bibfnamefont {F.}~\bibnamefont {{De
  Colle}}}, \ and\ \bibinfo {author} {\bibfnamefont {W.~H.}\ \bibnamefont
  {Lee}},\ }\href {\doibase 10.1088/2041-8205/788/1/L8} {\bibfield  {journal}
  {\bibinfo  {journal} {Astrophys. J.}\ }\textbf {\bibinfo {volume} {788}},\
  \bibinfo {pages} {L8} (\bibinfo {year} {2014})}\BibitemShut {NoStop}%
\bibitem [{\citenamefont {Duffell}\ \emph {et~al.}(2015)\citenamefont
  {Duffell}, \citenamefont {Quataert},\ and\ \citenamefont
  {MacFadyen}}]{duffell_qm2015}%
  \BibitemOpen
  \bibfield  {author} {\bibinfo {author} {\bibfnamefont {P.~C.}\ \bibnamefont
  {Duffell}}, \bibinfo {author} {\bibfnamefont {E.}~\bibnamefont {Quataert}}, \
  and\ \bibinfo {author} {\bibfnamefont {A.~I.}\ \bibnamefont {MacFadyen}},\
  }\href {\doibase 10.1088/0004-637X/813/1/64} {\bibfield  {journal} {\bibinfo
  {journal} {Astrophys. J.}\ }\textbf {\bibinfo {volume} {813}},\ \bibinfo
  {pages} {64} (\bibinfo {year} {2015})}\BibitemShut {NoStop}%
\bibitem [{\citenamefont {Just}\ \emph {et~al.}(2016)\citenamefont {Just},
  \citenamefont {Obergaulinger}, \citenamefont {Janka}, \citenamefont
  {Bauswein},\ and\ \citenamefont {Schwarz}}]{just_ojbs2016}%
  \BibitemOpen
  \bibfield  {author} {\bibinfo {author} {\bibfnamefont {O.}~\bibnamefont
  {Just}}, \bibinfo {author} {\bibfnamefont {M.}~\bibnamefont {Obergaulinger}},
  \bibinfo {author} {\bibfnamefont {H.-T.}\ \bibnamefont {Janka}}, \bibinfo
  {author} {\bibfnamefont {A.}~\bibnamefont {Bauswein}}, \ and\ \bibinfo
  {author} {\bibfnamefont {N.}~\bibnamefont {Schwarz}},\ }\href {\doibase
  10.3847/2041-8205/816/2/L30} {\bibfield  {journal} {\bibinfo  {journal}
  {Astrophys. J.}\ }\textbf {\bibinfo {volume} {816}},\ \bibinfo {pages} {L30}
  (\bibinfo {year} {2016})}\BibitemShut {NoStop}%
\bibitem [{\citenamefont {Murguia-Berthier}\ \emph {et~al.}(2017)\citenamefont
  {Murguia-Berthier}, \citenamefont {Ramirez-Ruiz}, \citenamefont {Montes},
  \citenamefont {{De Colle}}, \citenamefont {Rezzolla}, \citenamefont
  {Rosswog}, \citenamefont {Takami}, \citenamefont {Perego},\ and\
  \citenamefont {Lee}}]{murguiabertheir_rmdrrtpl2017}%
  \BibitemOpen
  \bibfield  {author} {\bibinfo {author} {\bibfnamefont {A.}~\bibnamefont
  {Murguia-Berthier}}, \bibinfo {author} {\bibfnamefont {E.}~\bibnamefont
  {Ramirez-Ruiz}}, \bibinfo {author} {\bibfnamefont {G.}~\bibnamefont
  {Montes}}, \bibinfo {author} {\bibfnamefont {F.}~\bibnamefont {{De Colle}}},
  \bibinfo {author} {\bibfnamefont {L.}~\bibnamefont {Rezzolla}}, \bibinfo
  {author} {\bibfnamefont {S.}~\bibnamefont {Rosswog}}, \bibinfo {author}
  {\bibfnamefont {K.}~\bibnamefont {Takami}}, \bibinfo {author} {\bibfnamefont
  {A.}~\bibnamefont {Perego}}, \ and\ \bibinfo {author} {\bibfnamefont {W.~H.}\
  \bibnamefont {Lee}},\ }\href {\doibase 10.3847/2041-8213/aa5b9e} {\bibfield
  {journal} {\bibinfo  {journal} {Astrophys. J.}\ }\textbf {\bibinfo {volume}
  {853}},\ \bibinfo {pages} {L34} (\bibinfo {year} {2017})}\BibitemShut
  {NoStop}%
\bibitem [{\citenamefont {{Abbott}}\ \emph
  {et~al.}(2017{\natexlab{d}})\citenamefont {{Abbott}}, \citenamefont
  {{Abbott}}, \citenamefont {{Abbott}}, \citenamefont {{Acernese}},
  \citenamefont {{Ackley}}, \citenamefont {{Adams}}, \citenamefont {{Adams}},
  \citenamefont {{Addesso}}, \citenamefont {{Adhikari}}, \citenamefont
  {{Adya}},\ and\ \citenamefont {et~al.}}]{ligovirgo2017-3}%
  \BibitemOpen
  \bibfield  {author} {\bibinfo {author} {\bibfnamefont {B.~P.}\ \bibnamefont
  {{Abbott}}}, \bibinfo {author} {\bibfnamefont {R.}~\bibnamefont {{Abbott}}},
  \bibinfo {author} {\bibfnamefont {T.~D.}\ \bibnamefont {{Abbott}}}, \bibinfo
  {author} {\bibfnamefont {F.}~\bibnamefont {{Acernese}}}, \bibinfo {author}
  {\bibfnamefont {K.}~\bibnamefont {{Ackley}}}, \bibinfo {author}
  {\bibfnamefont {C.}~\bibnamefont {{Adams}}}, \bibinfo {author} {\bibfnamefont
  {T.}~\bibnamefont {{Adams}}}, \bibinfo {author} {\bibfnamefont
  {P.}~\bibnamefont {{Addesso}}}, \bibinfo {author} {\bibfnamefont {R.~X.}\
  \bibnamefont {{Adhikari}}}, \bibinfo {author} {\bibfnamefont {V.~B.}\
  \bibnamefont {{Adya}}}, \ and\ \bibinfo {author} {\bibnamefont {et~al.}},\
  }\href {\doibase 10.1103/PhysRevLett.119.161101} {\bibfield  {journal}
  {\bibinfo  {journal} {Phys. Rev. Lett.}\ }\textbf {\bibinfo {volume} {119}},\
  \bibinfo {pages} {161101} (\bibinfo {year} {2017}{\natexlab{d}})}\BibitemShut
  {NoStop}%
\bibitem [{\citenamefont {{Abbott}}\ \emph
  {et~al.}(2017{\natexlab{e}})\citenamefont {{Abbott}}, \citenamefont
  {{Abbott}}, \citenamefont {{Abbott}}, \citenamefont {{Acernese}},
  \citenamefont {{Ackley}}, \citenamefont {{Adams}}, \citenamefont {{Adams}},
  \citenamefont {{Addesso}}, \citenamefont {{Adhikari}}, \citenamefont
  {{Adya}},\ and\ \citenamefont {et~al.}}]{ligovirgoem2017}%
  \BibitemOpen
  \bibfield  {author} {\bibinfo {author} {\bibfnamefont {B.~P.}\ \bibnamefont
  {{Abbott}}}, \bibinfo {author} {\bibfnamefont {R.}~\bibnamefont {{Abbott}}},
  \bibinfo {author} {\bibfnamefont {T.~D.}\ \bibnamefont {{Abbott}}}, \bibinfo
  {author} {\bibfnamefont {F.}~\bibnamefont {{Acernese}}}, \bibinfo {author}
  {\bibfnamefont {K.}~\bibnamefont {{Ackley}}}, \bibinfo {author}
  {\bibfnamefont {C.}~\bibnamefont {{Adams}}}, \bibinfo {author} {\bibfnamefont
  {T.}~\bibnamefont {{Adams}}}, \bibinfo {author} {\bibfnamefont
  {P.}~\bibnamefont {{Addesso}}}, \bibinfo {author} {\bibfnamefont {R.~X.}\
  \bibnamefont {{Adhikari}}}, \bibinfo {author} {\bibfnamefont {V.~B.}\
  \bibnamefont {{Adya}}}, \ and\ \bibinfo {author} {\bibnamefont {et~al.}},\
  }\href {\doibase 10.3847/2041-8213/aa91c9} {\bibfield  {journal} {\bibinfo
  {journal} {Astrophys. J.}\ }\textbf {\bibinfo {volume} {848}},\ \bibinfo
  {pages} {L12} (\bibinfo {year} {2017}{\natexlab{e}})}\BibitemShut {NoStop}%
\bibitem [{\citenamefont {{Abbott}}\ \emph
  {et~al.}(2017{\natexlab{f}})\citenamefont {{Abbott}}, \citenamefont
  {{Abbott}}, \citenamefont {{Abbott}}, \citenamefont {{Acernese}},
  \citenamefont {{Ackley}}, \citenamefont {{Adams}}, \citenamefont {{Adams}},
  \citenamefont {{Addesso}}, \citenamefont {{Adhikari}}, \citenamefont
  {{Adya}},\ and\ \citenamefont {et~al.}}]{ligovirgogamma2017}%
  \BibitemOpen
  \bibfield  {author} {\bibinfo {author} {\bibfnamefont {B.~P.}\ \bibnamefont
  {{Abbott}}}, \bibinfo {author} {\bibfnamefont {R.}~\bibnamefont {{Abbott}}},
  \bibinfo {author} {\bibfnamefont {T.~D.}\ \bibnamefont {{Abbott}}}, \bibinfo
  {author} {\bibfnamefont {F.}~\bibnamefont {{Acernese}}}, \bibinfo {author}
  {\bibfnamefont {K.}~\bibnamefont {{Ackley}}}, \bibinfo {author}
  {\bibfnamefont {C.}~\bibnamefont {{Adams}}}, \bibinfo {author} {\bibfnamefont
  {T.}~\bibnamefont {{Adams}}}, \bibinfo {author} {\bibfnamefont
  {P.}~\bibnamefont {{Addesso}}}, \bibinfo {author} {\bibfnamefont {R.~X.}\
  \bibnamefont {{Adhikari}}}, \bibinfo {author} {\bibfnamefont {V.~B.}\
  \bibnamefont {{Adya}}}, \ and\ \bibinfo {author} {\bibnamefont {et~al.}},\
  }\href {\doibase 10.3847/2041-8213/aa920c} {\bibfield  {journal} {\bibinfo
  {journal} {Astrophys. J.}\ }\textbf {\bibinfo {volume} {848}},\ \bibinfo
  {pages} {L13} (\bibinfo {year} {2017}{\natexlab{f}})}\BibitemShut {NoStop}%
\bibitem [{\citenamefont {{K. P. Mooley et al.}}(2017)}]{mooley_etal2017}%
  \BibitemOpen
  \bibfield  {author} {\bibinfo {author} {\bibnamefont {{K. P. Mooley et
  al.}}},\ }\href@noop {} {\  (\bibinfo {year} {2017})},\ \Eprint
  {http://arxiv.org/abs/arXiv:1711.11573} {arXiv:1711.11573} \BibitemShut
  {NoStop}%
\bibitem [{\citenamefont {Ruan}\ \emph {et~al.}(2017)\citenamefont {Ruan},
  \citenamefont {Nynka}, \citenamefont {Haggard}, \citenamefont {Kalogera},\
  and\ \citenamefont {Evans}}]{ruan_nhke2017}%
  \BibitemOpen
  \bibfield  {author} {\bibinfo {author} {\bibfnamefont {J.~J.}\ \bibnamefont
  {Ruan}}, \bibinfo {author} {\bibfnamefont {M.}~\bibnamefont {Nynka}},
  \bibinfo {author} {\bibfnamefont {D.}~\bibnamefont {Haggard}}, \bibinfo
  {author} {\bibfnamefont {V.}~\bibnamefont {Kalogera}}, \ and\ \bibinfo
  {author} {\bibfnamefont {P.}~\bibnamefont {Evans}},\ }\href@noop {} {\
  (\bibinfo {year} {2017})},\ \Eprint {http://arxiv.org/abs/arXiv:1712.02809}
  {arXiv:1712.02809} \BibitemShut {NoStop}%
\bibitem [{\citenamefont {Wanajo}\ \emph {et~al.}(2014)\citenamefont {Wanajo},
  \citenamefont {Sekiguchi}, \citenamefont {Nishimura}, \citenamefont {Kiuchi},
  \citenamefont {Kyutoku},\ and\ \citenamefont {Shibata}}]{wanajo_snkks2014}%
  \BibitemOpen
  \bibfield  {author} {\bibinfo {author} {\bibfnamefont {S.}~\bibnamefont
  {Wanajo}}, \bibinfo {author} {\bibfnamefont {Y.}~\bibnamefont {Sekiguchi}},
  \bibinfo {author} {\bibfnamefont {N.}~\bibnamefont {Nishimura}}, \bibinfo
  {author} {\bibfnamefont {K.}~\bibnamefont {Kiuchi}}, \bibinfo {author}
  {\bibfnamefont {K.}~\bibnamefont {Kyutoku}}, \ and\ \bibinfo {author}
  {\bibfnamefont {M.}~\bibnamefont {Shibata}},\ }\href {\doibase
  10.1088/2041-8205/789/2/L39} {\bibfield  {journal} {\bibinfo  {journal}
  {Astrophys. J.}\ }\textbf {\bibinfo {volume} {789}},\ \bibinfo {pages} {L39}
  (\bibinfo {year} {2014})}\BibitemShut {NoStop}%
\bibitem [{\citenamefont {Freiburghaus}\ \emph {et~al.}(1999)\citenamefont
  {Freiburghaus}, \citenamefont {Rosswog},\ and\ \citenamefont
  {Thielemann}}]{freiburghaus_rt1999}%
  \BibitemOpen
  \bibfield  {author} {\bibinfo {author} {\bibfnamefont {C.}~\bibnamefont
  {Freiburghaus}}, \bibinfo {author} {\bibfnamefont {S.}~\bibnamefont
  {Rosswog}}, \ and\ \bibinfo {author} {\bibfnamefont {F.-K.}\ \bibnamefont
  {Thielemann}},\ }\href {\doibase 10.1086/312343} {\bibfield  {journal}
  {\bibinfo  {journal} {Astrophys. J.}\ }\textbf {\bibinfo {volume} {525}},\
  \bibinfo {pages} {L121} (\bibinfo {year} {1999})}\BibitemShut {NoStop}%
\bibitem [{\citenamefont {Goriely}\ \emph {et~al.}(2011)\citenamefont
  {Goriely}, \citenamefont {Bauswein},\ and\ \citenamefont
  {Janka}}]{goriely_bj2011}%
  \BibitemOpen
  \bibfield  {author} {\bibinfo {author} {\bibfnamefont {S.}~\bibnamefont
  {Goriely}}, \bibinfo {author} {\bibfnamefont {A.}~\bibnamefont {Bauswein}}, \
  and\ \bibinfo {author} {\bibfnamefont {H.-T.}\ \bibnamefont {Janka}},\ }\href
  {\doibase 10.1088/2041-8205/738/2/L32} {\bibfield  {journal} {\bibinfo
  {journal} {Astrophys. J.}\ }\textbf {\bibinfo {volume} {738}},\ \bibinfo
  {pages} {L32} (\bibinfo {year} {2011})}\BibitemShut {NoStop}%
\bibitem [{\citenamefont {Korobkin}\ \emph {et~al.}(2012)\citenamefont
  {Korobkin}, \citenamefont {Rosswog}, \citenamefont {Arcones},\ and\
  \citenamefont {Winteler}}]{korobkin_raw2012}%
  \BibitemOpen
  \bibfield  {author} {\bibinfo {author} {\bibfnamefont {O.}~\bibnamefont
  {Korobkin}}, \bibinfo {author} {\bibfnamefont {S.}~\bibnamefont {Rosswog}},
  \bibinfo {author} {\bibfnamefont {A.}~\bibnamefont {Arcones}}, \ and\
  \bibinfo {author} {\bibfnamefont {C.}~\bibnamefont {Winteler}},\ }\href
  {\doibase 10.1111/j.1365-2966.2012.21859.x} {\bibfield  {journal} {\bibinfo
  {journal} {Mon. Not. R. Astron. Soc.}\ }\textbf {\bibinfo {volume} {426}},\
  \bibinfo {pages} {1940} (\bibinfo {year} {2012})}\BibitemShut {NoStop}%
\bibitem [{\citenamefont {Goriely}\ \emph {et~al.}(2015)\citenamefont
  {Goriely}, \citenamefont {Bauswein}, \citenamefont {Just}, \citenamefont
  {Pllumbi},\ and\ \citenamefont {Janka}}]{goriely_bjpj2015}%
  \BibitemOpen
  \bibfield  {author} {\bibinfo {author} {\bibfnamefont {S.}~\bibnamefont
  {Goriely}}, \bibinfo {author} {\bibfnamefont {A.}~\bibnamefont {Bauswein}},
  \bibinfo {author} {\bibfnamefont {O.}~\bibnamefont {Just}}, \bibinfo {author}
  {\bibfnamefont {E.}~\bibnamefont {Pllumbi}}, \ and\ \bibinfo {author}
  {\bibfnamefont {H.-T.}\ \bibnamefont {Janka}},\ }\href {\doibase
  10.1093/mnras/stv1526} {\bibfield  {journal} {\bibinfo  {journal} {Mon. Not.
  R. Astron. Soc.}\ }\textbf {\bibinfo {volume} {452}},\ \bibinfo {pages}
  {3894} (\bibinfo {year} {2015})}\BibitemShut {NoStop}%
\bibitem [{\citenamefont {Roberts}\ \emph {et~al.}(2017)\citenamefont
  {Roberts}, \citenamefont {Lippuner}, \citenamefont {Duez}, \citenamefont
  {Faber}, \citenamefont {Foucart}, \citenamefont {{Lomberdi, Jr.}},
  \citenamefont {Ning}, \citenamefont {Ott},\ and\ \citenamefont
  {Ponce}}]{roberts_ldfflnop2017}%
  \BibitemOpen
  \bibfield  {author} {\bibinfo {author} {\bibfnamefont {L.~F.}\ \bibnamefont
  {Roberts}}, \bibinfo {author} {\bibfnamefont {J.}~\bibnamefont {Lippuner}},
  \bibinfo {author} {\bibfnamefont {M.~D.}\ \bibnamefont {Duez}}, \bibinfo
  {author} {\bibfnamefont {J.~A.}\ \bibnamefont {Faber}}, \bibinfo {author}
  {\bibfnamefont {F.}~\bibnamefont {Foucart}}, \bibinfo {author} {\bibfnamefont
  {J.~C.}\ \bibnamefont {{Lomberdi, Jr.}}}, \bibinfo {author} {\bibfnamefont
  {S.}~\bibnamefont {Ning}}, \bibinfo {author} {\bibfnamefont {C.~D.}\
  \bibnamefont {Ott}}, \ and\ \bibinfo {author} {\bibfnamefont
  {M.}~\bibnamefont {Ponce}},\ }\href {\doibase 10.1093/mnras/stw2622}
  {\bibfield  {journal} {\bibinfo  {journal} {Mon. Not. R. Astron. Soc.}\
  }\textbf {\bibinfo {volume} {464}},\ \bibinfo {pages} {3907} (\bibinfo {year}
  {2017})}\BibitemShut {NoStop}%
\bibitem [{\citenamefont {Sneden}\ \emph {et~al.}(2008)\citenamefont {Sneden},
  \citenamefont {Cowan},\ and\ \citenamefont {Gallino}}]{sneden_cg2008}%
  \BibitemOpen
  \bibfield  {author} {\bibinfo {author} {\bibfnamefont {C.}~\bibnamefont
  {Sneden}}, \bibinfo {author} {\bibfnamefont {J.~J.}\ \bibnamefont {Cowan}}, \
  and\ \bibinfo {author} {\bibfnamefont {R.}~\bibnamefont {Gallino}},\ }\href
  {\doibase 10.1146/annurev.astro.46.060407.145207} {\bibfield  {journal}
  {\bibinfo  {journal} {Annu. Rev. Astron. Astrophys.}\ }\textbf {\bibinfo
  {volume} {46}},\ \bibinfo {pages} {241} (\bibinfo {year} {2008})}\BibitemShut
  {NoStop}%
\bibitem [{\citenamefont {{Siqueira Mello}}\ \emph {et~al.}(2014)\citenamefont
  {{Siqueira Mello}}, \citenamefont {Hill}, \citenamefont {Barbuy},
  \citenamefont {Spite}, \citenamefont {Spite}, \citenamefont {Beers},
  \citenamefont {Caffau}, \citenamefont {Bonifacio}, \citenamefont {Cayrel},
  \citenamefont {Fran{\c c}ois}, \citenamefont {Schatz},\ and\ \citenamefont
  {Wanajo}}]{siqueiramello_etal2014}%
  \BibitemOpen
  \bibfield  {author} {\bibinfo {author} {\bibfnamefont {C.}~\bibnamefont
  {{Siqueira Mello}}}, \bibinfo {author} {\bibfnamefont {V.}~\bibnamefont
  {Hill}}, \bibinfo {author} {\bibfnamefont {B.}~\bibnamefont {Barbuy}},
  \bibinfo {author} {\bibfnamefont {M.}~\bibnamefont {Spite}}, \bibinfo
  {author} {\bibfnamefont {F.}~\bibnamefont {Spite}}, \bibinfo {author}
  {\bibfnamefont {T.~C.}\ \bibnamefont {Beers}}, \bibinfo {author}
  {\bibfnamefont {E.}~\bibnamefont {Caffau}}, \bibinfo {author} {\bibfnamefont
  {P.}~\bibnamefont {Bonifacio}}, \bibinfo {author} {\bibfnamefont
  {R.}~\bibnamefont {Cayrel}}, \bibinfo {author} {\bibfnamefont
  {P.}~\bibnamefont {Fran{\c c}ois}}, \bibinfo {author} {\bibfnamefont
  {H.}~\bibnamefont {Schatz}}, \ and\ \bibinfo {author} {\bibfnamefont
  {S.}~\bibnamefont {Wanajo}},\ }\href {\doibase 10.1051/0004-6361/201423826/}
  {\bibfield  {journal} {\bibinfo  {journal} {Astron. Astrophys.}\ }\textbf
  {\bibinfo {volume} {565}},\ \bibinfo {pages} {A93} (\bibinfo {year}
  {2014})}\BibitemShut {NoStop}%
\bibitem [{\citenamefont {Dessart}\ \emph {et~al.}(2009)\citenamefont
  {Dessart}, \citenamefont {Ott}, \citenamefont {Burrows}, \citenamefont
  {Rosswog},\ and\ \citenamefont {Livne}}]{dessart_obrl2009}%
  \BibitemOpen
  \bibfield  {author} {\bibinfo {author} {\bibfnamefont {L.}~\bibnamefont
  {Dessart}}, \bibinfo {author} {\bibfnamefont {C.~D.}\ \bibnamefont {Ott}},
  \bibinfo {author} {\bibfnamefont {A.}~\bibnamefont {Burrows}}, \bibinfo
  {author} {\bibfnamefont {S.}~\bibnamefont {Rosswog}}, \ and\ \bibinfo
  {author} {\bibfnamefont {E.}~\bibnamefont {Livne}},\ }\href {\doibase
  10.1088/0004-637X/690/2/1681} {\bibfield  {journal} {\bibinfo  {journal}
  {Astrophys. J.}\ }\textbf {\bibinfo {volume} {690}},\ \bibinfo {pages} {1681}
  (\bibinfo {year} {2009})}\BibitemShut {NoStop}%
\bibitem [{\citenamefont {Perego}\ \emph {et~al.}(2014)\citenamefont {Perego},
  \citenamefont {Rosswog}, \citenamefont {Cabez{\'o}n}, \citenamefont
  {Korobkin}, \citenamefont {K{\"a}ppeli}, \citenamefont {Arcones},\ and\
  \citenamefont {Liebend{\"o}rfer}}]{perego_rckkal2014}%
  \BibitemOpen
  \bibfield  {author} {\bibinfo {author} {\bibfnamefont {A.}~\bibnamefont
  {Perego}}, \bibinfo {author} {\bibfnamefont {S.}~\bibnamefont {Rosswog}},
  \bibinfo {author} {\bibfnamefont {R.}~\bibnamefont {Cabez{\'o}n}}, \bibinfo
  {author} {\bibfnamefont {O.}~\bibnamefont {Korobkin}}, \bibinfo {author}
  {\bibfnamefont {R.}~\bibnamefont {K{\"a}ppeli}}, \bibinfo {author}
  {\bibfnamefont {A.}~\bibnamefont {Arcones}}, \ and\ \bibinfo {author}
  {\bibfnamefont {M.}~\bibnamefont {Liebend{\"o}rfer}},\ }\href {\doibase
  10.1093/mnras/stu1352} {\bibfield  {journal} {\bibinfo  {journal} {Mon. Not.
  R. Astron. Soc.}\ }\textbf {\bibinfo {volume} {443}},\ \bibinfo {pages}
  {3134} (\bibinfo {year} {2014})}\BibitemShut {NoStop}%
\bibitem [{\citenamefont {Martin}\ \emph {et~al.}(2015)\citenamefont {Martin},
  \citenamefont {Perego}, \citenamefont {Arcones}, \citenamefont {Thielemann},
  \citenamefont {Korobkin},\ and\ \citenamefont {Rosswog}}]{martin_patkr2015}%
  \BibitemOpen
  \bibfield  {author} {\bibinfo {author} {\bibfnamefont {D.}~\bibnamefont
  {Martin}}, \bibinfo {author} {\bibfnamefont {A.}~\bibnamefont {Perego}},
  \bibinfo {author} {\bibfnamefont {A.}~\bibnamefont {Arcones}}, \bibinfo
  {author} {\bibfnamefont {F.-K.}\ \bibnamefont {Thielemann}}, \bibinfo
  {author} {\bibfnamefont {O.}~\bibnamefont {Korobkin}}, \ and\ \bibinfo
  {author} {\bibfnamefont {S.}~\bibnamefont {Rosswog}},\ }\href {\doibase
  10.1088/0004-637X/813/1/2} {\bibfield  {journal} {\bibinfo  {journal}
  {Astrophys. J.}\ }\textbf {\bibinfo {volume} {813}},\ \bibinfo {pages} {2}
  (\bibinfo {year} {2015})}\BibitemShut {NoStop}%
\bibitem [{\citenamefont {Fujibayashi}\ \emph {et~al.}(2017)\citenamefont
  {Fujibayashi}, \citenamefont {Sekiguchi}, \citenamefont {Kiuchi},\ and\
  \citenamefont {Shibata}}]{fujibayashi_sks2017}%
  \BibitemOpen
  \bibfield  {author} {\bibinfo {author} {\bibfnamefont {S.}~\bibnamefont
  {Fujibayashi}}, \bibinfo {author} {\bibfnamefont {Y.}~\bibnamefont
  {Sekiguchi}}, \bibinfo {author} {\bibfnamefont {K.}~\bibnamefont {Kiuchi}}, \
  and\ \bibinfo {author} {\bibfnamefont {M.}~\bibnamefont {Shibata}},\ }\href
  {\doibase 10.3847/1538-4357/aa8039} {\bibfield  {journal} {\bibinfo
  {journal} {Astrophys. J.}\ }\textbf {\bibinfo {volume} {846}},\ \bibinfo
  {pages} {114} (\bibinfo {year} {2017})}\BibitemShut {NoStop}%
\bibitem [{\citenamefont {Metzger}\ and\ \citenamefont
  {Fern{\'a}ndez}(2014)}]{metzger_fernandez2014}%
  \BibitemOpen
  \bibfield  {author} {\bibinfo {author} {\bibfnamefont {B.~D.}\ \bibnamefont
  {Metzger}}\ and\ \bibinfo {author} {\bibfnamefont {R.}~\bibnamefont
  {Fern{\'a}ndez}},\ }\href {\doibase 10.1093/mnras/stu802} {\bibfield
  {journal} {\bibinfo  {journal} {Mon. Not. R. Astron. Soc.}\ }\textbf
  {\bibinfo {volume} {441}},\ \bibinfo {pages} {3444} (\bibinfo {year}
  {2014})}\BibitemShut {NoStop}%
\bibitem [{\citenamefont {Foucart}\ \emph
  {et~al.}(2016{\natexlab{b}})\citenamefont {Foucart}, \citenamefont {Haas},
  \citenamefont {Duez}, \citenamefont {O'Connor}, \citenamefont {Ott},
  \citenamefont {Roberts}, \citenamefont {Kidder}, \citenamefont {Lippuner},
  \citenamefont {Pfeiffer},\ and\ \citenamefont
  {Scheel}}]{foucart_hdoorklps2016}%
  \BibitemOpen
  \bibfield  {author} {\bibinfo {author} {\bibfnamefont {F.}~\bibnamefont
  {Foucart}}, \bibinfo {author} {\bibfnamefont {R.}~\bibnamefont {Haas}},
  \bibinfo {author} {\bibfnamefont {M.~D.}\ \bibnamefont {Duez}}, \bibinfo
  {author} {\bibfnamefont {E.}~\bibnamefont {O'Connor}}, \bibinfo {author}
  {\bibfnamefont {C.~D.}\ \bibnamefont {Ott}}, \bibinfo {author} {\bibfnamefont
  {L.}~\bibnamefont {Roberts}}, \bibinfo {author} {\bibfnamefont {L.~E.}\
  \bibnamefont {Kidder}}, \bibinfo {author} {\bibfnamefont {J.}~\bibnamefont
  {Lippuner}}, \bibinfo {author} {\bibfnamefont {H.~P.}\ \bibnamefont
  {Pfeiffer}}, \ and\ \bibinfo {author} {\bibfnamefont {M.~A.}\ \bibnamefont
  {Scheel}},\ }\href {\doibase 10.1103/PhysRevD.93.044019} {\bibfield
  {journal} {\bibinfo  {journal} {Phys. Rev. D}\ }\textbf {\bibinfo {volume}
  {93}},\ \bibinfo {pages} {044019} (\bibinfo {year}
  {2016}{\natexlab{b}})}\BibitemShut {NoStop}%
\bibitem [{\citenamefont {Wu}\ \emph {et~al.}(2016)\citenamefont {Wu},
  \citenamefont {Fern{\'a}ndez}, \citenamefont {Mart{\'i}nez-Pinedo},\ and\
  \citenamefont {Metzger}}]{wu_fmm2016}%
  \BibitemOpen
  \bibfield  {author} {\bibinfo {author} {\bibfnamefont {M.-R.}\ \bibnamefont
  {Wu}}, \bibinfo {author} {\bibfnamefont {R.}~\bibnamefont {Fern{\'a}ndez}},
  \bibinfo {author} {\bibfnamefont {G.}~\bibnamefont {Mart{\'i}nez-Pinedo}}, \
  and\ \bibinfo {author} {\bibfnamefont {B.~D.}\ \bibnamefont {Metzger}},\
  }\href {\doibase 10.1093/mnras/stw2156} {\bibfield  {journal} {\bibinfo
  {journal} {Mon. Not. R. Astron. Soc.}\ }\textbf {\bibinfo {volume} {463}},\
  \bibinfo {pages} {2323} (\bibinfo {year} {2016})}\BibitemShut {NoStop}%
\bibitem [{\citenamefont {Foucart}\ \emph {et~al.}(2015)\citenamefont
  {Foucart}, \citenamefont {O'Connor}, \citenamefont {Roberts}, \citenamefont
  {Duez}, \citenamefont {Haas}, \citenamefont {Kidder}, \citenamefont {Ott},
  \citenamefont {Pfeiffer}, \citenamefont {Scheel},\ and\ \citenamefont
  {Szilagyi}}]{foucart_etal2015}%
  \BibitemOpen
  \bibfield  {author} {\bibinfo {author} {\bibfnamefont {F.}~\bibnamefont
  {Foucart}}, \bibinfo {author} {\bibfnamefont {E.}~\bibnamefont {O'Connor}},
  \bibinfo {author} {\bibfnamefont {L.}~\bibnamefont {Roberts}}, \bibinfo
  {author} {\bibfnamefont {M.~D.}\ \bibnamefont {Duez}}, \bibinfo {author}
  {\bibfnamefont {R.}~\bibnamefont {Haas}}, \bibinfo {author} {\bibfnamefont
  {L.~E.}\ \bibnamefont {Kidder}}, \bibinfo {author} {\bibfnamefont {C.~D.}\
  \bibnamefont {Ott}}, \bibinfo {author} {\bibfnamefont {H.~P.}\ \bibnamefont
  {Pfeiffer}}, \bibinfo {author} {\bibfnamefont {M.~A.}\ \bibnamefont
  {Scheel}}, \ and\ \bibinfo {author} {\bibfnamefont {B.}~\bibnamefont
  {Szilagyi}},\ }\href {\doibase 10.1103/PhysRevD.91.124021} {\bibfield
  {journal} {\bibinfo  {journal} {Phys. Rev. D}\ }\textbf {\bibinfo {volume}
  {91}},\ \bibinfo {pages} {124021} (\bibinfo {year} {2015})}\BibitemShut
  {NoStop}%
\bibitem [{\citenamefont {Steiner}\ \emph {et~al.}(2013)\citenamefont
  {Steiner}, \citenamefont {Hempel},\ and\ \citenamefont
  {Fischer}}]{steiner_hf2013}%
  \BibitemOpen
  \bibfield  {author} {\bibinfo {author} {\bibfnamefont {A.~W.}\ \bibnamefont
  {Steiner}}, \bibinfo {author} {\bibfnamefont {M.}~\bibnamefont {Hempel}}, \
  and\ \bibinfo {author} {\bibfnamefont {T.}~\bibnamefont {Fischer}},\ }\href
  {\doibase 10.1088/0004-637X/774/1/17} {\bibfield  {journal} {\bibinfo
  {journal} {Astrophys. J.}\ }\textbf {\bibinfo {volume} {774}},\ \bibinfo
  {pages} {17} (\bibinfo {year} {2013})}\BibitemShut {NoStop}%
\bibitem [{\citenamefont {Banik}\ \emph {et~al.}(2014)\citenamefont {Banik},
  \citenamefont {Hempel},\ and\ \citenamefont {Bandyopadhyay}}]{banik_hb2014}%
  \BibitemOpen
  \bibfield  {author} {\bibinfo {author} {\bibfnamefont {S.}~\bibnamefont
  {Banik}}, \bibinfo {author} {\bibfnamefont {M.}~\bibnamefont {Hempel}}, \
  and\ \bibinfo {author} {\bibfnamefont {D.}~\bibnamefont {Bandyopadhyay}},\
  }\href {\doibase 10.1088/0067-0049/214/2/22} {\bibfield  {journal} {\bibinfo
  {journal} {Astrophys. J. Suppl.}\ }\textbf {\bibinfo {volume} {214}},\
  \bibinfo {pages} {22} (\bibinfo {year} {2014})}\BibitemShut {NoStop}%
\bibitem [{\citenamefont {Hempel}\ \emph {et~al.}(2012)\citenamefont {Hempel},
  \citenamefont {Fischer}, \citenamefont {Schaffner-Bielich},\ and\
  \citenamefont {Liebend{\"o}rfer}}]{hempel_fsl2012}%
  \BibitemOpen
  \bibfield  {author} {\bibinfo {author} {\bibfnamefont {M.}~\bibnamefont
  {Hempel}}, \bibinfo {author} {\bibfnamefont {T.}~\bibnamefont {Fischer}},
  \bibinfo {author} {\bibfnamefont {J.}~\bibnamefont {Schaffner-Bielich}}, \
  and\ \bibinfo {author} {\bibfnamefont {M.}~\bibnamefont {Liebend{\"o}rfer}},\
  }\href {\doibase 10.1088/0004-637X/748/1/70} {\bibfield  {journal} {\bibinfo
  {journal} {Astrophys. J.}\ }\textbf {\bibinfo {volume} {748}},\ \bibinfo
  {pages} {70} (\bibinfo {year} {2012})}\BibitemShut {NoStop}%
\bibitem [{\citenamefont {Hebeler}\ \emph {et~al.}(2010)\citenamefont
  {Hebeler}, \citenamefont {Lattimer}, \citenamefont {Pethick},\ and\
  \citenamefont {Schwenk}}]{hebeler_lps2010}%
  \BibitemOpen
  \bibfield  {author} {\bibinfo {author} {\bibfnamefont {K.}~\bibnamefont
  {Hebeler}}, \bibinfo {author} {\bibfnamefont {J.~M.}\ \bibnamefont
  {Lattimer}}, \bibinfo {author} {\bibfnamefont {C.~J.}\ \bibnamefont
  {Pethick}}, \ and\ \bibinfo {author} {\bibfnamefont {A.}~\bibnamefont
  {Schwenk}},\ }\href {\doibase 10.1103/PhysRevLett.105.161102} {\bibfield
  {journal} {\bibinfo  {journal} {Phys. Rev. Lett.}\ }\textbf {\bibinfo
  {volume} {105}},\ \bibinfo {pages} {161102} (\bibinfo {year}
  {2010})}\BibitemShut {NoStop}%
\bibitem [{\citenamefont {Tews}\ \emph {et~al.}(2017)\citenamefont {Tews},
  \citenamefont {Lattimer}, \citenamefont {Ohnishi},\ and\ \citenamefont
  {Kolomeitsev}}]{tews_lok2017}%
  \BibitemOpen
  \bibfield  {author} {\bibinfo {author} {\bibfnamefont {I.}~\bibnamefont
  {Tews}}, \bibinfo {author} {\bibfnamefont {J.~M.}\ \bibnamefont {Lattimer}},
  \bibinfo {author} {\bibfnamefont {A.}~\bibnamefont {Ohnishi}}, \ and\
  \bibinfo {author} {\bibfnamefont {E.~E.}\ \bibnamefont {Kolomeitsev}},\
  }\href {\doibase 10.3847/1538-4357/aa8db9} {\bibfield  {journal} {\bibinfo
  {journal} {Astrophys. J.}\ }\textbf {\bibinfo {volume} {848}},\ \bibinfo
  {pages} {105} (\bibinfo {year} {2017})}\BibitemShut {NoStop}%
\bibitem [{\citenamefont {Demorest}\ \emph {et~al.}(2010)\citenamefont
  {Demorest}, \citenamefont {Pennucci}, \citenamefont {Ransom}, \citenamefont
  {Roberts},\ and\ \citenamefont {Hessels}}]{demorest_prrh2010}%
  \BibitemOpen
  \bibfield  {author} {\bibinfo {author} {\bibfnamefont {P.}~\bibnamefont
  {Demorest}}, \bibinfo {author} {\bibfnamefont {T.}~\bibnamefont {Pennucci}},
  \bibinfo {author} {\bibfnamefont {S.}~\bibnamefont {Ransom}}, \bibinfo
  {author} {\bibfnamefont {M.}~\bibnamefont {Roberts}}, \ and\ \bibinfo
  {author} {\bibfnamefont {J.}~\bibnamefont {Hessels}},\ }\href {\doibase
  10.1038/nature09466} {\bibfield  {journal} {\bibinfo  {journal} {Nature
  (London)}\ }\textbf {\bibinfo {volume} {467}},\ \bibinfo {pages} {1081}
  (\bibinfo {year} {2010})}\BibitemShut {NoStop}%
\bibitem [{\citenamefont {{J. Antoniadis et al.}}(2013)}]{antoniadis_etal2013}%
  \BibitemOpen
  \bibfield  {author} {\bibinfo {author} {\bibnamefont {{J. Antoniadis et
  al.}}},\ }\href {\doibase 10.1126/science.1233232} {\bibfield  {journal}
  {\bibinfo  {journal} {Science}\ }\textbf {\bibinfo {volume} {340}},\ \bibinfo
  {pages} {1233232} (\bibinfo {year} {2013})}\BibitemShut {NoStop}%
\bibitem [{\citenamefont {Kyutoku}\ \emph {et~al.}(2009)\citenamefont
  {Kyutoku}, \citenamefont {Shibata},\ and\ \citenamefont
  {Taniguchi}}]{kyutoku_st2009}%
  \BibitemOpen
  \bibfield  {author} {\bibinfo {author} {\bibfnamefont {K.}~\bibnamefont
  {Kyutoku}}, \bibinfo {author} {\bibfnamefont {M.}~\bibnamefont {Shibata}}, \
  and\ \bibinfo {author} {\bibfnamefont {K.}~\bibnamefont {Taniguchi}},\ }\href
  {\doibase 10.1103/PhysRevD.79.124018} {\bibfield  {journal} {\bibinfo
  {journal} {Phys. Rev. D}\ }\textbf {\bibinfo {volume} {79}},\ \bibinfo
  {pages} {124018} (\bibinfo {year} {2009})}\BibitemShut {NoStop}%
\bibitem [{\citenamefont {Kreidberg}\ \emph {et~al.}(2012)\citenamefont
  {Kreidberg}, \citenamefont {Bailyn}, \citenamefont {Farr},\ and\
  \citenamefont {Kalogera}}]{kreidberg_bfk2012}%
  \BibitemOpen
  \bibfield  {author} {\bibinfo {author} {\bibfnamefont {L.}~\bibnamefont
  {Kreidberg}}, \bibinfo {author} {\bibfnamefont {C.~D.}\ \bibnamefont
  {Bailyn}}, \bibinfo {author} {\bibfnamefont {W.~M.}\ \bibnamefont {Farr}}, \
  and\ \bibinfo {author} {\bibfnamefont {V.}~\bibnamefont {Kalogera}},\ }\href
  {\doibase 10.1088/0004-637X/757/1/36} {\bibfield  {journal} {\bibinfo
  {journal} {Astrophys. J.}\ }\textbf {\bibinfo {volume} {757}},\ \bibinfo
  {pages} {36} (\bibinfo {year} {2012})}\BibitemShut {NoStop}%
\bibitem [{\citenamefont {Shibata}\ and\ \citenamefont
  {Nakamura}(1995)}]{shibata_nakamura1995}%
  \BibitemOpen
  \bibfield  {author} {\bibinfo {author} {\bibfnamefont {M.}~\bibnamefont
  {Shibata}}\ and\ \bibinfo {author} {\bibfnamefont {T.}~\bibnamefont
  {Nakamura}},\ }\href {\doibase 10.1103/PhysRevD.52.5428} {\bibfield
  {journal} {\bibinfo  {journal} {Phys. Rev. D}\ }\textbf {\bibinfo {volume}
  {52}},\ \bibinfo {pages} {5428} (\bibinfo {year} {1995})}\BibitemShut
  {NoStop}%
\bibitem [{\citenamefont {Baumgarte}\ and\ \citenamefont
  {Shapiro}(1998)}]{baumgarte_shapiro1998}%
  \BibitemOpen
  \bibfield  {author} {\bibinfo {author} {\bibfnamefont {T.~W.}\ \bibnamefont
  {Baumgarte}}\ and\ \bibinfo {author} {\bibfnamefont {S.~L.}\ \bibnamefont
  {Shapiro}},\ }\href {\doibase 10.1103/PhysRevD.59.024007} {\bibfield
  {journal} {\bibinfo  {journal} {Phys. Rev. D}\ }\textbf {\bibinfo {volume}
  {59}},\ \bibinfo {pages} {024007} (\bibinfo {year} {1998})}\BibitemShut
  {NoStop}%
\bibitem [{\citenamefont {Campanelli}\ \emph {et~al.}(2006)\citenamefont
  {Campanelli}, \citenamefont {Lousto}, \citenamefont {Marronetti},\ and\
  \citenamefont {Zlochower}}]{campanelli_lmz2006}%
  \BibitemOpen
  \bibfield  {author} {\bibinfo {author} {\bibfnamefont {M.}~\bibnamefont
  {Campanelli}}, \bibinfo {author} {\bibfnamefont {C.~O.}\ \bibnamefont
  {Lousto}}, \bibinfo {author} {\bibfnamefont {P.}~\bibnamefont {Marronetti}},
  \ and\ \bibinfo {author} {\bibfnamefont {Y.}~\bibnamefont {Zlochower}},\
  }\href {\doibase 10.1103/PhysRevLett.96.111101} {\bibfield  {journal}
  {\bibinfo  {journal} {Phys. Rev. Lett.}\ }\textbf {\bibinfo {volume} {96}},\
  \bibinfo {pages} {111101} (\bibinfo {year} {2006})}\BibitemShut {NoStop}%
\bibitem [{\citenamefont {Baker}\ \emph {et~al.}(2006)\citenamefont {Baker},
  \citenamefont {Centrella}, \citenamefont {Choi}, \citenamefont {Koppitz},\
  and\ \citenamefont {van Meter}}]{baker_cckm2006}%
  \BibitemOpen
  \bibfield  {author} {\bibinfo {author} {\bibfnamefont {J.~G.}\ \bibnamefont
  {Baker}}, \bibinfo {author} {\bibfnamefont {J.}~\bibnamefont {Centrella}},
  \bibinfo {author} {\bibfnamefont {D.-I.}\ \bibnamefont {Choi}}, \bibinfo
  {author} {\bibfnamefont {M.}~\bibnamefont {Koppitz}}, \ and\ \bibinfo
  {author} {\bibfnamefont {J.}~\bibnamefont {van Meter}},\ }\href {\doibase
  10.1103/PhysRevLett.96.111102} {\bibfield  {journal} {\bibinfo  {journal}
  {Phys. Rev. Lett.}\ }\textbf {\bibinfo {volume} {96}},\ \bibinfo {pages}
  {111102} (\bibinfo {year} {2006})}\BibitemShut {NoStop}%
\bibitem [{\citenamefont {Marronetti}\ \emph {et~al.}(2008)\citenamefont
  {Marronetti}, \citenamefont {Tichy}, \citenamefont {Br{\"u}gmann},
  \citenamefont {Gonz{\'a}lez},\ and\ \citenamefont
  {Sperhake}}]{marronetti_tbgs2008}%
  \BibitemOpen
  \bibfield  {author} {\bibinfo {author} {\bibfnamefont {P.}~\bibnamefont
  {Marronetti}}, \bibinfo {author} {\bibfnamefont {W.}~\bibnamefont {Tichy}},
  \bibinfo {author} {\bibfnamefont {B.}~\bibnamefont {Br{\"u}gmann}}, \bibinfo
  {author} {\bibfnamefont {J.~A.}\ \bibnamefont {Gonz{\'a}lez}}, \ and\
  \bibinfo {author} {\bibfnamefont {U.}~\bibnamefont {Sperhake}},\ }\href
  {\doibase 10.1103/PhysRevD.77.064010} {\bibfield  {journal} {\bibinfo
  {journal} {Phys. Rev. D}\ }\textbf {\bibinfo {volume} {77}},\ \bibinfo
  {pages} {064010} (\bibinfo {year} {2008})}\BibitemShut {NoStop}%
\bibitem [{\citenamefont {Thorne}(1981)}]{thorne1981}%
  \BibitemOpen
  \bibfield  {author} {\bibinfo {author} {\bibfnamefont {K.~S.}\ \bibnamefont
  {Thorne}},\ }\href {\doibase 10.1093/mnras/194.2.439} {\bibfield  {journal}
  {\bibinfo  {journal} {Mon. Not. R. Astron. Soc.}\ }\textbf {\bibinfo {volume}
  {194}},\ \bibinfo {pages} {439} (\bibinfo {year} {1981})}\BibitemShut
  {NoStop}%
\bibitem [{\citenamefont {Shibata}\ \emph {et~al.}(2011)\citenamefont
  {Shibata}, \citenamefont {Kiuchi}, \citenamefont {Sekiguchi},\ and\
  \citenamefont {Suwa}}]{shibata_kss2011}%
  \BibitemOpen
  \bibfield  {author} {\bibinfo {author} {\bibfnamefont {M.}~\bibnamefont
  {Shibata}}, \bibinfo {author} {\bibfnamefont {K.}~\bibnamefont {Kiuchi}},
  \bibinfo {author} {\bibfnamefont {Y.}~\bibnamefont {Sekiguchi}}, \ and\
  \bibinfo {author} {\bibfnamefont {Y.}~\bibnamefont {Suwa}},\ }\href {\doibase
  10.1143/PTP.125.1255} {\bibfield  {journal} {\bibinfo  {journal} {Prog.
  Theor. Phys.}\ }\textbf {\bibinfo {volume} {125}},\ \bibinfo {pages} {1255}
  (\bibinfo {year} {2011})}\BibitemShut {NoStop}%
\bibitem [{\citenamefont {Sekiguchi}(2010)}]{sekiguchi2010}%
  \BibitemOpen
  \bibfield  {author} {\bibinfo {author} {\bibfnamefont {Y.}~\bibnamefont
  {Sekiguchi}},\ }\href {\doibase 10.1143/PTP.124.331} {\bibfield  {journal}
  {\bibinfo  {journal} {Prog. Theor. Phys.}\ }\textbf {\bibinfo {volume}
  {124}},\ \bibinfo {pages} {331} (\bibinfo {year} {2010})}\BibitemShut
  {NoStop}%
\bibitem [{\citenamefont {Sekiguchi}\ and\ \citenamefont
  {Shibata}(2011)}]{sekiguchi_shibata2011}%
  \BibitemOpen
  \bibfield  {author} {\bibinfo {author} {\bibfnamefont {Y.}~\bibnamefont
  {Sekiguchi}}\ and\ \bibinfo {author} {\bibfnamefont {M.}~\bibnamefont
  {Shibata}},\ }\href {\doibase 10.1088/0004-637X/737/1/6} {\bibfield
  {journal} {\bibinfo  {journal} {Astrophys. J.}\ }\textbf {\bibinfo {volume}
  {737}},\ \bibinfo {pages} {6} (\bibinfo {year} {2011})}\BibitemShut {NoStop}%
\bibitem [{\citenamefont {Sekiguchi}\ \emph {et~al.}(2012)\citenamefont
  {Sekiguchi}, \citenamefont {Kiuchi}, \citenamefont {Kyutoku},\ and\
  \citenamefont {Shibata}}]{sekiguchi_kks2012}%
  \BibitemOpen
  \bibfield  {author} {\bibinfo {author} {\bibfnamefont {Y.}~\bibnamefont
  {Sekiguchi}}, \bibinfo {author} {\bibfnamefont {K.}~\bibnamefont {Kiuchi}},
  \bibinfo {author} {\bibfnamefont {K.}~\bibnamefont {Kyutoku}}, \ and\
  \bibinfo {author} {\bibfnamefont {M.}~\bibnamefont {Shibata}},\ }\href
  {\doibase 10.1093/ptep/pts011} {\bibfield  {journal} {\bibinfo  {journal}
  {Prog. Theor. Exp. Phys.}\ }\textbf {\bibinfo {volume} {2012}},\ \bibinfo
  {pages} {01A304} (\bibinfo {year} {2012})}\BibitemShut {NoStop}%
\bibitem [{\citenamefont {Kyutoku}\ \emph {et~al.}(2010)\citenamefont
  {Kyutoku}, \citenamefont {Shibata},\ and\ \citenamefont
  {Taniguchi}}]{kyutoku_st2010}%
  \BibitemOpen
  \bibfield  {author} {\bibinfo {author} {\bibfnamefont {K.}~\bibnamefont
  {Kyutoku}}, \bibinfo {author} {\bibfnamefont {M.}~\bibnamefont {Shibata}}, \
  and\ \bibinfo {author} {\bibfnamefont {K.}~\bibnamefont {Taniguchi}},\ }\href
  {\doibase 10.1103/PhysRevD.82.044049} {\bibfield  {journal} {\bibinfo
  {journal} {Phys. Rev. D}\ }\textbf {\bibinfo {volume} {82}},\ \bibinfo
  {pages} {044049} (\bibinfo {year} {2010})}\BibitemShut {NoStop}%
\bibitem [{\citenamefont {Kyutoku}\ \emph
  {et~al.}(2011{\natexlab{b}})\citenamefont {Kyutoku}, \citenamefont
  {Shibata},\ and\ \citenamefont {Taniguchi}}]{kyutoku_st2010e}%
  \BibitemOpen
  \bibfield  {author} {\bibinfo {author} {\bibfnamefont {K.}~\bibnamefont
  {Kyutoku}}, \bibinfo {author} {\bibfnamefont {M.}~\bibnamefont {Shibata}}, \
  and\ \bibinfo {author} {\bibfnamefont {K.}~\bibnamefont {Taniguchi}},\ }\href
  {\doibase 10.1103/PhysRevD.84.049902} {\bibfield  {journal} {\bibinfo
  {journal} {Phys. Rev. D}\ }\textbf {\bibinfo {volume} {84}},\ \bibinfo
  {pages} {049902(E)} (\bibinfo {year} {2011}{\natexlab{b}})}\BibitemShut
  {NoStop}%
\bibitem [{\citenamefont {Foucart}\ \emph {et~al.}(2011)\citenamefont
  {Foucart}, \citenamefont {Duez}, \citenamefont {Kidder},\ and\ \citenamefont
  {Teukolsky}}]{foucart_dkt2011}%
  \BibitemOpen
  \bibfield  {author} {\bibinfo {author} {\bibfnamefont {F.}~\bibnamefont
  {Foucart}}, \bibinfo {author} {\bibfnamefont {M.~D.}\ \bibnamefont {Duez}},
  \bibinfo {author} {\bibfnamefont {L.~E.}\ \bibnamefont {Kidder}}, \ and\
  \bibinfo {author} {\bibfnamefont {S.~A.}\ \bibnamefont {Teukolsky}},\ }\href
  {\doibase 10.1103/PhysRevD.83.024005} {\bibfield  {journal} {\bibinfo
  {journal} {Phys. Rev. D}\ }\textbf {\bibinfo {volume} {83}},\ \bibinfo
  {pages} {024005} (\bibinfo {year} {2011})}\BibitemShut {NoStop}%
\bibitem [{\citenamefont {Shibata}\ and\ \citenamefont
  {Taniguchi}(2011)}]{shibata_taniguchi2011}%
  \BibitemOpen
  \bibfield  {author} {\bibinfo {author} {\bibfnamefont {M.}~\bibnamefont
  {Shibata}}\ and\ \bibinfo {author} {\bibfnamefont {K.}~\bibnamefont
  {Taniguchi}},\ }\href {\doibase 10.12942/lrr-2011-6} {\bibfield  {journal}
  {\bibinfo  {journal} {Living Rev. Relativity}\ }\textbf {\bibinfo {volume}
  {14}},\ \bibinfo {pages} {6} (\bibinfo {year} {2011})}\BibitemShut {NoStop}%
\bibitem [{\citenamefont {Fern{\'a}ndez}\ and\ \citenamefont
  {Metzger}(2013)}]{fernandez_metzger2013}%
  \BibitemOpen
  \bibfield  {author} {\bibinfo {author} {\bibfnamefont {R.}~\bibnamefont
  {Fern{\'a}ndez}}\ and\ \bibinfo {author} {\bibfnamefont {B.~D.}\ \bibnamefont
  {Metzger}},\ }\href {\doibase 10.1093/mnras/stt1312} {\bibfield  {journal}
  {\bibinfo  {journal} {Mon. Not. R. Astron. Soc.}\ }\textbf {\bibinfo {volume}
  {435}},\ \bibinfo {pages} {502} (\bibinfo {year} {2013})}\BibitemShut
  {NoStop}%
\bibitem [{\citenamefont {Fern{\'a}ndez}\ \emph {et~al.}(2015)\citenamefont
  {Fern{\'a}ndez}, \citenamefont {Kasen}, \citenamefont {Metzger},\ and\
  \citenamefont {Quataert}}]{fernandez_kmq2015}%
  \BibitemOpen
  \bibfield  {author} {\bibinfo {author} {\bibfnamefont {R.}~\bibnamefont
  {Fern{\'a}ndez}}, \bibinfo {author} {\bibfnamefont {D.}~\bibnamefont
  {Kasen}}, \bibinfo {author} {\bibfnamefont {B.~D.}\ \bibnamefont {Metzger}},
  \ and\ \bibinfo {author} {\bibfnamefont {E.}~\bibnamefont {Quataert}},\
  }\href {\doibase 10.1093/mnras/stu2112} {\bibfield  {journal} {\bibinfo
  {journal} {Mon. Not. R. Astron. Soc.}\ }\textbf {\bibinfo {volume} {446}},\
  \bibinfo {pages} {750} (\bibinfo {year} {2015})}\BibitemShut {NoStop}%
\bibitem [{\citenamefont {Siegel}\ and\ \citenamefont
  {Metzger}(2017)}]{siegel_metzger2017}%
  \BibitemOpen
  \bibfield  {author} {\bibinfo {author} {\bibfnamefont {D.~M.}\ \bibnamefont
  {Siegel}}\ and\ \bibinfo {author} {\bibfnamefont {B.~D.}\ \bibnamefont
  {Metzger}},\ }\href {\doibase 10.1103/PhysRevLett.119.231102} {\bibfield
  {journal} {\bibinfo  {journal} {Phys. Rev. Lett.}\ }\textbf {\bibinfo
  {volume} {119}},\ \bibinfo {pages} {231102} (\bibinfo {year}
  {2017})}\BibitemShut {NoStop}%
\bibitem [{\citenamefont {Fern{\'a}ndez}\ \emph {et~al.}(2017)\citenamefont
  {Fern{\'a}ndez}, \citenamefont {Foucart}, \citenamefont {Kasen},
  \citenamefont {Lippuner}, \citenamefont {Desai},\ and\ \citenamefont
  {Roberts}}]{fernandez_fkldr2017}%
  \BibitemOpen
  \bibfield  {author} {\bibinfo {author} {\bibfnamefont {R.}~\bibnamefont
  {Fern{\'a}ndez}}, \bibinfo {author} {\bibfnamefont {F.}~\bibnamefont
  {Foucart}}, \bibinfo {author} {\bibfnamefont {D.}~\bibnamefont {Kasen}},
  \bibinfo {author} {\bibfnamefont {J.}~\bibnamefont {Lippuner}}, \bibinfo
  {author} {\bibfnamefont {D.}~\bibnamefont {Desai}}, \ and\ \bibinfo {author}
  {\bibfnamefont {L.~F.}\ \bibnamefont {Roberts}},\ }\href {\doibase
  10.1088/1361-6382/aa7a77} {\bibfield  {journal} {\bibinfo  {journal} {Class.
  Quantum Grav.}\ }\textbf {\bibinfo {volume} {34}},\ \bibinfo {pages} {154001}
  (\bibinfo {year} {2017})}\BibitemShut {NoStop}%
\bibitem [{\citenamefont {Ruffert}\ \emph {et~al.}(1997)\citenamefont
  {Ruffert}, \citenamefont {Janka}, \citenamefont {Takahashi},\ and\
  \citenamefont {Sch{\"a}fer}}]{ruffert_jts1997}%
  \BibitemOpen
  \bibfield  {author} {\bibinfo {author} {\bibfnamefont {M.}~\bibnamefont
  {Ruffert}}, \bibinfo {author} {\bibfnamefont {H.-T.}\ \bibnamefont {Janka}},
  \bibinfo {author} {\bibfnamefont {K.}~\bibnamefont {Takahashi}}, \ and\
  \bibinfo {author} {\bibfnamefont {G.}~\bibnamefont {Sch{\"a}fer}},\
  }\href@noop {} {\bibfield  {journal} {\bibinfo  {journal} {Astron.
  Astrophys.}\ }\textbf {\bibinfo {volume} {319}},\ \bibinfo {pages} {122}
  (\bibinfo {year} {1997})}\BibitemShut {NoStop}%
\bibitem [{\citenamefont {Rosswog}\ and\ \citenamefont
  {Liebend{\"o}rfer}(2003)}]{rosswog_liebendorfer2003}%
  \BibitemOpen
  \bibfield  {author} {\bibinfo {author} {\bibfnamefont {S.}~\bibnamefont
  {Rosswog}}\ and\ \bibinfo {author} {\bibfnamefont {M.}~\bibnamefont
  {Liebend{\"o}rfer}},\ }\href {\doibase 10.1046/j.1365-8711.2003.06579.x}
  {\bibfield  {journal} {\bibinfo  {journal} {Mon. Not. R. Astron. Soc.}\
  }\textbf {\bibinfo {volume} {342}},\ \bibinfo {pages} {673} (\bibinfo {year}
  {2003})}\BibitemShut {NoStop}%
\bibitem [{\citenamefont {Sekiguchi}\ \emph
  {et~al.}(2011{\natexlab{a}})\citenamefont {Sekiguchi}, \citenamefont
  {Kiuchi}, \citenamefont {Kyutoku},\ and\ \citenamefont
  {Shibata}}]{sekiguchi_kks2011}%
  \BibitemOpen
  \bibfield  {author} {\bibinfo {author} {\bibfnamefont {Y.}~\bibnamefont
  {Sekiguchi}}, \bibinfo {author} {\bibfnamefont {K.}~\bibnamefont {Kiuchi}},
  \bibinfo {author} {\bibfnamefont {K.}~\bibnamefont {Kyutoku}}, \ and\
  \bibinfo {author} {\bibfnamefont {M.}~\bibnamefont {Shibata}},\ }\href
  {\doibase 10.1103/PhysRevLett.107.051102} {\bibfield  {journal} {\bibinfo
  {journal} {Phys. Rev. Lett.}\ }\textbf {\bibinfo {volume} {107}},\ \bibinfo
  {pages} {051102} (\bibinfo {year} {2011}{\natexlab{a}})}\BibitemShut
  {NoStop}%
\bibitem [{\citenamefont {Janka}\ \emph {et~al.}(1999)\citenamefont {Janka},
  \citenamefont {Eberl}, \citenamefont {Ruffert},\ and\ \citenamefont
  {Fryer}}]{janka_erf1999}%
  \BibitemOpen
  \bibfield  {author} {\bibinfo {author} {\bibfnamefont {H.-T.}\ \bibnamefont
  {Janka}}, \bibinfo {author} {\bibfnamefont {T.}~\bibnamefont {Eberl}},
  \bibinfo {author} {\bibfnamefont {M.}~\bibnamefont {Ruffert}}, \ and\
  \bibinfo {author} {\bibfnamefont {C.~L.}\ \bibnamefont {Fryer}},\ }\href
  {\doibase 10.1086/312397} {\bibfield  {journal} {\bibinfo  {journal}
  {Astrophys. J.}\ }\textbf {\bibinfo {volume} {527}},\ \bibinfo {pages} {L39}
  (\bibinfo {year} {1999})}\BibitemShut {NoStop}%
\bibitem [{\citenamefont {Foucart}(2012)}]{foucart2012}%
  \BibitemOpen
  \bibfield  {author} {\bibinfo {author} {\bibfnamefont {F.}~\bibnamefont
  {Foucart}},\ }\href {\doibase 10.1103/PhysRevD.86.124007} {\bibfield
  {journal} {\bibinfo  {journal} {Phys. Rev. D}\ }\textbf {\bibinfo {volume}
  {86}},\ \bibinfo {pages} {124007} (\bibinfo {year} {2012})}\BibitemShut
  {NoStop}%
\bibitem [{\citenamefont {Pannarale}(2013)}]{pannarale2013}%
  \BibitemOpen
  \bibfield  {author} {\bibinfo {author} {\bibfnamefont {F.}~\bibnamefont
  {Pannarale}},\ }\href {\doibase 10.1103/PhysRevD.88.104025} {\bibfield
  {journal} {\bibinfo  {journal} {Phys. Rev. D}\ }\textbf {\bibinfo {volume}
  {88}},\ \bibinfo {pages} {104025} (\bibinfo {year} {2013})}\BibitemShut
  {NoStop}%
\bibitem [{\citenamefont {Sekiguchi}\ \emph
  {et~al.}(2011{\natexlab{b}})\citenamefont {Sekiguchi}, \citenamefont
  {Kiuchi}, \citenamefont {Kyutoku},\ and\ \citenamefont
  {Shibata}}]{sekiguchi_kks2011-2}%
  \BibitemOpen
  \bibfield  {author} {\bibinfo {author} {\bibfnamefont {Y.}~\bibnamefont
  {Sekiguchi}}, \bibinfo {author} {\bibfnamefont {K.}~\bibnamefont {Kiuchi}},
  \bibinfo {author} {\bibfnamefont {K.}~\bibnamefont {Kyutoku}}, \ and\
  \bibinfo {author} {\bibfnamefont {M.}~\bibnamefont {Shibata}},\ }\href
  {\doibase 10.1103/PhysRevLett.107.211101} {\bibfield  {journal} {\bibinfo
  {journal} {Phys. Rev. Lett.}\ }\textbf {\bibinfo {volume} {107}},\ \bibinfo
  {pages} {211101} (\bibinfo {year} {2011}{\natexlab{b}})}\BibitemShut
  {NoStop}%
\bibitem [{\citenamefont {Kato}\ \emph {et~al.}(2008)\citenamefont {Kato},
  \citenamefont {Fukue},\ and\ \citenamefont {Mineshige}}]{kato_fm}%
  \BibitemOpen
  \bibfield  {author} {\bibinfo {author} {\bibfnamefont {S.}~\bibnamefont
  {Kato}}, \bibinfo {author} {\bibfnamefont {J.}~\bibnamefont {Fukue}}, \ and\
  \bibinfo {author} {\bibfnamefont {S.}~\bibnamefont {Mineshige}},\ }\href@noop
  {} {\emph {\bibinfo {title} {Black-hole accretion disks --- towards a new
  paradigm ---}}}\ (\bibinfo  {publisher} {Kyoto University Press},\ \bibinfo
  {year} {2008})\BibitemShut {NoStop}%
\bibitem [{\citenamefont {Popham}\ \emph {et~al.}(1999)\citenamefont {Popham},
  \citenamefont {Woosley},\ and\ \citenamefont {Fryer}}]{popham_wf1999}%
  \BibitemOpen
  \bibfield  {author} {\bibinfo {author} {\bibfnamefont {R.}~\bibnamefont
  {Popham}}, \bibinfo {author} {\bibfnamefont {S.~E.}\ \bibnamefont {Woosley}},
  \ and\ \bibinfo {author} {\bibfnamefont {C.}~\bibnamefont {Fryer}},\ }\href
  {\doibase 10.1086/307259} {\bibfield  {journal} {\bibinfo  {journal}
  {Astrophys. J.}\ }\textbf {\bibinfo {volume} {518}},\ \bibinfo {pages} {356}
  (\bibinfo {year} {1999})}\BibitemShut {NoStop}%
\bibitem [{\citenamefont {Kohri}\ and\ \citenamefont
  {Mineshige}(2002)}]{kohri_mineshige2002}%
  \BibitemOpen
  \bibfield  {author} {\bibinfo {author} {\bibfnamefont {K.}~\bibnamefont
  {Kohri}}\ and\ \bibinfo {author} {\bibfnamefont {S.}~\bibnamefont
  {Mineshige}},\ }\href {\doibase 10.1086/342166} {\bibfield  {journal}
  {\bibinfo  {journal} {Astrophys. J.}\ }\textbf {\bibinfo {volume} {577}},\
  \bibinfo {pages} {311} (\bibinfo {year} {2002})}\BibitemShut {NoStop}%
\bibitem [{\citenamefont {Shibata}\ \emph {et~al.}(2007)\citenamefont
  {Shibata}, \citenamefont {Sekiguchi},\ and\ \citenamefont
  {Takahashi}}]{shibata_st2007}%
  \BibitemOpen
  \bibfield  {author} {\bibinfo {author} {\bibfnamefont {M.}~\bibnamefont
  {Shibata}}, \bibinfo {author} {\bibfnamefont {Y.-i.}\ \bibnamefont
  {Sekiguchi}}, \ and\ \bibinfo {author} {\bibfnamefont {R.}~\bibnamefont
  {Takahashi}},\ }\href {\doibase 10.1143/PTP.118.257} {\bibfield  {journal}
  {\bibinfo  {journal} {Prog. Theor. Phys.}\ }\textbf {\bibinfo {volume}
  {118}},\ \bibinfo {pages} {257} (\bibinfo {year} {2007})}\BibitemShut
  {NoStop}%
\bibitem [{\citenamefont {Shibata}\ \emph {et~al.}(2017)\citenamefont
  {Shibata}, \citenamefont {Kiuchi},\ and\ \citenamefont
  {Sekiguchi}}]{shibata_ks2017}%
  \BibitemOpen
  \bibfield  {author} {\bibinfo {author} {\bibfnamefont {M.}~\bibnamefont
  {Shibata}}, \bibinfo {author} {\bibfnamefont {K.}~\bibnamefont {Kiuchi}}, \
  and\ \bibinfo {author} {\bibfnamefont {Y.-i.}\ \bibnamefont {Sekiguchi}},\
  }\href {\doibase 10.1103/PhysRevD.95.083005} {\bibfield  {journal} {\bibinfo
  {journal} {Phys. Rev. D}\ }\textbf {\bibinfo {volume} {95}},\ \bibinfo
  {pages} {083005} (\bibinfo {year} {2017})}\BibitemShut {NoStop}%
\bibitem [{\citenamefont {Shibata}\ and\ \citenamefont
  {Kiuchi}(2017)}]{shibata_kiuchi2017}%
  \BibitemOpen
  \bibfield  {author} {\bibinfo {author} {\bibfnamefont {M.}~\bibnamefont
  {Shibata}}\ and\ \bibinfo {author} {\bibfnamefont {K.}~\bibnamefont
  {Kiuchi}},\ }\href {\doibase 10.1103/PhysRevD.95.123003} {\bibfield
  {journal} {\bibinfo  {journal} {Phys. Rev. D}\ }\textbf {\bibinfo {volume}
  {95}},\ \bibinfo {pages} {123003} (\bibinfo {year} {2017})}\BibitemShut
  {NoStop}%
\bibitem [{\citenamefont {Lattimer}\ and\ \citenamefont
  {Prakash}(2001)}]{lattimer_prakash2001}%
  \BibitemOpen
  \bibfield  {author} {\bibinfo {author} {\bibfnamefont {J.~M.}\ \bibnamefont
  {Lattimer}}\ and\ \bibinfo {author} {\bibfnamefont {M.}~\bibnamefont
  {Prakash}},\ }\href {\doibase 10.1086/319702} {\bibfield  {journal} {\bibinfo
   {journal} {Astrophys. J.}\ }\textbf {\bibinfo {volume} {550}},\ \bibinfo
  {pages} {426} (\bibinfo {year} {2001})}\BibitemShut {NoStop}%
\bibitem [{\citenamefont {Richers}\ \emph {et~al.}(2015)\citenamefont
  {Richers}, \citenamefont {Kasen}, \citenamefont {O'Connor}, \citenamefont
  {Fern{\'a}ndez},\ and\ \citenamefont {Ott}}]{richers_kofo2015}%
  \BibitemOpen
  \bibfield  {author} {\bibinfo {author} {\bibfnamefont {S.}~\bibnamefont
  {Richers}}, \bibinfo {author} {\bibfnamefont {D.}~\bibnamefont {Kasen}},
  \bibinfo {author} {\bibfnamefont {E.}~\bibnamefont {O'Connor}}, \bibinfo
  {author} {\bibfnamefont {R.}~\bibnamefont {Fern{\'a}ndez}}, \ and\ \bibinfo
  {author} {\bibfnamefont {C.~D.}\ \bibnamefont {Ott}},\ }\href {\doibase
  10.1088/0004-637X/813/1/38} {\bibfield  {journal} {\bibinfo  {journal}
  {Astrophys. J.}\ }\textbf {\bibinfo {volume} {813}},\ \bibinfo {pages} {38}
  (\bibinfo {year} {2015})}\BibitemShut {NoStop}%
\bibitem [{\citenamefont {Tanaka}\ and\ \citenamefont
  {Hotokezaka}(2013)}]{tanaka_hotokezaka2013}%
  \BibitemOpen
  \bibfield  {author} {\bibinfo {author} {\bibfnamefont {M.}~\bibnamefont
  {Tanaka}}\ and\ \bibinfo {author} {\bibfnamefont {K.}~\bibnamefont
  {Hotokezaka}},\ }\href {\doibase 10.1088/0004-637X/775/2/113} {\bibfield
  {journal} {\bibinfo  {journal} {Astrophys. J.}\ }\textbf {\bibinfo {volume}
  {775}},\ \bibinfo {pages} {113} (\bibinfo {year} {2013})}\BibitemShut
  {NoStop}%
\bibitem [{\citenamefont {Kasen}\ \emph {et~al.}(2013)\citenamefont {Kasen},
  \citenamefont {Badnell},\ and\ \citenamefont {Barnes}}]{kasen_bb2013}%
  \BibitemOpen
  \bibfield  {author} {\bibinfo {author} {\bibfnamefont {D.}~\bibnamefont
  {Kasen}}, \bibinfo {author} {\bibfnamefont {N.~R.}\ \bibnamefont {Badnell}},
  \ and\ \bibinfo {author} {\bibfnamefont {J.}~\bibnamefont {Barnes}},\ }\href
  {\doibase 10.1088/0004-637X/774/1/25} {\bibfield  {journal} {\bibinfo
  {journal} {Astrophys. J.}\ }\textbf {\bibinfo {volume} {774}},\ \bibinfo
  {pages} {25} (\bibinfo {year} {2013})}\BibitemShut {NoStop}%
\bibitem [{\citenamefont {Hotokezaka}\ \emph {et~al.}(2016)\citenamefont
  {Hotokezaka}, \citenamefont {Wanajo}, \citenamefont {Tanaka}, \citenamefont
  {Banba}, \citenamefont {Terado},\ and\ \citenamefont
  {Piran}}]{hotokezaka_wtbtp2016}%
  \BibitemOpen
  \bibfield  {author} {\bibinfo {author} {\bibfnamefont {K.}~\bibnamefont
  {Hotokezaka}}, \bibinfo {author} {\bibfnamefont {S.}~\bibnamefont {Wanajo}},
  \bibinfo {author} {\bibfnamefont {M.}~\bibnamefont {Tanaka}}, \bibinfo
  {author} {\bibfnamefont {A.}~\bibnamefont {Banba}}, \bibinfo {author}
  {\bibfnamefont {Y.}~\bibnamefont {Terado}}, \ and\ \bibinfo {author}
  {\bibfnamefont {T.}~\bibnamefont {Piran}},\ }\href {\doibase
  10.1093/mnras/stw404} {\bibfield  {journal} {\bibinfo  {journal} {Mon. Not.
  R. Astron. Soc.}\ }\textbf {\bibinfo {volume} {459}},\ \bibinfo {pages} {35}
  (\bibinfo {year} {2016})}\BibitemShut {NoStop}%
\bibitem [{\citenamefont {Barnes}\ \emph {et~al.}(2016)\citenamefont {Barnes},
  \citenamefont {Kasen}, \citenamefont {Wu},\ and\ \citenamefont
  {Mart{\'i}nez-Pinedo}}]{barnes_kwm2016}%
  \BibitemOpen
  \bibfield  {author} {\bibinfo {author} {\bibfnamefont {J.}~\bibnamefont
  {Barnes}}, \bibinfo {author} {\bibfnamefont {D.}~\bibnamefont {Kasen}},
  \bibinfo {author} {\bibfnamefont {M.-R.}\ \bibnamefont {Wu}}, \ and\ \bibinfo
  {author} {\bibfnamefont {G.}~\bibnamefont {Mart{\'i}nez-Pinedo}},\ }\href
  {\doibase 10.3847/0004-637X/829/2/110} {\bibfield  {journal} {\bibinfo
  {journal} {Astrophys. J.}\ }\textbf {\bibinfo {volume} {829}},\ \bibinfo
  {pages} {110} (\bibinfo {year} {2016})}\BibitemShut {NoStop}%
\bibitem [{\citenamefont {Goodman}\ \emph {et~al.}(1987)\citenamefont
  {Goodman}, \citenamefont {Dar},\ and\ \citenamefont
  {Nussinov}}]{goodman_dn1987}%
  \BibitemOpen
  \bibfield  {author} {\bibinfo {author} {\bibfnamefont {J.}~\bibnamefont
  {Goodman}}, \bibinfo {author} {\bibfnamefont {A.}~\bibnamefont {Dar}}, \ and\
  \bibinfo {author} {\bibfnamefont {S.}~\bibnamefont {Nussinov}},\ }\href
  {\doibase 10.1086/184840} {\bibfield  {journal} {\bibinfo  {journal}
  {Astrophys. J.}\ }\textbf {\bibinfo {volume} {314}},\ \bibinfo {pages} {L7}
  (\bibinfo {year} {1987})}\BibitemShut {NoStop}%
\bibitem [{\citenamefont {Eichler}\ \emph {et~al.}(1989)\citenamefont
  {Eichler}, \citenamefont {Livio}, \citenamefont {Piran},\ and\ \citenamefont
  {Schramm}}]{eichler_lps1989}%
  \BibitemOpen
  \bibfield  {author} {\bibinfo {author} {\bibfnamefont {D.}~\bibnamefont
  {Eichler}}, \bibinfo {author} {\bibfnamefont {M.}~\bibnamefont {Livio}},
  \bibinfo {author} {\bibfnamefont {T.}~\bibnamefont {Piran}}, \ and\ \bibinfo
  {author} {\bibfnamefont {D.~N.}\ \bibnamefont {Schramm}},\ }\href {\doibase
  10.1038/340126a0} {\bibfield  {journal} {\bibinfo  {journal} {Nature
  (London)}\ }\textbf {\bibinfo {volume} {340}},\ \bibinfo {pages} {126}
  (\bibinfo {year} {1989})}\BibitemShut {NoStop}%
\bibitem [{\citenamefont {Mochkovitch}\ \emph {et~al.}(1993)\citenamefont
  {Mochkovitch}, \citenamefont {Hernanz}, \citenamefont {Isern},\ and\
  \citenamefont {Martin}}]{mochkovitch_him1993}%
  \BibitemOpen
  \bibfield  {author} {\bibinfo {author} {\bibfnamefont {R.}~\bibnamefont
  {Mochkovitch}}, \bibinfo {author} {\bibfnamefont {M.}~\bibnamefont
  {Hernanz}}, \bibinfo {author} {\bibfnamefont {J.}~\bibnamefont {Isern}}, \
  and\ \bibinfo {author} {\bibfnamefont {X.}~\bibnamefont {Martin}},\ }\href
  {\doibase 10.1038/361236a0} {\bibfield  {journal} {\bibinfo  {journal}
  {Nature (London)}\ }\textbf {\bibinfo {volume} {361}},\ \bibinfo {pages}
  {236} (\bibinfo {year} {1993})}\BibitemShut {NoStop}%
\bibitem [{\citenamefont {Paschalidis}\ \emph {et~al.}(2015)\citenamefont
  {Paschalidis}, \citenamefont {Ruiz},\ and\ \citenamefont
  {Shapiro}}]{paschalidis_rs2015}%
  \BibitemOpen
  \bibfield  {author} {\bibinfo {author} {\bibfnamefont {V.}~\bibnamefont
  {Paschalidis}}, \bibinfo {author} {\bibfnamefont {M.}~\bibnamefont {Ruiz}}, \
  and\ \bibinfo {author} {\bibfnamefont {S.~L.}\ \bibnamefont {Shapiro}},\
  }\href {\doibase 10.1088/2041-8205/806/1/L14} {\bibfield  {journal} {\bibinfo
   {journal} {Astrophys. J.}\ }\textbf {\bibinfo {volume} {806}},\ \bibinfo
  {pages} {L14} (\bibinfo {year} {2015})}\BibitemShut {NoStop}%
\bibitem [{\citenamefont {Kawanaka}\ \emph {et~al.}(2013)\citenamefont
  {Kawanaka}, \citenamefont {Piran},\ and\ \citenamefont
  {Krolik}}]{kawanaka_pk2013}%
  \BibitemOpen
  \bibfield  {author} {\bibinfo {author} {\bibfnamefont {N.}~\bibnamefont
  {Kawanaka}}, \bibinfo {author} {\bibfnamefont {T.}~\bibnamefont {Piran}}, \
  and\ \bibinfo {author} {\bibfnamefont {J.~H.}\ \bibnamefont {Krolik}},\
  }\href {\doibase 10.1088/0004-637X/766/1/31} {\bibfield  {journal} {\bibinfo
  {journal} {Astrophys. J.}\ }\textbf {\bibinfo {volume} {766}},\ \bibinfo
  {pages} {31} (\bibinfo {year} {2013})}\BibitemShut {NoStop}%
\bibitem [{\citenamefont {Lee}\ \emph {et~al.}(2005)\citenamefont {Lee},
  \citenamefont {Ramirez-Ruiz},\ and\ \citenamefont {Page}}]{lee_rp2005}%
  \BibitemOpen
  \bibfield  {author} {\bibinfo {author} {\bibfnamefont {W.~H.}\ \bibnamefont
  {Lee}}, \bibinfo {author} {\bibfnamefont {E.}~\bibnamefont {Ramirez-Ruiz}}, \
  and\ \bibinfo {author} {\bibfnamefont {D.}~\bibnamefont {Page}},\ }\href
  {\doibase 10.1086/432373} {\bibfield  {journal} {\bibinfo  {journal}
  {Astrophys. J.}\ }\textbf {\bibinfo {volume} {632}},\ \bibinfo {pages} {421}
  (\bibinfo {year} {2005})}\BibitemShut {NoStop}%
\bibitem [{\citenamefont {Setiawan}\ \emph {et~al.}(2006)\citenamefont
  {Setiawan}, \citenamefont {Ruffert},\ and\ \citenamefont
  {Janka}}]{setiawan_rj2006}%
  \BibitemOpen
  \bibfield  {author} {\bibinfo {author} {\bibfnamefont {S.}~\bibnamefont
  {Setiawan}}, \bibinfo {author} {\bibfnamefont {M.}~\bibnamefont {Ruffert}}, \
  and\ \bibinfo {author} {\bibfnamefont {H.-T.}\ \bibnamefont {Janka}},\ }\href
  {\doibase 10.1051/0004-6361:20054193} {\bibfield  {journal} {\bibinfo
  {journal} {Astron. Astrophys.}\ }\textbf {\bibinfo {volume} {458}},\ \bibinfo
  {pages} {553} (\bibinfo {year} {2006})}\BibitemShut {NoStop}%
\bibitem [{\citenamefont {Kagawa}\ \emph {et~al.}(2015)\citenamefont {Kagawa},
  \citenamefont {Yonetoku}, \citenamefont {Sawano}, \citenamefont {Toyanago},
  \citenamefont {Nakamura}, \citenamefont {Takahashi}, \citenamefont
  {Kashiyama},\ and\ \citenamefont {Ioka}}]{kagawa_etal2015}%
  \BibitemOpen
  \bibfield  {author} {\bibinfo {author} {\bibfnamefont {Y.}~\bibnamefont
  {Kagawa}}, \bibinfo {author} {\bibfnamefont {D.}~\bibnamefont {Yonetoku}},
  \bibinfo {author} {\bibfnamefont {T.}~\bibnamefont {Sawano}}, \bibinfo
  {author} {\bibfnamefont {A.}~\bibnamefont {Toyanago}}, \bibinfo {author}
  {\bibfnamefont {T.}~\bibnamefont {Nakamura}}, \bibinfo {author}
  {\bibfnamefont {K.}~\bibnamefont {Takahashi}}, \bibinfo {author}
  {\bibfnamefont {K.}~\bibnamefont {Kashiyama}}, \ and\ \bibinfo {author}
  {\bibfnamefont {K.}~\bibnamefont {Ioka}},\ }\href {\doibase
  10.1088/0004-637X/811/1/4} {\bibfield  {journal} {\bibinfo  {journal}
  {Astrophys. J.}\ }\textbf {\bibinfo {volume} {811}},\ \bibinfo {pages} {4}
  (\bibinfo {year} {2015})}\BibitemShut {NoStop}%
\end{thebibliography}
%
\end{document}